\documentclass[oneside,12pt,fleqn]{extarticle}

\usepackage{amsmath,amsfonts,amssymb,latexsym,ctable}
\usepackage{theorem}
\usepackage{verbatim}
\usepackage{dcolumn}
\usepackage{setspace}
\usepackage[USenglish]{babel}
\usepackage{fullpage}
\usepackage{caption}
\usepackage{subfig,graphicx}
\usepackage{rotating}
\usepackage{lscape,epsfig}

\usepackage{lineno}
\usepackage{appendix}
\usepackage{siunitx}
\usepackage{dcolumn}
\newcolumntype{.}{D{.}{.}{-1}}

\usepackage[pdftex,colorlinks=true,linkcolor=blue, citecolor=blue, urlcolor=blue]{hyperref}
\usepackage{natbib}

\usepackage{longtable}
\usepackage{hhline}
\usepackage{multirow}
\usepackage{setspace}

\newcommand{\lambdab}{\mbox{\boldmath \(\lambda\)}}
\newcommand{\xib}{\mbox{\boldmath \(\xi\)}}
\newcommand{\nub}{\mbox{\boldmath \(\nu\)}}

\renewcommand{\baselinestretch}{1.5}

\DeclareRobustCommand\IPCClongname{}

\begin{document}


\title{Climate-Related Disasters and the Death Toll\thanks{\footnotesize V. Chavez-Demoulin: Faculty of Business and Economics (HEC Lausanne) and Enterprise for Society (E4S) Center, University of Lausanne, CH-1015 Lausanne, Switzerland (e-mail: valerie.chavez@unil.ch). E. Jondeau: Faculty of Business and Economics (HEC Lausanne), University of Lausanne, Swiss Finance Institute, and Enterprise for Society (E4S) Center, CH-1015 Lausanne, Switzerland (e-mail: eric.jondeau@unil.ch). Linda Mhalla: Faculty of Business and Economics (HEC Lausanne), University of Lausanne, CH-1015 Lausanne, Switzerland (e-mail: linda.mhalla@unil.ch). The Enterprise for Society (E4S) Center, a joint initiative of EPFL, HEC Lausanne, and IMD, provided generous financial support. The authors thank Kevin Billy for the preparation of the database and preliminary work.}}

\date{September 2021}

\author{Val\'{e}rie Chavez-Demoulin, Eric Jondeau, and Linda Mhalla}

\maketitle
\bigskip

\begin{abstract}

\noindent With climate change accelerating, the frequency of climate disasters is expected to increase in the decades to come. There is ongoing debate as to how different climatic regions will be affected by such an acceleration. In this paper, we describe a model for predicting the frequency of climate disasters and the severity of the resulting number of deaths. The frequency of disasters is described as a Poisson process driven by aggregate CO$_2$ emissions. The severity of disasters is described using a generalized Pareto distribution driven by the trend in regional real gross domestic product (GDP) per capita. We predict the death toll for different types of climate disasters based on the projections made by the Intergovernmental Panel on Climate Change for the population, the regional real GDP per capita, and aggregate CO$_2$ emissions in the ``sustainable'' and ``business-as-usual'' baseline scenarios.

\renewcommand{\baselinestretch}{1.2}
\small

\noindent 


\noindent \textbf{Acknowledgment: } 

\end{abstract}

\newpage


\section{Introduction}\label{sec: Introduction}

The Intergovernmental Panel on Climate Change (\citealp{IPCC2021}) projects an increase in the frequency of climate-related disasters, as a result of climate change acceleration. The number of climate disasters has already increased to 6,641 in the last 20 years (2000--2019) from 3,656 in the previous 20 years (1980--1999) and 1,171 in the two decades before that (1960--1979). Most of the increase in the number of disasters corresponds to the increased number of floods and storms, although the frequency of landslides, heat waves, and wildfires also increased substantially over this period. In contrast, except for wildfires and heat waves, the number of deaths per disaster dramatically decreased over the same period. Consequently, the overall annual number of deaths due to climate related disasters has decreased to $26,779$ in the last 20 years from $33,006$ in the previous 20 years and $77,062$ in the two decades before that. 

In the long term, two components contribute to the death toll due to human-driven climate disasters. First, the frequency of climate-related disasters tends to increase over time because of climate change. Second, the severity of disasters results from the combination of two opposite trends. The increase in population contributes to the increase in the number of deaths when a disaster occurs. However, more developed countries are more likely to invest in disaster risk management, which helps reduce the number of deaths per disaster. As a result, the overall impact of climate change and economic development is not clear in the long run. Over our sample period (1960--2019), the annual number of disasters and world CO$_2$ emissions per capita exhibit a correlation of $56\%$; the number of deaths per disaster and world real gross domestic product (GDP) per capita exhibit a correlation of $-22\%$; however, the annual number of deaths has no significant correlation with either world CO$_2$ emissions or real GDP per capita.

The objective of this paper is twofold. First, we model the death toll due to different types of climate disasters over the last 60 years across the main geographical areas. We decompose the evolution of the number of deaths into the frequency and severity of the disasters using a state-of-the-art econometric methodology. This approach allows us to rely on well-established statistical models. We describe the number of disasters as a Poisson process, in which the frequency parameter is driven by aggregate CO$_2$ emissions per capita, which we use as a proxy for climate change. We describe the number of deaths per disaster, corrected for population growth, as drawn from a generalized Pareto distribution (GPD), whose parameters are driven by the regional real GDP per capita, which we use as a proxy for disaster risk management. Our analysis is based on disaster data from the Emergency Events Database (EM-DAT) \citep{GuhaSapirBelowHoyois2009}. We focus on six types of climate disasters (floods, storms, landslides, wildfires, heat waves, and cold waves) for seven regions defined by the World Bank. We also use demographic and economic data in addition to CO$_2$ emissions from the World Bank. We find that the sensitivities of the frequency and severity of disasters to their covariates depend on the disaster type and the affected region.  



Second, armed with this model, we project the number of deaths due to climate-related disasters. Our projections of covariates rely on Shared Socioeconomic Pathways (SSP) scenarios elaborated by the IPCC for its 6th Assessment Report (\citealp{IPCC6_2017, IPCC2021}). These scenarios describe some representative and consistent evolutions of demographic and economic drivers (population, GDP per capita) and infer the possible impact on climate of various mitigation scenarios (through the dynamics of CO$_2$ emissions). We focus on two particularly important scenarios. The ``sustainable'' Scenario SSP1 (``Sustainability -- Taking the Green Road'') is associated with low population growth, a high increase in GDP per capita, and a low increase in CO$_2$ emissions. In this scenario, all covariates point toward a decrease in the frequency and severity of disasters. In contrast, the ``business-as-usual" Scenario SSP3 (``Regional Rivalry -- A Rocky Road'') is associated with high population growth, low economic growth, and high CO$_2$ emissions. 

Our framework illustrates how policy decisions can help reduce the impact of climate change on the death toll through two mechanisms. On the one hand, a voluntary reduction in CO$_2$ emissions would be a way to mitigate climate change and reduce the number of climate-related disasters. On the other hand, real GDP per capita reflects the ability of authorities to adapt their country to new climate conditions. Better disaster risk management would help reduce the number of deaths per disaster. Our projections would therefore reflect the mitigation and adaptation strategies that could be implemented to address climate-change challenges. In the ``business-as-usual" scenario the overall number of disasters would increase by $33\%$ in 2040 and $60\%$ in 2100. The annual number of deaths due to climate-related disasters would increase by $50\%$ in 2040 and $33\%$ in 2100. The annual death toll in the ``sustainable'' scenario would be approximately 30 times lower than in the ``business-as-usual" scenario in 2100.


\medskip

In our model, we assume two relationships between the frequency and severity of climate-related disasters and covariates. On the one hand, we relate the frequency of disasters to climate change and more specifically to CO$_2$ emissions. This relationship has been made clear by \cite{IPCC2012}, \cite{ThomasAlbertPerez2013}, and \cite{IPCC2014} through three connections. First, greenhouse gas (including CO$_2$) emissions affect climate dynamics, especially temperature and the water cycle. Second, climate change impacts the frequency of climate-related hazards. Third, the frequency of climate-related hazards affects the risk of natural disasters. \cite{ThomasLopez2015} provide several references for all three linkages. The advantage of working with carbon emissions instead of, for instance, temperature is that policies and human behavior can directly alter carbon emissions. \cite{AndersonBausch2006} predict that a doubling of atmospheric CO$_2$ concentrations will triple the number of storms with the highest intensity. \cite{CaiEtAlii2014} present climate modeling evidence for a doubling in the occurrence of El Ni\~{n}o events in the future in response to greenhouse warming. \cite{MittnikSemmlerHaider2020} report empirical evidence that the sum of increases in CO$_2$ concentration has a positive and significant effect on the frequency of climate-related disasters.


On the other hand, we relate the severity of disasters in a given region to the GDP per capita of this region. The relationship between a country's development level and the ways in which the country is affected by disasters is analyzed in several papers. See, among others, \cite{Twigg2004}, \cite{KellettCaravaniPichon2014}, and \cite{KellenbergMobarak2011} for a survey of the literature. \cite{Kahn2005} finds that richer nations do not experience fewer natural disasters than poorer nations but richer nations do suffer less death from disasters. Thus, real GDP per capita has a negative and significant effect on the death toll from natural disasters for most of the considered disaster types. The interpretation is that richer countries can provide implicit disaster insurance through effective regulation and planning and by providing quality infrastructure. \cite{EscalerasRegister2008} study 146 tsunamis between 1966 and 2004 and find that GDP per capita captures the ability of the country to protect local populations through better-developed building codes or hazard-sensitive zoning. \cite{ToyaSkidmore2007} provide evidence that higher educational attainment, greater economic openness, and more complete financial systems have a negative effect on the number of human losses by natural disasters. Related to our study, \cite{FranzkeSentelles2020} investigate trends in continentally aggregated fatality data. They examined whether modes of climate variability affect the propensity of fatalities and find statistically significant increasing trends for heat waves and floods for worldwide aggregated data. Their methodology relies on a generalized Pareto distribution (GPD) model for fatalities where they identify covariates that affect the number of fatalities aggregated over all hazard types. They find no evidence for a significant direct impact from socioeconomic indicators but find that the number of fatalities is affected by tropical cyclone activity, sea surface temperature anomalies, and atmospheric teleconnection patterns.\footnote{Our projection analysis differs from the one performed by \cite{FranzkeSentelles2020}. They describe the number of deaths per year using a GPD model with climate-related covariates, such as tropical cyclone activity and atmospheric teleconnection patterns. Then, they project the number of deaths for long horizons assuming that the covariates do not change over time. In contrast, our projections rely on scenarios provided by \cite{IPCC6_2017}, which describe a consistent evolution of the covariates.}

\medskip

The remainder of the paper is organized as follows. In Section \ref{sec: Data}, we present the data used in our empirical analysis. In Section \ref{sec: Modeling}, we describe the methodology that we use for modeling the frequency and severity of climate disasters and present the estimation results. In Section \ref{sec: Projections}, we present the projections of the number of deaths due to climate disasters and discuss some of the potential issues raised by our approach. Section \ref{sec: Conclusion} concludes the paper.

\section{Data and Preliminary Analysis}\label{sec: Data}

\subsection{Data}

To measure the frequency and severity of climate-related disasters, we use the Emergency Events Database (EM-DAT), launched in 1988 by the Centre for Research on the Epidemiology of Disasters (CRED) (see \citealp{GuhaSapirBelowHoyois2009}).\footnote{The main objective of this database is to support humanitarian actions at national and international levels by rationalizing decision making for disaster preparedness, as well as providing a base for vulnerability assessment and priority setting. It has been used in several academic studies, such as \cite{Kahn2005}, \cite{FranzkeSentelles2020} and \cite{MittnikSemmlerHaider2020}. See \cite{BakkensenXiangyingZurita2018} and \cite{PanwarSen2020} for a comparison with alternative databases.} The database contains core data on the occurrence and impact of over 22,000 disasters since the start of the twentieth century.  For each disaster, it reports the number of deaths, the number of affected, the financial losses in U.S. dollars, and the insured losses in U.S. dollars. A disaster is reported in the database if it meets at least one of the following criteria: 10 or more people dead, 100 or more people affected, the declaration of a state of emergency, or a call for international assistance. 

The EM-DAT database contains the following disaster types, classified into five main categories (with the main subtypes in parentheses): geophysical (earthquakes, mass movements, volcanic activity), meteorological (storms, heat waves, cold waves), hydrological (floods, landslides), climatological (droughts, wildfires), and biological (epidemics). We focus on meteorological, hydrological, and climatological events, as they are directly related to climate change. We end up with two categories of disasters: those related to the impacts of climate change on the water cycle (storms, floods, and landslides) and those related to direct impacts on temperature (wildfires, heat waves, and cold waves).\footnote{We do not consider droughts in our analysis, although the occurrence of droughts may increase in the future in the context of climate change. The main reason why they are not included is the fact that droughts are far from punctual events and may last for several years in some particularly severe cases. Such events cannot be considered punctual events, and recording the number of deaths attributed to droughts is therefore a difficult task.}

Regarding the location of these disasters, we use the seven regions defined by the World Bank: East Asia and Pacific, Europe and Central Asia, Latin America and Caribbean, Middle East and North Africa, North America, South Asia, and Sub-Saharan Africa.\footnote{The map representing the seven regions is available on the following link: \url{https://datatopics.worldbank.org/world-development-indicators/the-world-by-income-and-region.html}.} As we cover a relatively long time period (1960--2019), we take the population evolution into account.\footnote{We start in 1960 to avoid any reporting bias: records of disasters become scarcer as we go back in time, as only the most severe events are systematically recorded.} All death numbers are adjusted for the population of the country where the disaster occurred. This is done by dividing the number of deaths of a given event by the country's population at the time the disaster occurred and then multiplying it by the most recent estimate of this country's population at our disposal (end of 2019). All population figures are also taken from the World Bank (code SP.POP.TOTL). We use piecewise cubic spline interpolation to transform annual data to a daily frequency so that it matches the exact dates of our punctual events. Then, we rescale the number of deaths corresponding to the population at the end of 2019.

We measure the level of development of a given region using the GDP per capita based on purchasing power parity (PPP) in constant 2017 USD (code NY.GDP.PCAP.PP.KD). The data are aggregated at a regional level.\footnote{PPP GDP is the gross domestic product converted to international dollars using purchasing power parity rates. An international dollar has the same purchasing power over GDP as the U.S. dollar has in the United States. We use GDP per capita at PPP in constant 2017 USD to be consistent with the projections provided by the IPCC. In the World Bank database, these data are only available from 1990 onward. Before 1990, we backfilled these series using the GDP per capita in constant 2010 USD.} CO$_2$ emissions per capita (code EN.ATM.CO2E.PC) are those stemming from the burning of fossil fuels and the manufacturing of cement. These emissions include carbon dioxide produced during the consumption of solid, liquid, and gas fuels and gas flaring. They are measured in metric tons per capita.

Our projections are based on Shared Socioeconomic Pathways prepared by the IPCC for its 6th Assessment Report (\citealp{IPCC6_2017}). SSP projections of population are described in \cite{KCLutz2017}. They use a multidimensional demographic model to project national populations based on alternative assumptions on future fertility, mortality, migration, and educational transitions. These projections are designed to be consistent with five SSP storylines. There are three sets of economic (GDP per capita) projections for each SSP (\citealp{CrespoCuaresma2017}, \citealp{DellinkEtAlii2017}, \citealp{LeimbachEtAlii2017}), which were developed together with the demographic projections to maintain consistency with the education and ageing assumptions. We use the projections provided by \cite{DellinkEtAlii2017} for all SSPs to ensure consistency. Scenarios SSP1, SSP2, and SSP3 correspond to low, medium, and high challenges to mitigation and adaptation, respectively. SSP4 and SSP5 are intermediary scenarios, corresponding to low challenges to mitigation, high challenges to adaptation and to high challenges to mitigation and low challenges to adaptation, respectively. All five scenarios are summarized by \cite{IPCC6_2017}.\footnote{These scenarios are described in detail by \cite{vanVuurenEtAlii2017}, \cite{FrickoEtAlii2017}, \cite{FujimoriEtAlii2017}, \cite{CalvinEtAlii2017}, and \cite{KrieglerEtAlii2017}, respectively. Projections of population, real GDP per capita, and CO$_2$ emissions were obtained from the SSP Database, available on the International Institute for Applied System Analysis website, \url{https://tntcat.iiasa.ac.at/SspDb}.} The demographic, socioeconomic, and carbon emission scenarios underlying the IPCC's 6th Assessment Report were established based on data back to 2010 or so. As we now have access to data up to 2019, we adapt these projections with our most recent data. 

\subsection{Historical Frequency and Severity}\label{sec: Historical}

We characterize the human impact on climate-related disasters through two variables: the frequency of disasters per year and the number of deaths per disaster. The number of deaths is made comparable over time by correcting for the change in a country's population. We define the rescaled number of deaths that occurred due to disaster $i$ in country $c$ and year $t$ as $\tilde{D}_{i,t}^{(c)}=D_{i,t}^{(c)} \times \frac{P_{T}^{(c)}}{P_{t}^{(c)}}$, where $D_{i,t}^{(c)}$ is the actual number of deaths during the disaster and $P_{T}^{(c)}/P_{t}^{(c)}$ measures the increase in the country's population between years $t$ and $T=2019$. To define the number of deaths in a given region $r$, we group disasters for all countries in the same region and use the notation $\tilde{D}_{i,t}^{(r)}$ to define the rescaled number of deaths due to disaster $i$ in region $r$ and year $t$. Then, the rescaled number of deaths in region $r$ and year $t$ can be described as:
\begin{equation}
\tilde{S}_{t}^{(r)} = \sum_{i=1}^{N_{t}^{(r)}} \tilde{D}_{i,t}^{(r)}, \label{PGPD}
\end{equation}
where $N_{t}^{(r)}$ is the number of disasters of a given type in the corresponding region and year.

Tables \ref{tab: Stats 1} and \ref{tab: Stats 2} report the annual frequency and severity of disasters by region and by category over the full sampling period (1960--2019) and over the three subsamples of 20 years each. For the severity measure, we report the average (rescaled) number of deaths per disaster and the annual number of deaths, which combines the first two metrics. Table \ref{tab: Stats 1} provides measures for floods, storms, and landslides, while Table \ref{tab: Stats 2} shows the values for wildfires, heat waves, and cold waves. The tables clearly demonstrate an increase in the number of disasters and a decrease in the number of deaths per disaster over the sampling period. Even if the data may suffer from some underreporting at the beginning of the period, these changes are particularly pronounced for all disaster types, with the exception of wildfires and heat waves. 

Disasters related to the water cycle are by far the most lethal, as they represent $92.5\%$ of the death toll over the sampling period (Table \ref{tab: Stats 1}). All three types of disasters are characterized by a substantial increase in the annual number of disasters and a decrease in the number of deaths per disaster. Overall, the annual number of deaths has dramatically decreased over the sampling period, so that they correspond to $67\%$ of the death toll in 2000--2019.

Floods, which consist of riverine, coastal, and flash floods, represent $47\%$ of all climate-related disasters between 1960 and 2019. The total number of floods increased from $21$ per year over the 1960--1979 period to $161$ over the 2000--2019 period. East Asia and Pacific is the most often affected region, with on average of $20.7$ floods per year. In contrast, the average number of deaths due to floods decreased from $493$ over the 1960--1979 period to $37$ over the 2000--2019 period. The most impacted region is South Asia with $4,260$ deaths per year on average over the sampling period.

Storms, which also include hurricanes and cyclones, represent $37\%$ of all disasters over the sampling period. The number of storms increased from $25$ per year over the 1960--1979 period to $102$ over the 2000--2019 period. The East Asia and Pacific region is again the most often affected. The average number of deaths due to storms substantially decreased from $2,585$ over the 1960--1979 period to 108 over the 2000--2019 period. South Asia was by far the most impacted region between 1960 and 1999, as it accounted for more than $85\%$ of the death toll. In the last 20-year period, most deaths from storms were in East Asia and Pacific.

Landslides include mudslides, avalanches, and rockfalls but exclude dry mass movement. They correspond to $6.6\%$ of all climate-related disasters over the full sampling period. The number of deaths due to landslides also decreased from $367$ per disaster in 1960--1979 to $55$ in 2000--2019.

The number of disasters directly related to temperature also experienced an increase, although they started from a much lower number. Importantly, the number of deaths per disaster increased or barely decreased over the sampling period, so that overall, the death toll due to these disasters increased over time.

Wildfires, which include both forest and land fires, constitute a less frequent ($4.1\%$ of all disasters) and impactful disaster type. However, in recent years, several large events have brought them more to the public's attention. In the past two decades, $30\%$ of wildfire events (and $69\%$ of economic losses due to wildfires) occurred in North America and $30\%$ have occurred in Europe and Central Asia. In addition, the number of deaths per disaster remained stable over time and the average death toll was higher in 2000--2019 than in 1960--1979. 

Extreme temperature events include heat and cold waves. They represent only $1.9\%$ and $3.5\%$ of all disasters, respectively, although their frequency has increased over the sampling period. The average annual number of deaths has substantially increased for heat waves, from $147$ over the 1960--1979 period to $7,870$ over the 2000--2019 period. This increase is partly due to the heat waves in Europe in 2003 and in Russia in 2010, which caused $72,210$ and $55,736$ deaths, respectively. 

The frequency of cold waves has also increased, from less than one event per year and per region in 1960--1979 to more than 14 in 2000--2019. Even if the number of deaths per disaster has slightly decreased in the recent period, the overall severity of cold waves has increased in the recent period. These aggregate results are in contrast to those reported by \cite{GasparriniEtAlii2015}, who report that many more temperature-attributable deaths were caused by cold ($7.29\%$) than by heat ($0.42\%$).

We compute the annual number of deaths as the product of the number of disasters and the average number of deaths per disaster. For floods, storms, and landslides, we find that despite the increase in the frequency of disasters, the annual number of deaths decreases because of the decrease in severity. In contrast, for wildfires, heat waves, and cold waves, we observe an increase in the annual number of deaths.\footnote{This empirical evidence differs from the results in \cite{UNDRR2020}. The reason for this difference is because this report included earthquakes in its analysis.}


\begin{center}
[Insert Tables \ref{tab: Stats 1} to \ref{tab: Stats 2} here]
\end{center}

As this preliminary analysis reveals, the total death toll results from the combination of two opposite trends: except for wildfires and heat waves, the increase in the frequency of disasters, probably due to climate change, is partly compensated for by the decrease in the severity of the disasters, which can be explained by improved risk management. As an expected increase in the world population in most scenarios would positively affect the number of deaths, it is not clear what the overall impact would be on the number of deaths. The model described in the next section aims to disentangle the effect of the various drivers.

\section{Modeling of the Frequency and Severity of Disasters}\label{sec: Modeling}

\subsection{The Model}\label{sec: PGPD}

The first objective of our paper is to disentangle the evolution of the rescaled death toll per disaster $\tilde{D}_{i,t}^{(r)}$ and the evolution of the number of disasters per year $N_t^{(r)}$ at the regional level and for a given disaster type. The combination of these two numbers informs us about the annual number $\tilde{S}_t^{(r)}$ of deaths due to disasters. 



The rescaled number of deaths is a highly asymmetric process, usually displaying a heavy right tail because a few disasters may result in a very large number of deaths. For this reason, we describe the properties in the number of deaths using extreme value theory (EVT).\footnote{Although the number of deaths is discrete by nature, in practice it can be considered a continuous process, in particular because it is rescaled to the population size. Therefore, the theoretical developments of EVT for continuous distributions are still useful to describe the tail of discrete but large data values.}

There are two main approaches to describe the properties and dynamics of extreme events: (1) The magnitude of events over subsamples (say, over one-year sampling intervals) can be described using the block maxima approach. In this approach, we consider a sample of maxima over one-year intervals and apply the generalized extreme value (GEV) distribution to the resulting sample of maxima. The GEV is the limiting distribution of the rescaled maximum of a sequence of independent and identically distributed (iid) variables coming from most common continuous distributions (Fisher-Tippett theorem). (2) One can also consider the worst events over the full sampling period by fixing a high threshold over which the exceedance points define the extreme values. In this peaks-over-threshold, or POT method, the exceedance sizes or excesses (i.e., the data value minus the threshold) form a sample that can be supposed to come from the generalized Pareto distribution (GPD) (\citealp{BalkemaDeHaan1974} and \citealp{Pickands1975} theorem). The models in both approaches are different but result in the same measure of tail heaviness described by the tail index. See \cite{DavisonSmith1990}, \cite{Chavez-DemoulinDavison2005}, and \cite{ChavezDemoulinEmbrechtsHofert2014}.

As our sample covers a large number of years but a limited number of events per year, we focus on the second approach, the  POT method. Following \cite{BalkemaDeHaan1974} and \cite{Pickands1975} theorem, the exceedance sizes $\tilde{D}_{1,t}^{(r)}-u, \cdots, \tilde{D}_{N_t^{(r)},t}^{(r)}-u$ over a given threshold $u$ approximately follow (independent of $N_t^{(r)}$) a GPD with tail index $\xi_t^{(r)} > -1$ and scale parameter $\beta_t^{(r)}>0$. The associated distribution function, allowing for time-dependent parameters, is given by
\begin{equation}
G_{\xi_t^{(r)},\beta_t^{(r)}}(y)= \left\{
\begin{array}{ll}
1-(1+\xi_t^{(r)} y/\beta_t^{(r)})^{-1/\xi_t^{(r)}} & \text{if } \xi_t^{(r)} \neq 0 \\
1-\exp(-\xi_t^{(r)} y/\beta_t^{(r)}) & \text{if } \xi_t^{(r)} = 0 
\end{array}
\right.
\end{equation}
for $y \geq 0$, if $\xi_t^{(r)} \geq 0$, and $y \in [0,-\beta_t^{(r)}/\xi_t^{(r)}]$, if $\xi_t^{(r)}<0$. As our database contains only disasters that are extreme events of interest, the threshold $u$ is naturally set at $u=0$. Therefore, any disaster resulting in at least one death is considered an extreme event, and we model its associated death toll. The scale parameter $\beta_t^{(r)}$ describes the dispersion of the extreme realizations, while $\xi_t^{(r)}$ controls the shape of the distribution and describes the heaviness of the right tail. The GPD is a valid distribution for exceedances provided $\xi_t^{(r)} >-1$. When $\xi_t^{(r)} \geq 1$, the process has an infinite mean. To ensure that parameters used in the estimation of the GPD are orthogonal with respect to the Fisher information matrix, we reparametrize $\beta_t^{(r)}$ by $\nu_t^{(r)}=\log((1+\xi_t^{(r)})\beta_t^{(r)})$, which is orthogonal to $\xi_t^{(r)}$ (see \citealp{ChavezDemoulinEmbrechtsHofert2014}).

The counting process $N_t^{(r)}$ is typically described by a Poisson process $Poi(\lambda_t^{(r)})$, where $\lambda_t^{(r)}>0$ denotes the intensity parameter. Intensity $\lambda_t^{(r)}$ may be homogeneous if it is constant, or nonhomogeneous if it varies over time. The probability function is defined by $\Pr[N_t^{(r)}=n]=\frac{\lambda_t^{(r)n}}{n!}\exp(-\lambda_t^{(r)})$.


The assumption that the frequency and severity of disasters are stationary is unlikely to hold over a long period of time, even once the population dynamics are taken into account. In fact, as we discussed in the introduction, there are two potentially contradicting tensions. First, the frequency of disasters may increase over time because of climate change. Second, we expect disaster risk management to improve over time as countries become richer; therefore, the number of deaths per disaster should decrease over time. The resulting annual death toll will depend on the strength of these two forces. For these reasons, we allow the parameters of the model to vary over time according to a set $\mathbf{x}_t=\{x_{1,t},\ldots,x_{q,t}\}$ of $q$ covariates.

For each disaster type, we assume that $\lambda_t$, $\xi_t$, and $\nu_t$ are of the form:
\begin{eqnarray}
\log(\lambda_t) &=& \lambda_{0} + \lambdab_{1}^{\top} \mathbf{x}_t + \epsilon_{\lambda,t}, \label{lambda_t}\\
\xi_t &=&  \xi_{0} + \xib_{1}^{\top} \mathbf{x}_t + \epsilon_{\xi,t},\label{xi_t}\\
\nu_t &=& \nu_{0} + \nub_{1}^{\top} \mathbf{x}_t + \epsilon_{\nu,t},\label{nu_t}
\end{eqnarray}
where $\epsilon_{\lambda,t}$, $\epsilon_{\xi,t}$, and $\epsilon_{\nu,t}$ are Gaussian noises with zero mean.\footnote{We could alternatively consider nonparametric models to describe the dynamics of $\lambda_t$, $\xi_t$, and $\nu_t$. However, this approach would be less suited to predict the future evolution of the severity and frequency of disasters.} The scale parameter $\beta_t$ is recovered from $\beta_t=\exp(\nu_t)/(1+\xi_t)$. 

In our empirical analysis, we consider a few variables for $\mathbf{x}_t$. The number of candidates is limited by our objective to project the death toll for alternative IPCC scenarios. To explain the increase in the frequency of disasters, we consider the (aggregate or regional) CO$_2$ emissions per capita, in level or in growth rate (possibly smoothed over several years, as in \citealp{MittnikSemmlerHaider2020}). To explain the severity of disasters, we consider the (aggregate or regional) real GDP per capita, in level or in growth rate (possibly smoothed over several years). In the selection process, we also impose some socioeconomic restrictions. In particular, we do not consider specifications where the severity would increase for richer countries, as previous empirical evidence suggests the opposite. Eventually, we use $\log($world CO$_2$ emissions per capita$)$ as a covariate for the frequency parameter $\lambda_t$ and $\log($regional real GDP per capita$)$ as covariates for the severity parameters $\xi_t$ and $\nu_t$. The regional effect is captured by region-dependent parameters, as indicated in Equations (\ref{lambda_t})--(\ref{nu_t}). We note that as covariates are defined annually, so are the parameters $\lambda_t$, $\xi_t$, and $\nu_t$. Estimates of the coefficients are obtained by maximum likelihood. We apply this approach for each disaster type while pooling data for all regions. 

The final step of our analysis consists in using the model estimates and covariate projections to infer the expected number of disasters and the expected number of deaths (per disaster and per year) for a given time horizon $h$, say 20 years. With Equations (\ref{lambda_t})--(\ref{nu_t}), we project our model's parameters, which we use to project the number of disasters and the number of deaths per disaster. For a given region $r$, when $\xi_{T+h}^{(r)}<1$, we compute the expected value of the number of deaths per disaster as
\begin{equation}
E_T[\tilde{D}_{T+h}^{(r)}]= \frac{\beta_{T+h}^{(r)}}{1-\xi_{T+h}^{(r)}}. \label{eq: Expected value}
\end{equation}
When $\xi_{T+h}^{(r)} \geq 1$, the expected value of the GPD is infinite and Equation (\ref{eq: Expected value}) no longer applies. In such a case, we use the median number of deaths.

Once the rescaled expected number of deaths in year $T+h$, $E_T[\tilde{D}_{T+h}^{(r)}]$, is obtained, we transform the prediction to account for the predicted population in the region, consistent with the given IPCC scenario. Eventually, we define the projection of the number of deaths in year $T+h$ as $E_T[D_{T+h}^{(r)}]=E_T[\tilde{D}_{T+h}^{(r)}]\times \frac{P_{T+h}^{(r)}}{P_{T}^{(r)}}$, where $P_{T+h}^{(r)}$ denotes the projection of the population of region $r$ in year $T+h$.\footnote{In our projections, we assume that the population in all countries in a given region has the same growth as the population of the region projected by the IPCC.} Finally, the projection of the annual number of deaths is obtained as: $E_T[S_{T+h}^{(r)}]= E_T[N_{T+h}^{(r)}] \times E_T[D_{T+h}^{(r)}]$.

\medskip

All our projections and the associated uncertainties are based on a parametric bootstrap procedure. For one simulation $b$, for $b=1,\ldots,B$, we draw from a Poisson distribution (with intensity $\hat{\lambda}_{t}^{(r)}$ in region $r$) a sample of 60 annual numbers of disasters, $\{n_{t,b}^{(r)}\}_{t=1960}^T$. Then for each year $t$, we draw from a GPD (with parameters $\hat{\beta}_{t}^{(r)}$ and $\hat{\xi}_{t}^{(r)}$) a sample of (rescaled) number of deaths $\{\tilde{d}_{i,t,b}^{(r)}\}_{i=1}^{n_{t,b}^{(r)}}$. Then, based on this $b$-th bootstrap sample, we reestimate the regression model parameters and get estimates $\hat{\lambda}_{t,b}^{(r)}$, $\hat{\beta}_{t,b}^{(r)}$, and $\hat{\xi}_{t,b}^{(r)}$. Relying on the projections of the covariates by the IPCC, we finally predict the expected number of disasters $E_T[N_{T+h,b}^{(r)}]=\hat{\lambda}_{T+h,b}^{(r)}$ and the expected number of deaths $E_T[D_{T+h,b}^{(r)}]=\hat{\beta}_{T+h,b}^{(r)}/(1-\hat{\xi}_{T+h,b}^{(r)}) \times P_{T+h}^{(r)}/P_{T}^{(r)}$, where $\hat{\lambda}_{T+h,b}^{(r)}$,  $\hat{\xi}_{T+h,b}$, and $\hat{\beta}_{T+h,b}$ are the predicted parameters. 
We repeat this procedure many times ($B=10,000$) to obtain an uncertainty interval for all our projections. In the tables, we report the median values over the $B$ simulations of the various numbers.

\subsection{Estimation Results}\label{sec: EstRes}

We now present and comment on the model estimates for the frequency (Poisson distribution) and severity (GPD) of the death process. These results are reported in Tables \ref{tab: Parameter_Estimates_1} and \ref{tab: Parameter_Estimates_2}, where the first table corresponds to floods, storms, and landslides, while the second table corresponds to wildfires, heat waves, and cold waves. For each disaster type, we analyzed several specifications and selected the model with the best fit according to likelihood ratio tests. 

We start with the estimation of the Poisson model for the number of disasters per year. In this generalized linear model, the log-intensity depends linearly on the log of worldwide CO$_2$ emissions, as suggested, for instance, by \cite{ThomasLopez2015} and \cite{MittnikSemmlerHaider2020}. The increase in $\lambda_t^{(r)}$ occurring over a time horizon of $h$ years and for region $r$ is computed as $\lambda_{T+h}^{(r)}/\lambda_T^{(r)}=(CO_{2,T+h}/CO_{2,T})^{\lambda_{1,r}}$. This expression allows us to compare the predictions of the model with sample measures. For instance, in the case of floods in East Asia and Pacific, the annual number of disasters is $17.3$ in the period 1980--1999, with average aggregate CO$_2$ emissions of $4.11$ metric tons per capita. In the subsequent period 2000--2019, emissions increased to $4.61$ metric tons, which predicts a number of disasters of $17.3 \times (4.61/4.11)^{5.61}=33.3$. The realized number of disasters is equal to $39.2$, which represents an underestimation by $15\%$. As the adjusted $R^2$ reported in the table reveals, the quality of the fit for the Poisson model is relatively good for floods, storms, and landslides, with values between $38\%$ and $58\%$. For wildfires, heat waves, and cold waves, given the large number of years in which no disaster occurred, the fit quality is lower, with an adjusted $R^2$ ranging between $15\%$ and $27\%$.

The table also suggests that the sensitivities of disaster frequency to aggregate CO$_2$ emissions are almost always positive and significant, with only two exceptions for wildfires. Sensitivities are also higher for floods and cold waves, which are the two disaster types with the largest increase in the number of events over the sampling period. Sensitivities for storms and landslides indicate that the number of disasters is driven by the same trend in most regions, although to a lesser extent for landslides. Last, our estimates demonstrate that wildfires and heat waves are least affected by increased in CO$_2$ emissions. In particular, the number of disasters in Middle East and North Africa and in Sub-Saharan Africa are essentially insensitive to CO$_2$ emissions. This result is confirmed by the very low frequency of such disasters in these regions.

We now turn to the modeling of the severity of disasters, i.e., the estimation of the GPD parameters. This step is more challenging, as it is based on extreme realizations, which are by nature difficult to describe. Not surprisingly, our preferred specifications of the scale parameter $\nu_t^{(r)}$ and the tail index $\xi_t^{(r)}$ include more covariates for floods, storms, and landslides because these disaster types have more events ($3,658$, $2,926$, and $672$ observations, respectively) than wildfires, heat waves, and cold waves ($178$, $176$, and $294$, respectively). For each model, we systematically test the null hypothesis of a less sophisticated specification. See Appendix \ref{app: LRTests} for details on the LR tests.

Regarding the modeling of the scale parameter $\nu_t^{(r)}$, we find that the regional GDP per capita has a negative and significant coefficient in most disaster types and regions. This result is consistent with the evidence reported by \cite{Kahn2005}, which again suggests that richer countries suffer fewer deaths from disasters. For wildfires and heat waves, the sensitivities are negative but in general insignificant. These estimates suggest that an improvement in economic conditions has not been sufficient to reduce the severity of these disaster types. Worldwide, the annual number of deaths due to heat waves has increased over the last 40 years.

The modeling of the tail index $\xi_t^{(r)}$ is even more challenging, as it corresponds to the fundamental properties of extreme events. As we introduce covariates in the dynamics of the GPD parameters, we ensure that $\xi_t^{(r)}$ is always larger than $-1$ and that the log-likelihood is always well defined.\footnote{For cold waves, we use the log-transform $\log(1+\xi_t^{(r)})$ to ensure that $\xi_t^{(r)}>-1$ for all $t$.} As we do not impose any upper limit on $\xi_t^{(r)}$, the tail index is sometimes larger than $1$ in our sampling period, implying an infinite expected number of deaths. Such estimates typically occur in cases when a few events occurred with many deaths, such as in South Asia for storms at the beginning of the sampling period or in Europe and Central Asia for heat waves at the end of the sampling period. For floods, storms, and landslides, we find that $\xi_t^{(r)}$ depends negatively on the regional GDP per capita. As GDP per capita tends to increase in all regions in IPCC projections, it turns out that for all three disaster types, the predicted $\xi_{T+h}^{(r)}$ parameters are always smaller than 1, which allows us to compute an expected number of deaths. For wildfires, heat waves, and cold waves, only regional dummies were found to explain differences across regions. The only case with $\xi_{T+h}^{(r)}>1$ is the case of heat waves in Europe and Central Asia, with an estimated tail index equal to $0.581+1.964=2.545$, so that the expected number of deaths is infinite and we will report the median values instead.\footnote{In the projections that we report in the subsequent tables, we will indicate with a star cases where the median value is used.}

\section{Projections of the Death Toll}\label{sec: Projections}

Based on the models described in the previous section, we now discuss the projections of the covariates and the projections of the annual number of disasters, the number of deaths per disaster, and the annual number of deaths for the period from 2040 to 2100. 

\subsection{Covariates Projections}\label{sec: Covariates Projections}

\begin{figure}[h!]
\caption{World historical data and projections}
\vspace{-1cm}

\begin{center}
\bigskip
\includegraphics[scale=1,trim= 2cm 8.5cm 2cm 8cm,clip]{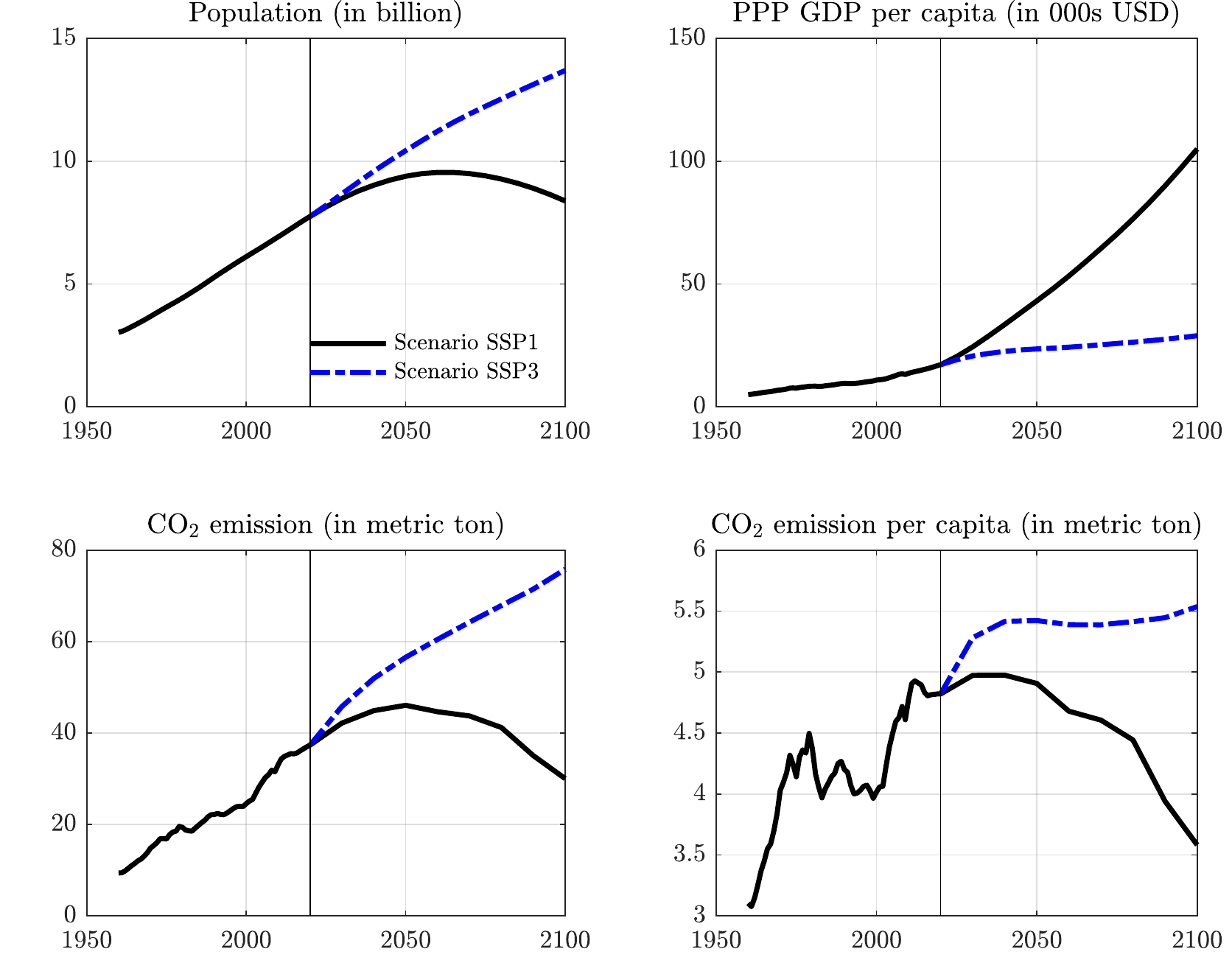} 
 \label{fig: World Covariates}
\end{center}

\end{figure}

As explained in Section \ref{sec: Data}, we rely on Scenarios SSP1 and SSP3 described in \cite{IPCC6_2017}. It is worth emphasizing that these scenarios have been defined using relatively old data: the scenarios for population and GDP per capita are based on data from 2010, with projections starting in 2015; the scenarios for CO$_2$ per capita are based on data from 2000, with projections starting in 2005. Our data end in 2019, and there is some discrepancy between the projections made by \cite{IPCC6_2017} and the actual data in 2019.\footnote{For instance, the world population in 2019 is $7.67$ billion people, while Scenarios SSP1 and SSP3 predicted a population in 2020 of $7.52$ and $7.69$ billion people, respectively. For GDP per capita, the PPP adjustment made by \cite{IPCC6_2017} is based on 2005 USD, while current World Bank estimates are based on 2017 USD.} To cope with this difference, we adapt the projections made in the three scenarios to the new data measured in 2019. Figure \ref{fig: World Covariates} displays the historical evolution and our projections in both scenarios for the world population, the aggregate real GDP per capita, and the aggregate CO$_2$ emissions per capita. The vertical line corresponds to 2020 when our own projections start. Population and GDP per capita can be compared with Figures 2-A  and 2-D and CO$_2$ emissions (in level) can be compared with Figure 5-A in \cite{IPCC6_2017}. Most of the discrepancies between the plots are due to using different starting points. 

In our predictive model, we use regional real GDP per capita, as we expect disaster risk management processes to be driven by regional economic development. The main characteristics of the narrative associated with the two scenarios are summarized in the figure. In Scenario SSP1, population growth is under control, and the fight against climate change is coordinated at an international level so that CO$_2$ emissions are substantially reduced after a peak in 2050. As a consequence, real GDP per capita greatly increases over the period, with emerging countries catching up to developed countries. This ``sustainable'' scenario is clearly very optimistic. In Scenario SSP3, in contrast, lack of cooperation and coordination is associated with high population growth in developing countries and nondecreasing CO$_2$ emissions per capita. As a result, GDP per capita remains close to its current level, reflecting a failure in adaptation and mitigation policies. This scenario broadly corresponds to the ``business-as-usual'' scenario.


\subsection{Regional Projections}\label{sec: Regional Projections}

We now consider, for each disaster type, the projections of the expected number of disasters, the expected number of deaths per disaster, and the annual expected number of deaths. Tables \ref{tab: Projections_1} and \ref{tab: Projections_2} are associated with the ``sustainable'' Scenario SSP1 (floods, storms, and landslides in Table \ref{tab: Projections_1} and wildfires, heat waves, and cold waves in Table \ref{tab: Projections_2}) and Tables \ref{tab: Projections_3} and \ref{tab: Projections_4} are associated with the ``business-as-usual'' Scenario SSP3 (floods, storms, and landslides in Table \ref{tab: Projections_3} and wildfires, heat waves, and cold waves in Table \ref{tab: Projections_4}). Panel A corresponds to the average over the most recent period (2015--2019), which we use as a reference level. Panels B to E correspond to projections from 2040 to 2100 with 20-year intervals.\footnote{The number of disasters and annual number of deaths for the world are computed by summing regional numbers. When the projected mean is replaced by the median, we use the median in the sum. The number of deaths per disaster is computed as the annual number of deaths divided by the number of disasters.} Appendix \ref{app: Regional Results} consists in bar charts that allow us to visualize the projections of the expected annual number of disasters and the expected annual number of deaths, for each region and each disaster type.

We start with the ``sustainable'' Scenario SSP1. As we identified in Section \ref{sec: EstRes}, for floods, storms, and landslides, the frequency of disasters depends positively on aggregate CO$_2$ emissions per capita. As a consequence, the number of disasters substantially reduces in Scenario SSP1, after the peak around 2050 for CO$_2$ emissions. In 2040, the number of disasters is expected to increase by approximately $55\%$ for floods and storms and by $10\%$ for landslides. As the number of floods is found to be more sensitive to CO$_2$ emissions, this number is expected to continue to increase until it reaches a peak and then decreases afterwards. By the end of the century, the number of floods is expected to be approximately one-fourth of the current number. We also note that, as the East Asia and Pacific region has the highest sensitivity of the number of disasters to CO$_2$ emissions for the three types of disasters, this region would be most affected in the next 20 years and would benefit the most from emission reduction in the subsequent period.

Regarding the severity of disasters, the two factors that affect the number of deaths (population growth and sensitivity to GDP per capita) are region dependent. For this reason, in some regions, the reduction in severity is not as pronounced as in other regions. For instance, in Sub-Saharan Africa, the number of deaths from storms would decrease relatively less because its population would increase more and the sensitivity to GDP per capita would be lower than in other regions. In contrast, the reduction would be substantial for the number of deaths due to landslides. An opposite dynamic is obtained for East Asia and Pacific. The decrease in the number of deaths would be larger for storms than for landslides.

In the ``sustainable'' Scenario SSP1, for floods, storms, and landslides, the annual number of deaths would benefit from a substantial decrease in the severity of disasters, which would compensate the increase in the number of disasters. The annual death toll would be reduced by approximately $60\%$. By the end of the century, the number of disasters would also decrease by $62\%$ relative to the 2015--2019 period due to the reduction in CO$_2$ emissions per capita. As the number of deaths per disaster would continue to decrease, the overall annual death toll would be reduced by $98\%$ for all three disaster types. 

For wildfires, heat waves, and cold waves, the projections in SSP1 are less favorable (Table \ref{tab: Projections_2}). The main reason is that the severity of these disaster types benefits less from the increase in regional GDP per capita. As the table reveals, in 2040, the number of disasters would increase by more than $100\%$ due to climate change, with a large increase in the number of cold waves. However, the number of deaths would benefit from the improvement in GDP per capita, with a substantial reduction in the number of deaths due to heat waves. However, the overall death toll would increase by almost $10\%$, mainly due to cold waves. At the end of the century, the number of disasters would decrease because of the decrease in CO$_2$ emissions, and the number of deaths per disaster would decrease because of the decrease in population. Overall, the annual number of deaths decreases by $88\%$, i.e., not as quickly as for the first three disaster types. In 2100, deaths due to heat waves would represent $63\%$ of the total death toll due to climate-related events.

\bigskip

The results for Scenario SSP3 reported in Tables \ref{tab: Projections_3} and \ref{tab: Projections_4} reflect the more pessimistic scenario. As CO$_2$ emissions do not decrease in this scenario, the number of disasters continues to increase for all disaster types. The expected increase in the number of disasters in 2040 is above $130\%$ on average, with an increase in all disaster types and regions. The number of disasters is expected to increase slightly for landslides and heat waves (below $100\%$), substantially for floods, storms, and wildfires (below $200\%$), and even more so for cold waves. The number of floods would be high in Latin America and Caribbean and in Sub-Saharan Africa; the number of storms would increase in East Asia and Pacific; the number of heat waves and cold waves would jump in Europe and Central Asia. By the end of the century, because of the continuous increase in CO$_2$ emissions, the number of disasters would remain high, approximately $150\%$ higher than over the 2015--2019 period. 

In addition, the expected number of deaths per disaster is affected by a larger increase in population and a lower increase in GDP per capita. As a consequence, the number of deaths would decrease but still contribute to a high annual death toll. The annual number of deaths in 2040 would decrease relative to the 2015--2019 period for landslides, it would increase slightly for floods and wildfires, and dramatically for storms, heat waves, and cold waves. Our projections reveal a large increase in the death toll due to floods in Sub-Saharan Africa, to storms in Latin America, and to heat waves in Europe and Central Asia. Overall, the number of deaths due to climate-related disasters would increase by $50\%$ compared to 2015--2019. The number of deaths would be $1.6$ times higher in Scenario SSP3 than in Scenario SSP1. 

Afterwards, the number of deaths per disaster would continue to decrease due to the improvement in disaster risk management. Our estimates of the annual number of deaths in 2100  would be of an order of magnitude comparable to the number of deaths in 2015--2019 for floods, storms, landslides, and wildfires. The annual death toll due to heat waves and cold waves would substantially exceed the level observed in the recent period.\footnote{The large increase in the number of deaths due to cold waves echoes the empirical evidence provided by \cite{GasparriniEtAlii2015}, who find that cold temperatures actually cause more fatalities than warm temperatures.} East Asia and Pacific would suffer from a high death toll due to floods and storms, Europe and Central Asia and South Asia would suffer from a high death toll due to heat waves. Overall, the annual death toll would represent more than $30\%$ of the 2015--2019 annual number of deaths, representing approximately 30 times the number of deaths in Scenario SSP1.

To summarize these results, several conclusions are worth noting. The projected death toll due to floods, storms, and landslides decreases substantially in Scenario SSP1, so most fatalities are due to heat and cold waves. This evidence clearly suggests that climate change is mainly characterized by an increase in temperature extremes, as emphasized in IPCC Assessment Reports (\citealp{IPCC2012,IPCC6_2017,IPCC2021}). In the ``business-as-usual" Scenario SSP3, where all covariates tend towards an increase in fatalities, all disaster types would contribute to the death toll, in particular floods, storms, and heat and cold waves, for an annual number of deaths above $11,000$ in 2100.

\subsection{Worldwide Projections}\label{sec: World Projections}

\begin{figure}[h!]
\caption{Number of disasters per year -- World -- Scenario SSP1} \label{fig: World numbers1}

\vspace{-1.25cm}

\begin{center}
\bigskip
\includegraphics[scale=.95,trim= 2.25cm 8.5cm 2.25cm 8.75cm,clip]{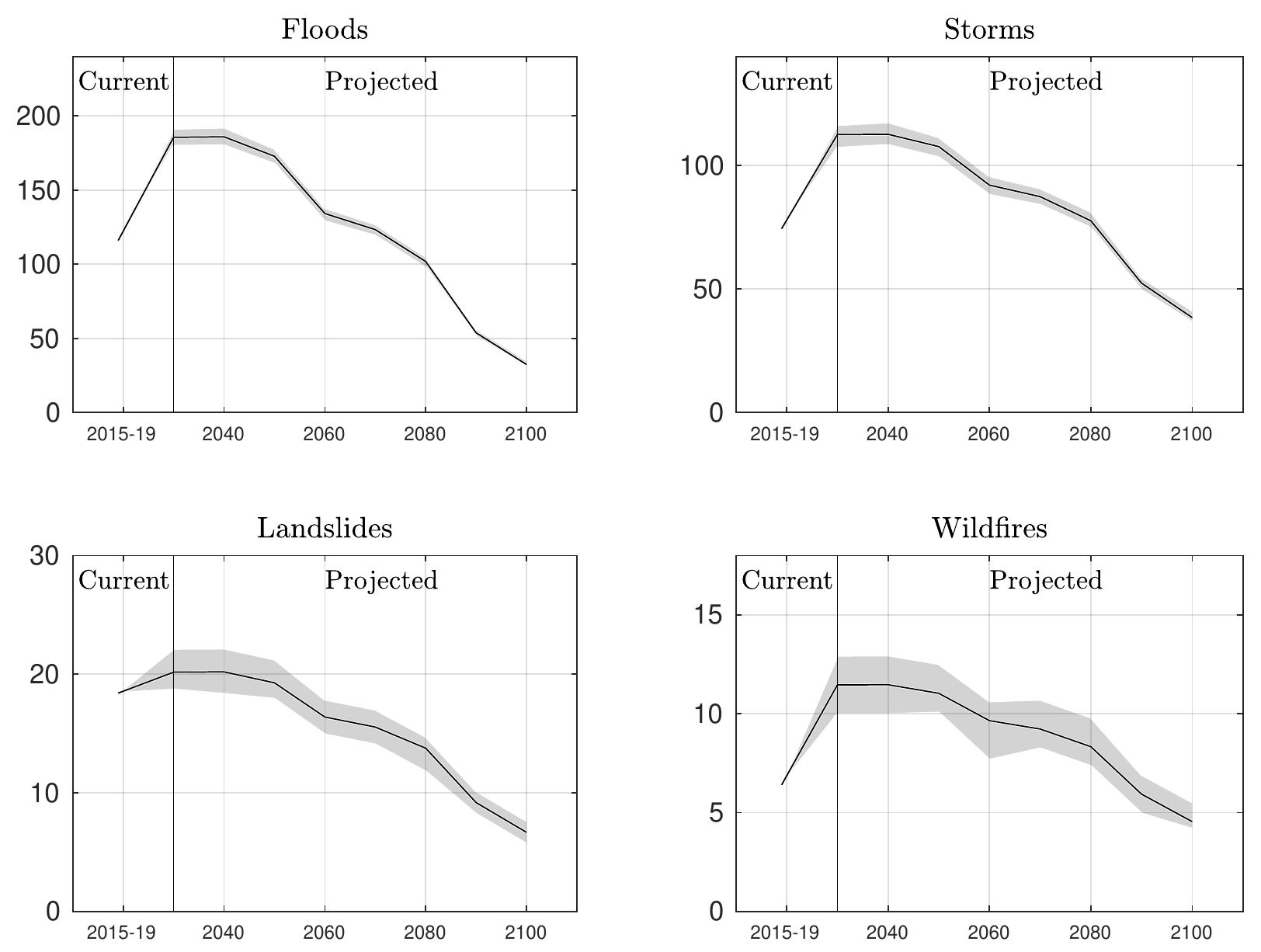} 
\includegraphics[scale=.95,trim= 2.25cm 15.5cm 2.25cm 7.25cm,clip]{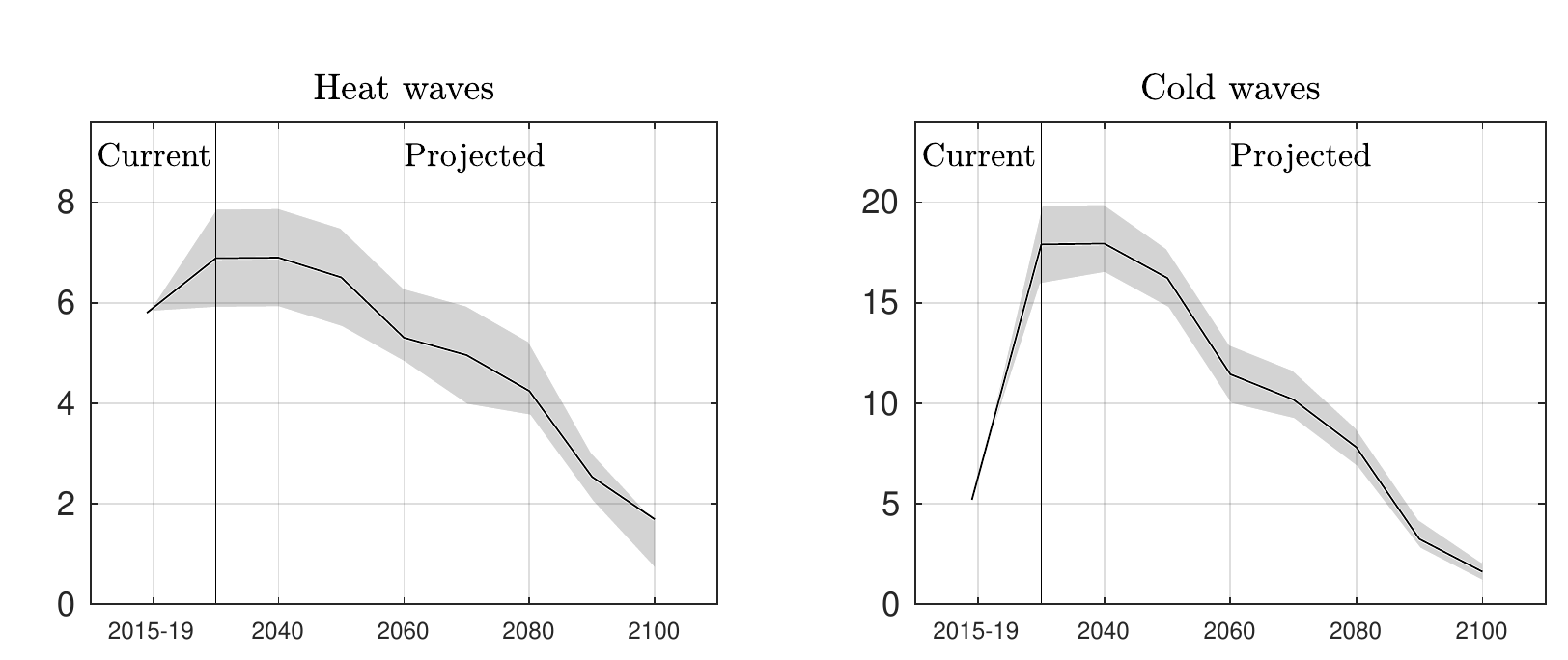} \end{center}

\end{figure}
\begin{figure}[h!]
\caption{Number of deaths per disaster -- World -- Scenario SSP1} \label{fig: World numbers2}

\vspace{-1.25cm}

\begin{center}
\bigskip
\includegraphics[scale=.95,trim= 2.25cm 8.5cm 2.25cm 8.75cm,clip]{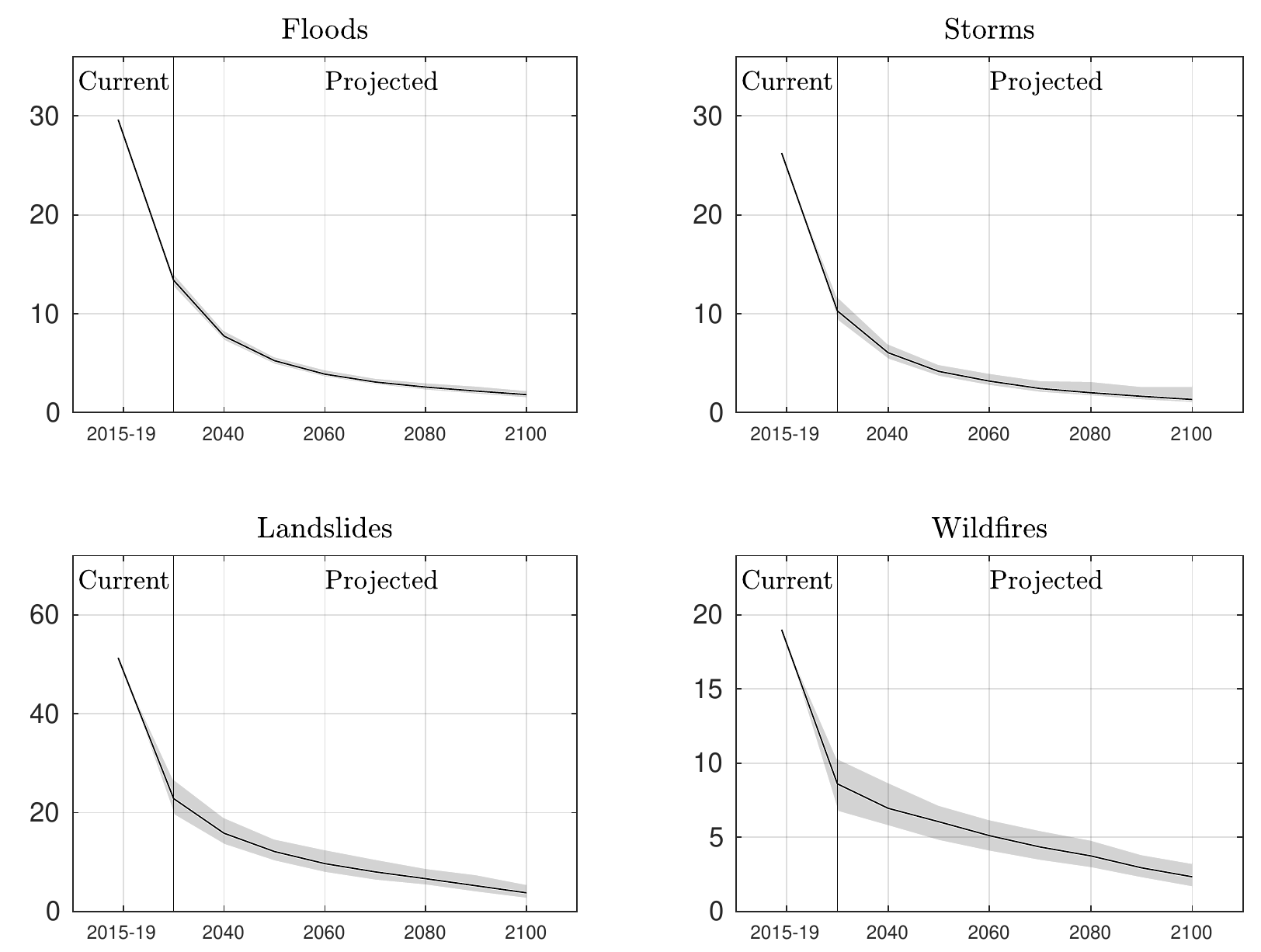} 
\includegraphics[scale=.95,trim= 2.25cm 15.5cm 2.25cm 7.25cm,clip]{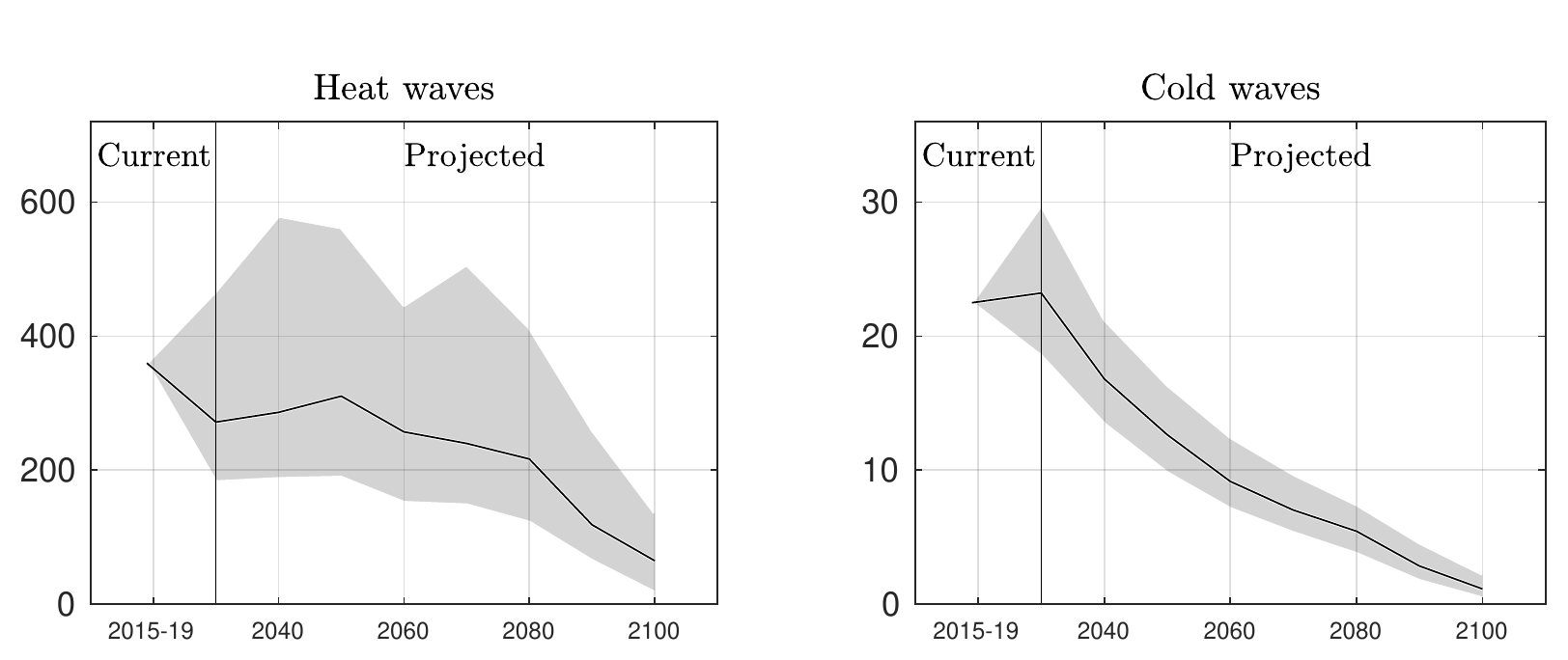} \end{center}

\end{figure}
\begin{figure}[h!]
\caption{Annual number of deaths -- World -- Scenario SSP1} \label{fig: World numbers3}

\vspace{-1.25cm}

\begin{center}
\bigskip
\includegraphics[scale=.95,trim= 2.25cm 8.5cm 2.25cm 8.75cm,clip]{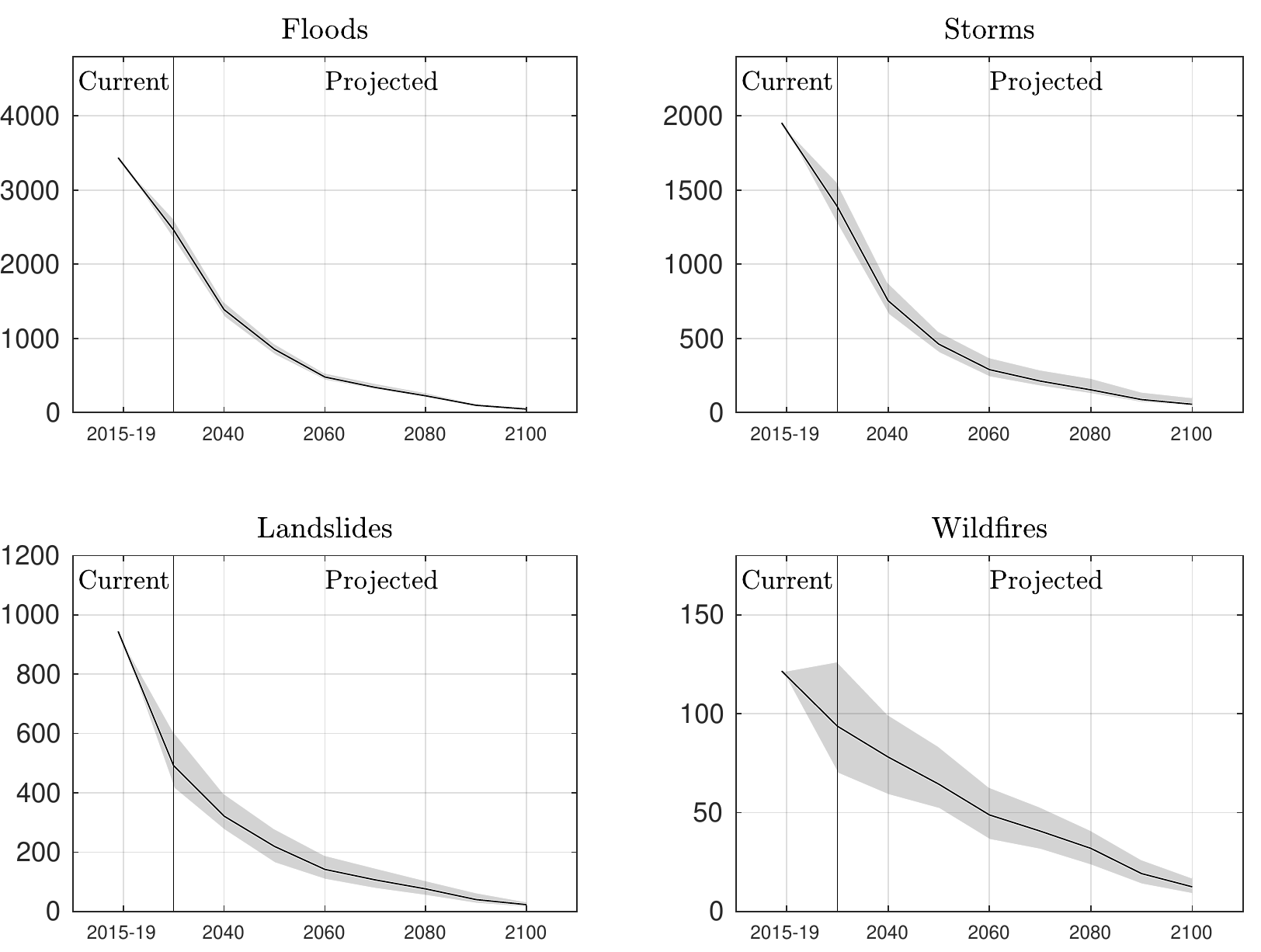} 
\includegraphics[scale=.95,trim= 2.25cm 15.5cm 2.25cm 7.25cm,clip]{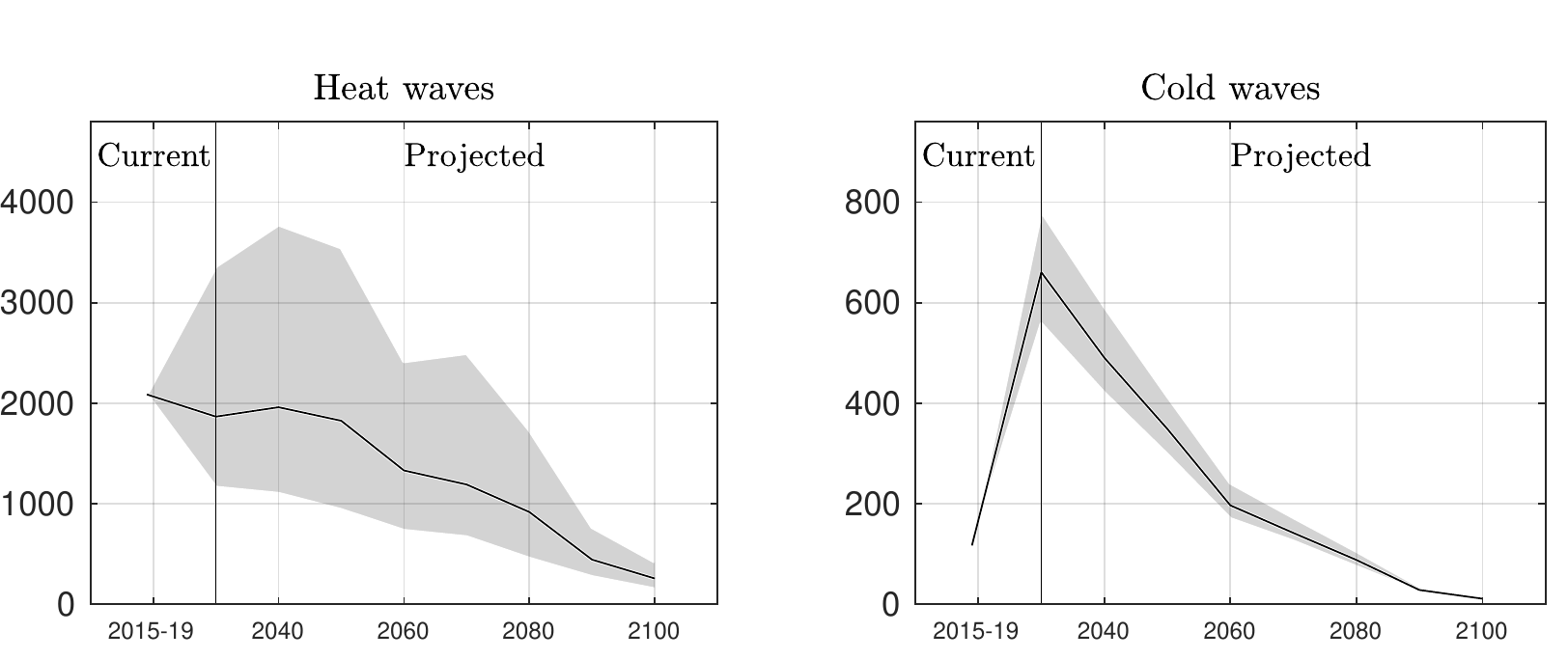} \end{center}

\end{figure}
\begin{figure}[h!]
\caption{Number of disasters per year -- World -- Scenario SSP3} \label{fig: World numbers4}

\vspace{-1.25cm}

\begin{center}
\bigskip
\includegraphics[scale=.95,trim= 2.25cm 8.5cm 2.25cm 8.75cm,clip]{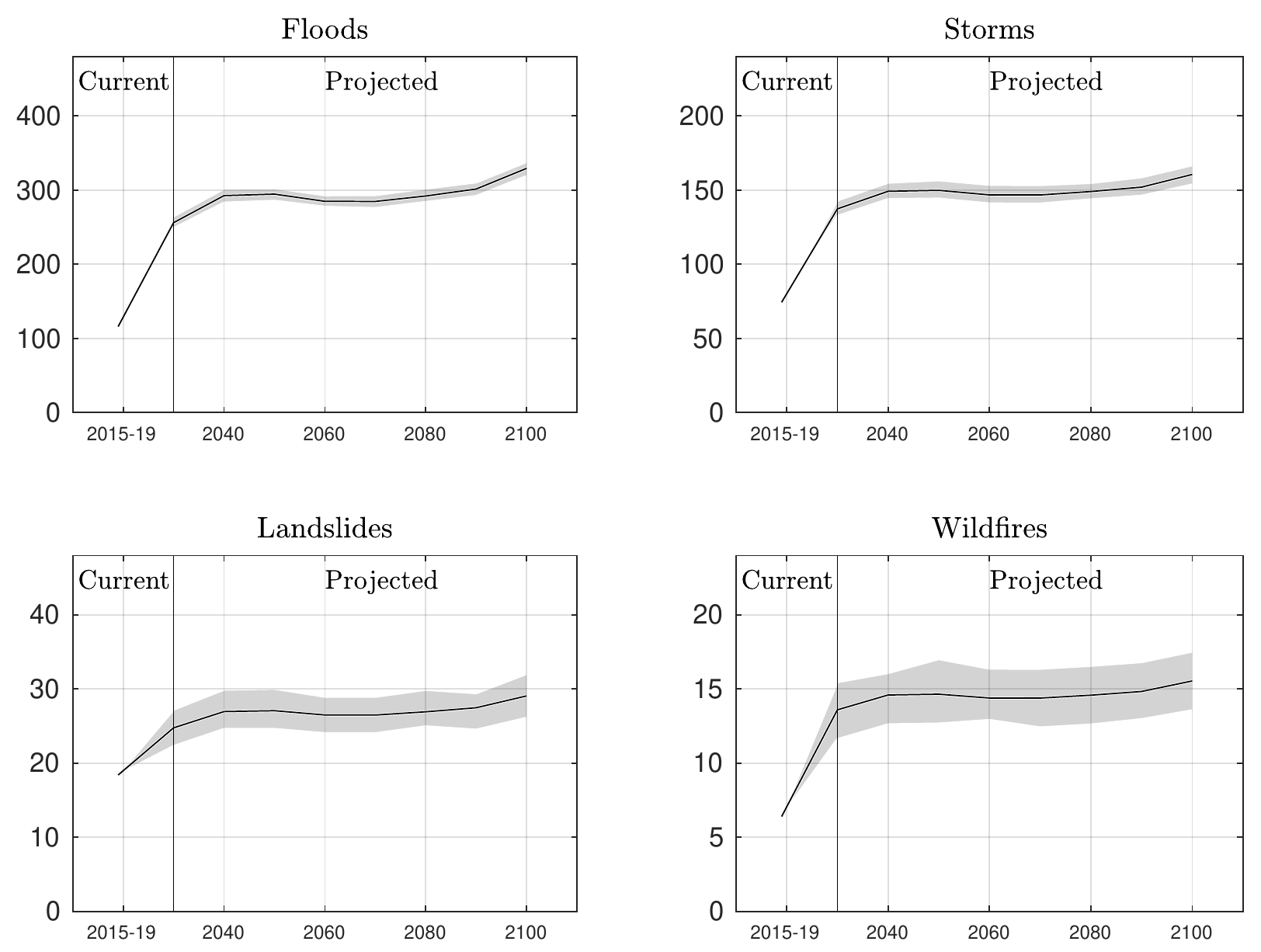} 
\includegraphics[scale=.95,trim= 2.25cm 15.5cm 2.25cm 7.25cm,clip]{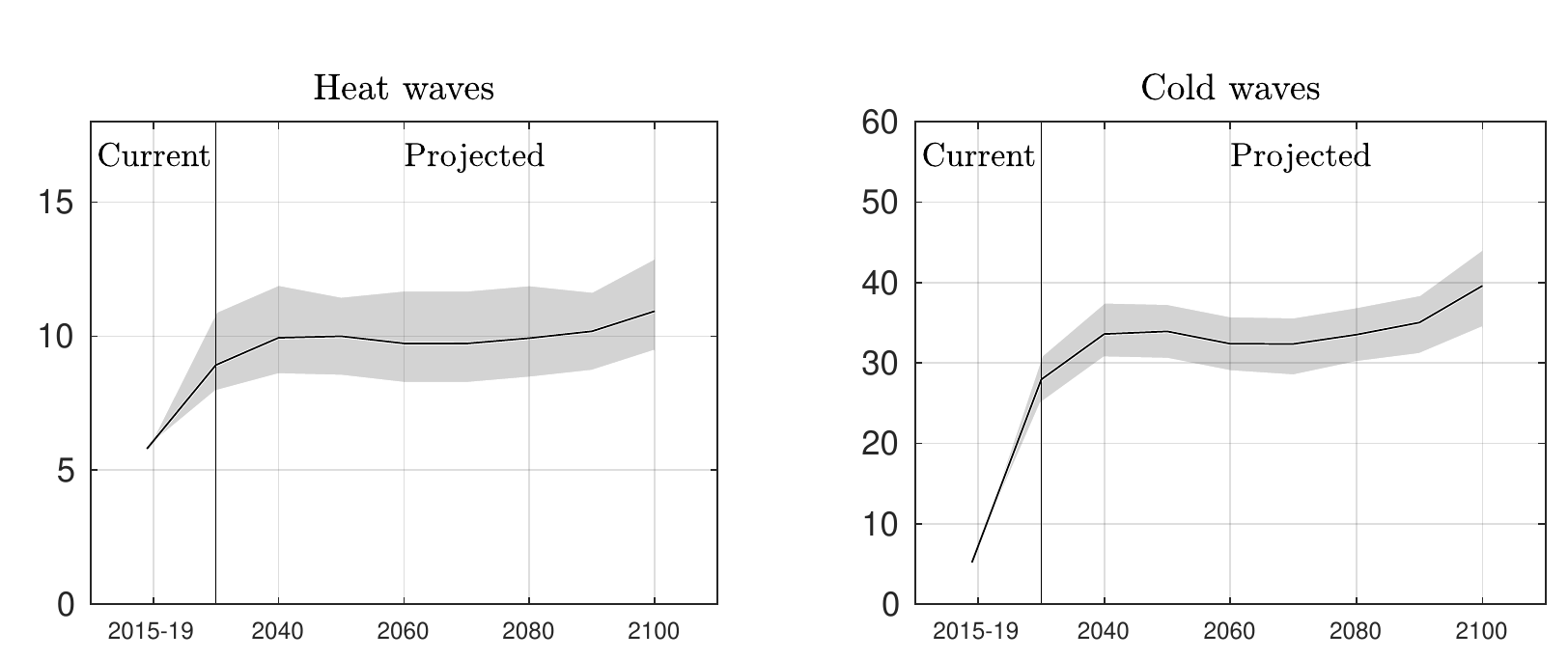} \end{center}

\end{figure}
\begin{figure}[h!]
\caption{Number of deaths per disaster -- World -- Scenario SSP3} \label{fig: World numbers5}

\vspace{-1.25cm}

\begin{center}
\bigskip
\includegraphics[scale=.95,trim= 2.25cm 8.5cm 2.25cm 8.75cm,clip]{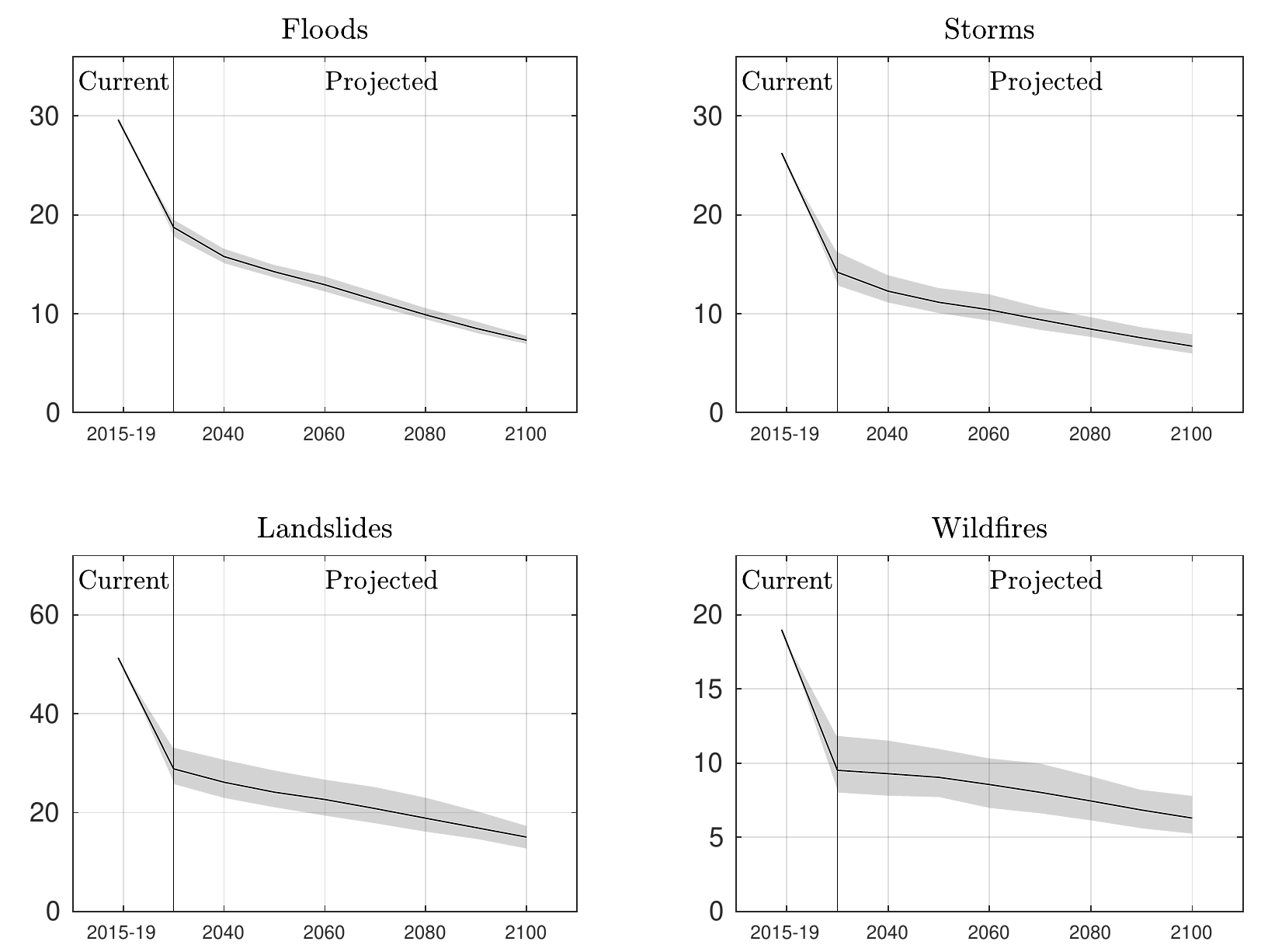} 
\includegraphics[scale=.95,trim= 2.25cm 15.5cm 2.25cm 7.25cm,clip]{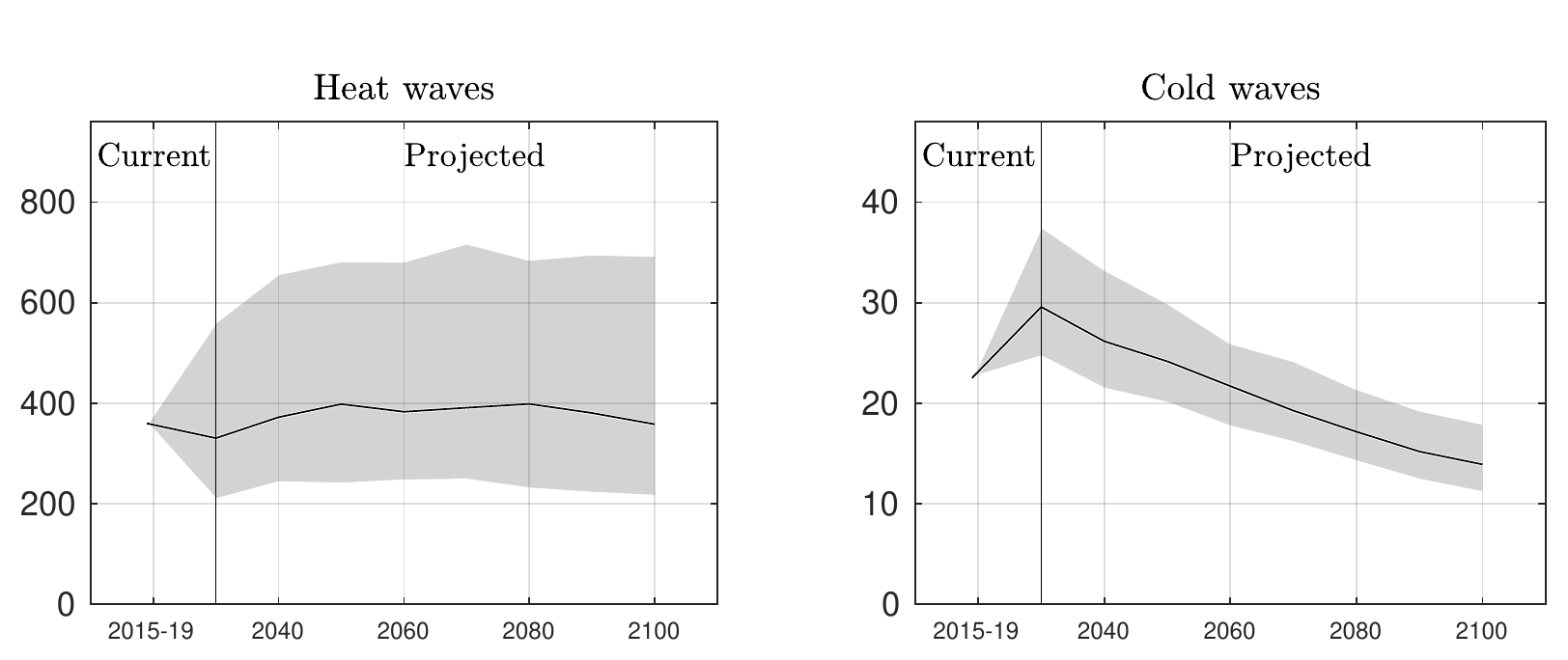} \end{center}

\end{figure}
\begin{figure}[h!]
\caption{Annual number of deaths -- World -- Scenario SSP3} \label{fig: World numbers6}

\vspace{-1.25cm}

\begin{center}
\bigskip
\includegraphics[scale=.95,trim= 2.25cm 8.5cm 2.25cm 8.75cm,clip]{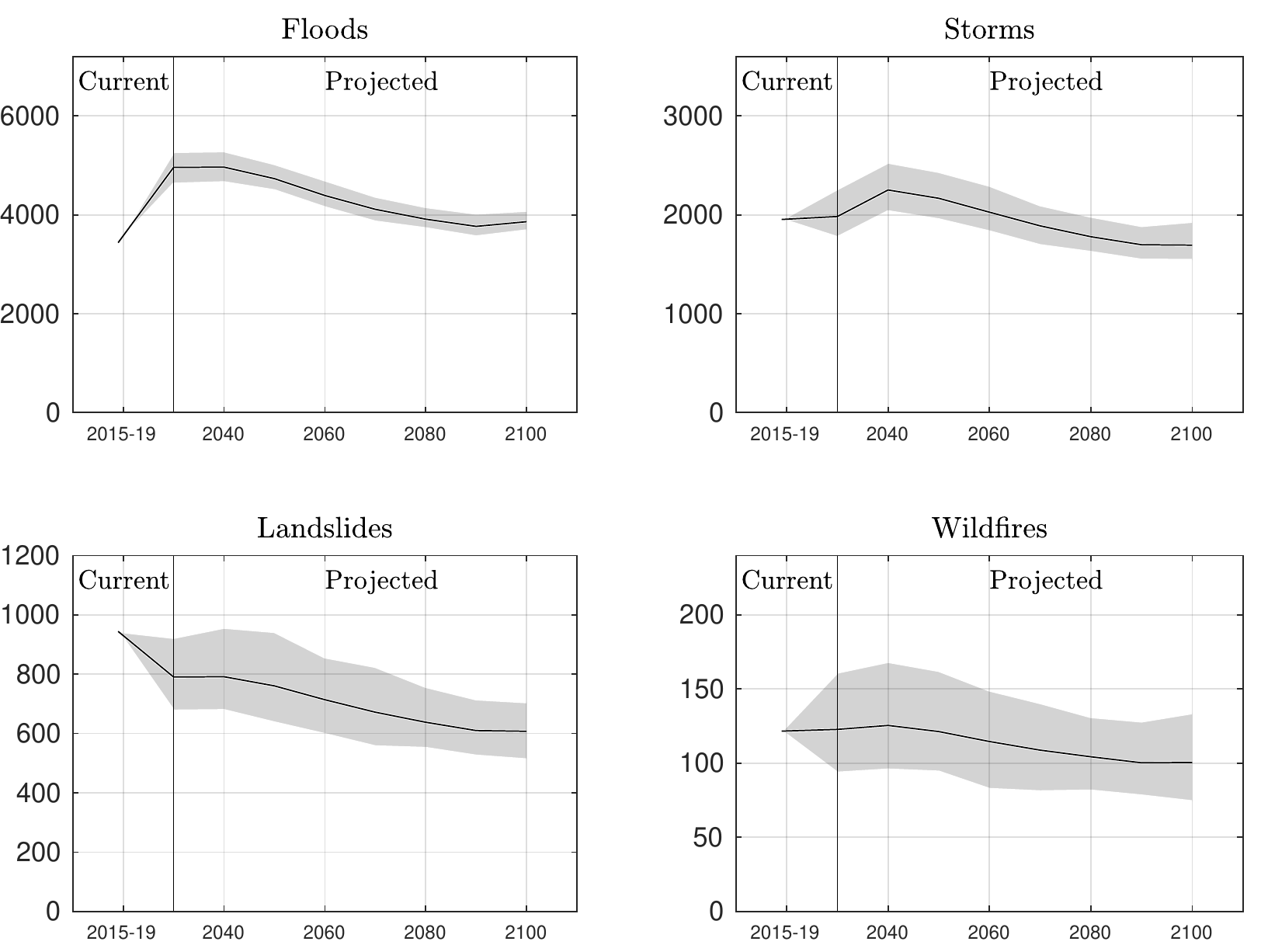} 
\includegraphics[scale=.95,trim= 2.25cm 15.5cm 2.25cm 7.25cm,clip]{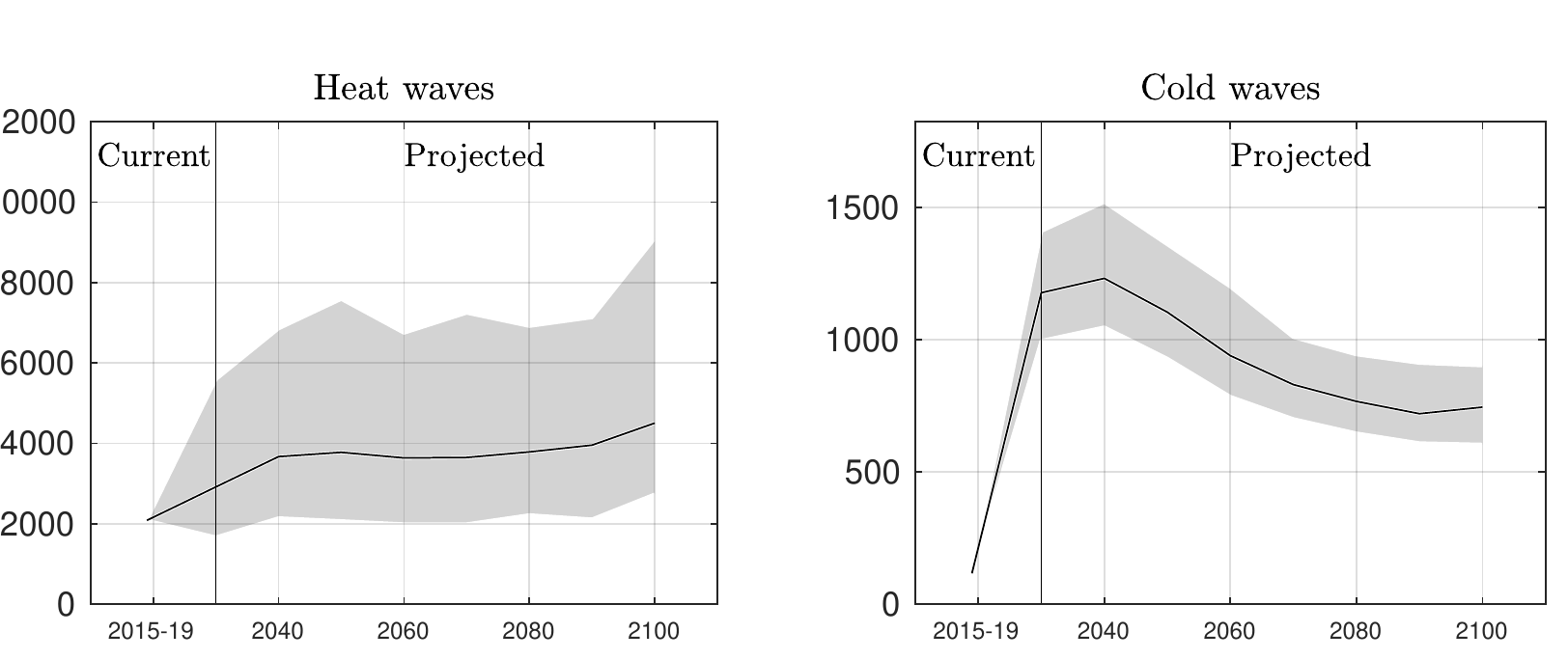} \end{center}

\end{figure}

In Figures \ref{fig: World numbers1} to \ref{fig: World numbers6}, we report the worldwide number of disasters, number of deaths per disaster, and annual number of deaths for the different disaster types for Scenarios SSP1 and SSP3. The objective of these figures is to visualize the current and projected world numbers, with an uncertainty interval for projected means. It is important to note that this interval is conditional on the IPCC scenarios for the projection of the covariates. Thus, projections of CO$_2$ emissions per capita, population, and real GDP per capita are taken as given in the simulation. As explained in Section \ref{sec: PGPD}, the uncertainty interval is obtained by bootstrapping as described in Section~\ref{sec: PGPD}. In this approach, we simulate the number of disasters and the number of deaths for each disaster type and each region. By summing these numbers, we obtain the projected worldwide numbers. From these simulations, we compute the uncertainty interval for the worldwide number of disasters and number of deaths as follows. We take 100 subsamples of 50 simulated numbers of disasters and compute the median number for each subsample. Then, we obtain the $95\%$ uncertainty interval over all these subsamples.

Figures \ref{fig: World numbers1} to \ref{fig: World numbers3} correspond to the number of disasters, number of deaths per disaster, and annual number of disasters for Scenario SSP1. The projections for the number of disasters clearly demonstrate the dependence on CO$_2$ emissions that we identify in the model. The annual number of disasters reaches a peak in approximately 2040 and then starts to decrease in the second half of the century. The figures also reveal that projections of the number of disasters are relatively accurate for floods and storms. This is due to a large number of these types of disasters and a good fit of the model. For other disaster types, uncertainty bands are approximately $20$--$30\%$ of the projections.

Regarding the death toll per disaster, the figure shows a negative trend for floods, storms, and landslides, which reflects the role of economic growth as a way to reduce the impact of climate-related disasters. In Scenario SSP1, as economies continue to grow in IPCC projections, we expect a substantial decrease in the death toll. For wildfires, heat waves, and cold waves, we did not identify a similar negative historical relationship between GDP per capita and the number of deaths. The decrease that we observe in the projections only reflects the decrease in the world population in this scenario. As we do not consider the uncertainty in the IPCC projections, the uncertainty bands are very narrow. We note that for heat waves, the particularly lethal events in Europe in 2003 and Russia in 2010 result in a tail index $\xi>1$ for Europe and Central Asia, so that the expected number of deaths in this region is replaced by the median. As a result, the projections are much lower than the death toll observed for 2000--2019. The large uncertainty bands for heat waves therefore reflect a relatively low quality of fit.

For Scenario SSP3, the main change is that for all disaster types, the number of disasters stabilize at or slightly increase to relatively high levels, due to the acceleration of CO$_2$ emissions in this scenario. Regarding the death toll, economic growth is positive but limited by international political tensions. In addition, improvement in disaster risk management is partly offset by faster population growth, so that projections in the number of deaths per disaster only slightly decrease. Overall, we do not project any substantial decrease in the annual number of deaths relative to the recent period, except for landslides. The number of deaths due to heat waves and cold waves is expected to increase dramatically, although the uncertainty bands are relatively wide.

\subsection{Discussion}\label{sec: Discussion}

In this section, we discuss two aspects of our modeling results that may have a substantial impact on the projected death toll.


First, our projections are based on a model that describes the number of disasters as a function of aggregate CO$_2$ emissions and the number of deaths per disaster as a function of regional real GDP growth. In the former relationship, CO$_2$ emissions represent a proxy for the increase in temperature and an intensification of the water cycle. In the latter, real GDP growth is a proxy for progress in disaster risk management. Our model implicitly assumes some proportionality between CO$_2$ emissions and aggregate temperature and between real GDP growth and disaster risk management. In both cases, this proportionality assumption may not be satisfied. On the one hand, some consequences of CO$_2$ emissions are irreversible on climate, especially those involving the water cycle (\citealp{IPCC2021}). On the other hand, the proportionality between economic growth and disaster risk management is not granted, as it mainly depends on political decisions. 

Related to this issue, in our modeling, we did not identify any clear relationship between climate change and the number of deaths per disaster. One reason for this is that in the most recent period, CO$_2$ emissions increased while the overall number of deaths per disaster decreased. It is possible that in the future, the death toll per disaster will increase with an increase in global temperature. This additional mechanism could compensate for the benefits of a better disaster risk management.




\bigskip

A second limitation of our analysis is that the projections discussed in Sections \ref{sec: Regional Projections} and \ref{sec: World Projections} are based on univariate models with covariates. However, there may be some dependence between the frequency or the severity of the various types of disasters. Failing to take this dependence into account may bias our projections.\footnote{Dependence between extreme climate events is related to the concept of compound (or multiple) events (\citealp{IPCC2012}). Such events correspond to situations where several disasters (possibly of different types) occur simultaneously or successively, with an amplifying effect on the death toll. For example, a combination of a flood and a storm or of a wildfire and a heat wave. The methodology for dealing with such compound events is described in \cite{LeonardEtAlii2014}.}

To address this possible issue, we filter out the effects of covariates and investigate the dependence between the resulting processes both for the frequency and the severity of disasters. For the frequency, we compute the correlation between the annual number of disasters for pairs of disaster types fitted by the univariate Poisson model. We proceed in the same way to measure the dependence between the number of deaths per disaster for pairs of disaster types fitted by the univariate GPD. 

Our results demonstrate that univariate models with covariates are sufficient to filter out the dependence between disaster types for most pairs.\footnote{The only exception is the flood/storm pair, for which we do observe a residual correlation between the annual number of disasters. This correlation is mostly driven by the high correlation in East Asia and Pacific and in Sub-Saharan Africa. However, for all pairs, the dependence between the number of deaths is correctly filtered out by the univariate model.} Our view is therefore that projections based on univariate models sufficiently capture the dependence between disaster types through the covariates that we include in the frequency and severity models. Details on the bivariate analysis are provided in Appendix \ref{app: Bivariate Analysis}.

%

\section{Conclusion}\label{sec: Conclusion}

The main objective of the paper is to predict the frequency and severity of climate-related disasters in relation to the scenarios established by the IPCC. We first describe a regional model for the frequency parameter (Poisson process) and severity parameters (GPD). We find that the frequency of extreme events increases with aggregate CO$_2$ emissions per capita, while the severity of these events can be mitigated by economic growth. As the covariates that we identified in our model are projected by the IPCC in its scenarios for the end of the century, we use these projections to predict the number of disasters and the number of deaths up to 2100.

We find substantial differences in our projections between the two contrasting scenarios. In the ``sustainable'' Scenario SSP1, both the frequency and severity of climate-related disasters substantially decrease, resulting in an impressive decrease in the projected death toll. In contrast, in the ``business-as-usual'' Scenario SSP3, an increase in CO$_2$ emissions, an increase in population, and lower economic growth combine to produce a marginal decrease in the death toll. In 2100, the annual death toll in Scenario SSP3 is predicted to be 30 times higher than the death toll in Scenario SSP1.

Our estimates suggest that in Scenario SSP3, in which CO$_2$ emissions continue to increase, the number of disasters would increase, but the number of deaths per disaster would not increase if disaster risk management continues to improve. In addition to the caveats mentioned in Section \ref{sec: Discussion}, it is worth noting that climate disasters are often accompanied by other dramatic consequences, including injuries and financial losses. Climate change could also cause massive human migrations because of the alteration in the natural environment through extreme temperatures or rises in sea level.


\clearpage\newpage

\renewcommand{\baselinestretch}{1.2} 
\selectfont

\DeclareRobustCommand\IPCClongname{, Intergovernmental Panel on Climate Change}

\bibliographystyle{natbib}
\makeatletter 
\makeatother

\bibliography{Disasters}


\newpage\clearpage

\begin{table}[h!]
\caption{Summary statistics on the frequency and severity of climate disasters }  \label{tab: Stats 1}
\begin{center}
\vspace{-0.5cm}

{\scalebox{0.75}[0.75]{   

\begin{tabular}{lrrr|rrr|rrr}
 \toprule
&	\multicolumn{3}{c|}{\textbf{Floods}}&\multicolumn{3}{c|}{\textbf{Storms}}&	\multicolumn{3}{c}{\textbf{Landslides}}	\\
	&	Nb of	&	Nb of	&	Ann. nb 	&	Nb of	&	Nb of	&	Ann. nb &	Nb of	&	Nb of	&	Ann. nb 	\\
		& disasters&deaths&of deaths 	& disasters&deaths&of deaths	& disasters&deaths&of deaths\\ \midrule
\textbf{Panel A: 1960--2019} &&&&&&&&& 	\\ [3pt]
East Asia - Pacific	&	20.7	&	106.4	&	2199.2	&	25.2	&	202.5	&	5095.9	&	4.0	&	67.0	&	264.5	\\
Europe - Central Asia	&	12.0	&	14.8	&	178.3	&	8.9	&	7.1	&	62.9	&	1.8	&	60.0	&	104.9	\\
Latin America - Caribbean	&	15.7	&	86.4	&	1356.6	&	10.1	&	170.0	&	1714.5	&	2.7	&	172.2	&	470.7	\\
Middle East - North Africa	&	5.0	&	74.9	&	374.3	&	1.2	&	31.6	&	38.5	&	0.2	&	44.7	&	8.9	\\
North America	&	3.7	&	12.0	&	43.9	&	10.5	&	28.8	&	301.1	&	0.1	&	126.7	&	12.7	\\
South Asia	&	11.5	&	369.8	&	4259.0	&	6.6	&	3736.8	&	24600.8	&	2.2	&	120.1	&	258.3	\\
Sub-Saharan Africa	&	15.4	&	38.6	&	593.9	&	4.2	&	45.8	&	192.2	&	0.9	&	87.6	&	78.8	\\
World	&	84.0	&	107.3	&	9005.3	&	66.6	&	480.6	&	32006.0	&	11.8	&	101.7	&	1198.9	\\ \midrule
\textbf{Panel B: 1960--1979} &&&&&&&&&	\\[3pt]
East Asia - Pacific	&	5.6	&	414.5	&	2300.5	&	11.1	&	275.2	&	3040.8	&	0.8	&	89.6	&	67.2	\\
Europe - Central Asia	&	1.7	&	100.6	&	166.1	&	1.8	&	24.9	&	44.9	&	1.0	&	162.5	&	162.5	\\
Latin America - Caribbean	&	5.5	&	142.3	&	775.4	&	4.0	&	693.6	&	2774.4	&	1.4	&	650.4	&	878.0	\\
Middle East - North Africa	&	2.1	&	165.9	&	340.0	&	0.3	&	153.4	&	38.3	&	0.1	&	93.9	&	9.4	\\
North America	&	0.8	&	86.1	&	64.6	&	3.1	&	112.9	&	350.0	&	0.1	&	338.7	&	33.9	\\
South Asia	&	3.5	&	1858.5	&	6411.8	&	3.9	&	15044.8	&	58674.9	&	0.6	&	413.1	&	227.2	\\
Sub-Saharan Africa	&	2.0	&	118.9	&	237.8	&	1.1	&	94.2	&	98.9	&	0.1	&	705.1	&	70.5	\\
World	&	20.9	&	492.6	&	10296.1	&	25.2	&	2585.4	&	65022.2	&	4.0	&	366.8	&	1448.7	\\ \midrule
\textbf{Panel C: 1980--1999} &&&&&&&&& 	\\[3pt]
East Asia - Pacific	&	17.3	&	161.6	&	2795.9	&	26.6	&	113.7	&	3025.5	&	4.0	&	69.8	&	279.0	\\
Europe - Central Asia	&	10.1	&	24.0	&	241.7	&	11.3	&	9.1	&	102.1	&	2.4	&	50.6	&	121.5	\\
Latin America - Caribbean	&	13.9	&	185.9	&	2575.0	&	9.8	&	187.9	&	1841.5	&	3.6	&	108.7	&	385.8	\\
Middle East - North Africa	&	4.5	&	117.6	&	529.0	&	0.8	&	46.2	&	36.9	&	0.2	&	24.6	&	4.9	\\
North America	&	4.2	&	8.5	&	35.6	&	13.6	&	18.3	&	248.5	&	0.1	&	0.0	&	0.0	\\
South Asia	&	10.2	&	383.6	&	3913.0	&	7.3	&	1985.6	&	14495.1	&	2.2	&	149.5	&	321.4	\\
Sub-Saharan Africa	&	9.4	&	71.4	&	667.6	&	3.5	&	66.6	&	233.0	&	0.4	&	43.4	&	15.2	\\
World	&	69.5	&	154.9	&	10757.8	&	72.9	&	274.3	&	19982.7	&	12.7	&	88.8	&	1127.9	\\ \midrule
\textbf{Panel D: 2000--2019}&&&&&&&&& 	\\[3pt]
East Asia - Pacific	&	39.2	&	38.3	&	1501.2	&	37.9	&	243.6	&	9221.4	&	7.1	&	63.0	&	447.4	\\
Europe - Central Asia	&	24.4	&	5.2	&	127.3	&	13.7	&	3.1	&	41.8	&	1.9	&	16.6	&	30.8	\\
Latin America - Caribbean	&	27.8	&	25.9	&	719.5	&	16.5	&	32.1	&	527.6	&	3.3	&	45.0	&	148.4	\\
Middle East - North Africa	&	8.5	&	30.0	&	253.7	&	2.6	&	15.4	&	40.1	&	0.3	&	41.7	&	12.5	\\
North America	&	6.1	&	5.2	&	31.7	&	14.7	&	20.8	&	304.8	&	0.2	&	27.6	&	4.1	\\
South Asia	&	20.9	&	117.3	&	2452.3	&	8.6	&	74.0	&	632.5	&	3.8	&	60.3	&	226.1	\\
Sub-Saharan Africa	&	34.8	&	25.2	&	876.4	&	8.1	&	30.4	&	244.8	&	2.3	&	67.0	&	150.8	\\
World	&	161.5	&	36.9	&	5962.0	&	101.8	&	108.2	&	11013.1	&	18.7	&	54.6	&	1020.2	\\ \bottomrule
\end{tabular}

}}

\end{center}
\small{
\noindent Note: This table reports the average number of disasters per year and the average and median number of deaths per disaster for floods, storms, and landslides. The number of deaths is rescaled for the population as of end of 2019.}
\end{table}


\newpage\clearpage

\begin{table}[h!]
\caption{Summary statistics on the frequency and severity of climate disasters (continued) }  \label{tab: Stats 2}
\begin{center}

{\scalebox{0.75}[0.75]{   

\begin{tabular}{lrrr|rrr|rrr}
 \toprule
&	\multicolumn{3}{c|}{\textbf{Wildfires	}}				&	\multicolumn{3}{c|}{\textbf{Heat waves}}					&	\multicolumn{3}{c}{\textbf{Cold waves}}					\\
	&	Nb of	&	Nb of	&	Ann. nb 	&	Nb of	&	Nb of	&	Ann. nb &	Nb of	&	Nb of	&	Ann. nb 	\\
		& disasters&deaths&of deaths 	& disasters&deaths&of deaths	& disasters&deaths&of deaths\\ \midrule
\textbf{Panel A: 1960--2019}&&&&&&&&& 	\\ [3pt]
East Asia - Pacific	&	1.82	&	13.3	&	24.1	&	0.53	&	59.6	&	31.8	&	0.32	&	20.4	&	6.5	\\
Europe - Central Asia	&	1.98	&	6.8	&	13.5	&	1.42	&	1720.3	&	2437.1	&	3.32	&	36.1	&	119.6	\\
Latin America - Caribbean	&	0.92	&	3.8	&	3.5	&	0.15	&	165.3	&	24.8	&	1.05	&	58.5	&	61.4	\\
Middle East - North Africa	&	0.17	&	9.8	&	1.6	&	0.13	&	34.3	&	4.6	&	0.17	&	5.2	&	0.9	\\
North America	&	1.73	&	3.2	&	5.5	&	0.37	&	213.6	&	78.3	&	0.23	&	38.7	&	9.0	\\
South Asia	&	0.15	&	18.4	&	2.8	&	0.72	&	433.8	&	310.9	&	1.18	&	219.0	&	259.2	\\
Sub-Saharan Africa	&	0.48	&	14.1	&	6.8	&	0.05	&	41.5	&	2.1	&	0.07	&	26.3	&	1.8	\\
World	&	7.25	&	8.0	&	57.9	&	3.37	&	858.3	&	2889.6	&	6.33	&	72.4	&	458.3	\\ \midrule
\textbf{Panel B: 1960--1979}&&&&&&&&& 	\\[3pt]
East Asia - Pacific	&	1.45	&	5.7	&	8.3	&	0.00	&	--	&	--	&	0.00	&	--	&	--	\\
Europe - Central Asia	&	0.05	&	28.8	&	1.4	&	0.00	&	--	&	--	&	0.15	&	190.0	&	28.5	\\
Latin America - Caribbean	&	0.10	&	0.0	&	0.0	&	0.25	&	164.3	&	41.1	&	0.05	&	138.2	&	6.9	\\
Middle East - North Africa	&	0.00	&	--	&	--	&	0.00	&	--	&	--	&	0.00	&	--	&	--	\\
North America	&	0.05	&	0.0	&	0.0	&	0.20	&	122.1	&	24.4	&	0.10	&	131.2	&	13.1	\\
South Asia	&	0.00	&	--	&	--	&	0.30	&	273.2	&	81.9	&	0.15	&	595.3	&	89.3	\\
Sub-Saharan Africa	&	0.10	&	3.4	&	0.3	&	0.00	&	--	&	--	&	0.00	&	--	&	--	\\
World	&	1.75	&	5.8	&	10.1	&	0.75	&	196.6	&	147.4	&	0.45	&	306.3	&	137.8	\\ \midrule
\textbf{Panel C: 1980--1999}&&&&&&&&& 	\\[3pt]
East Asia - Pacific	&	2.10	&	21.9	&	46.0	&	0.35	&	31.7	&	11.1	&	0.20	&	10.9	&	2.2	\\
Europe - Central Asia	&	2.40	&	5.5	&	13.3	&	0.65	&	105.9	&	68.8	&	1.55	&	35.9	&	55.7	\\
Latin America - Caribbean	&	1.20	&	4.9	&	5.8	&	0.10	&	305.9	&	30.6	&	0.70	&	66.1	&	46.3	\\
Middle East - North Africa	&	0.15	&	11.3	&	1.7	&	0.10	&	43.6	&	4.4	&	0.10	&	19.6	&	2.0	\\
North America	&	1.55	&	2.0	&	3.1	&	0.45	&	402.5	&	181.1	&	0.25	&	32.1	&	8.0	\\
South Asia	&	0.30	&	21.5	&	6.5	&	0.70	&	507.6	&	355.3	&	1.20	&	237.9	&	285.5	\\
Sub-Saharan Africa	&	0.45	&	15.6	&	7.0	&	0.00	&	--	&	--	&	0.15	&	26.3	&	3.9	\\
World	&	8.15	&	10.2	&	83.4	&	2.35	&	277.2	&	651.3	&	4.15	&	97.2	&	403.6	\\ \midrule
\textbf{Panel D: 2000--2019}&&&&&&&&&	\\[3pt]
East Asia - Pacific	&	1.90	&	9.5	&	18.0	&	1.25	&	67.5	&	84.3	&	0.75	&	22.9	&	17.2	\\
Europe - Central Asia	&	3.50	&	7.4	&	25.9	&	3.60	&	2011.8	&	7242.6	&	8.25	&	33.3	&	274.7	\\
Latin America - Caribbean	&	1.45	&	3.2	&	4.7	&	0.10	&	27.3	&	2.7	&	2.40	&	54.6	&	131.0	\\
Middle East - North Africa	&	0.35	&	9.1	&	3.2	&	0.30	&	31.2	&	9.4	&	0.40	&	1.6	&	0.6	\\
North America	&	3.60	&	3.7	&	13.5	&	0.45	&	65.4	&	29.4	&	0.35	&	16.9	&	5.9	\\
South Asia	&	0.15	&	12.1	&	1.8	&	1.15	&	430.8	&	495.4	&	2.20	&	183.0	&	402.7	\\
Sub-Saharan Africa	&	0.90	&	14.5	&	13.1	&	0.15	&	41.5	&	6.2	&	0.05	&	26.4	&	1.3	\\
World	&	11.85	&	6.8	&	80.2	&	7.00	&	1124.3	&	7870.0	&	14.40	&	57.9	&	833.4	\\ \bottomrule
\end{tabular}
}}

\end{center}
\small{
\noindent Note: This table reports the average number of disasters per year and the average and median number of deaths per disaster for wildfires, heat waves, and cold waves. The number of deaths is rescaled for the population as of end of 2019.}
\end{table}

\newpage\clearpage

\begin{table}[!ht]
\caption{Parameter estimates}  \label{tab: Parameter_Estimates_1}

\vspace{-0.5cm}

{\scalebox{0.8}[0.8]{   

\begin{tabular}{l r@{\hskip1.5pt}rr |r@{\hskip1.5pt}rr |r@{\hskip1.5pt}rr }
 \toprule
&	\multicolumn{3}{c|}{\textbf{Floods}}&\multicolumn{3}{c|}{\textbf{Storms}}&	\multicolumn{3}{c}{\textbf{Landslides}}	\\
&\multicolumn{2}{c}{Par. est.}& Std err.&\multicolumn{2}{c}{Par. est.}& Std err.&\multicolumn{2}{c}{Par. est.}& Std err.\\\midrule
\textbf{Panel A: Intensity parameter $\log(\lambda_t)$}	&		&		&		&		&		&		&		&		&		\\[5pt]	
Constant	&	-5.141	&	$^{(a)}$	&	0.228	&	-2.231	&	$^{(a)}$	&	0.233	&	-4.044	&	$^{(a)}$	&	0.557	\\	
$\log(CO2) \times$D(East Asia - Pacific)	&	5.610	&	$^{(a)}$	&	0.154	&	3.768	&	$^{(a)}$	&	0.159	&	3.747	&	$^{(a)}$	&	0.381	\\	
$\log(CO2) \times$D(Europe - Central Asia)	&	5.243	&	$^{(a)}$	&	0.155	&	3.062	&	$^{(a)}$	&	0.162	&	3.168	&	$^{(a)}$	&	0.386	\\	
$\log(CO2) \times$D(Latin America - Caribbean)	&	5.421	&	$^{(a)}$	&	0.155	&	3.145	&	$^{(a)}$	&	0.161	&	3.485	&	$^{(a)}$	&	0.383	\\	
$\log(CO2) \times$D(Middle East - North Africa)	&	4.644	&	$^{(a)}$	&	0.159	&	1.713	&	$^{(a)}$	&	0.179	&	1.699	&	$^{(a)}$	&	0.432	\\	
$\log(CO2) \times$D(North America)	&	4.439	&	$^{(a)}$	&	0.161	&	3.169	&	$^{(a)}$	&	0.161	&	1.217	&	$^{(c)}$	&	0.477	\\	
$\log(CO2) \times$D(South Asia)	&	5.214	&	$^{(a)}$	&	0.156	&	2.850	&	$^{(a)}$	&	0.163	&	3.338	&	$^{(a)}$	&	0.384	\\	
$\log(CO2) \times$D(Sub-Saharan Africa)	&	5.418	&	$^{(a)}$	&	0.155	&	2.556	&	$^{(a)}$	&	0.165	&	2.757	&	$^{(a)}$	&	0.392	\\[5pt]	
Adj. $R^2$	&	0.529	&		&		&	0.573	&		&		&	0.381	&		&		\\ \midrule	
\textbf{Panel B: Scale parameter $\nu_t$}	&	\multicolumn{3}{c}{(nb obs.: 3,658)}					&	\multicolumn{3}{c}{(nb obs.: 2,916)}					&	\multicolumn{3}{c}{(nb obs.: 672)}					\\[5pt]	
Constant	&	13.748	&	$^{(a)}$	&	1.364	&	14.536	&	$^{(a)}$	&	1.948	&	9.523	&	$^{(a)}$	&	1.954	\\	
D(Europe - Central Asia)	&	20.051	&	$^{(a)}$	&	5.977	&	17.899	&	$^{(c)}$	&	8.663	&	13.143	&	$^{(d)}$	&	7.533	\\	
D(Latin America - Caribbean)	&	23.051	&	$^{(a)}$	&	4.368	&	14.969	&		&	9.899	&	23.429	&	$^{(a)}$	&	6.880	\\	
D(Middle East - North Africa)	&	18.951	&	$^{(a)}$	&	5.500	&	15.205	&		&	30.056	&	6.216	&		&	10.769	\\	
D(North America)	&	22.508	&	$^{(b)}$	&	7.091	&	16.772	&	$^{(c)}$	&	6.706	&	47.377	&	$^{(a)}$	&	4.048	\\	
D(South Asia)	&	2.730	&		&	2.133	&	3.426	&		&	3.855	&	6.166	&	$^{(d)}$	&	3.188	\\	
D(Sub-Saharan Africa)	&	-0.113	&		&	5.785	&	-5.568	&		&	19.889	&	5.867	&		&	14.949	\\	
$\log(GDP) \times$D(East Asia - Pacific)	&	-1.090	&	$^{(a)}$	&	0.150	&	-1.222	&	$^{(a)}$	&	0.218	&	-0.613	&	$^{(b)}$	&	0.217	\\	
$\log(GDP) \times$D(Europe - Central Asia)	&	-3.079	&	$^{(a)}$	&	0.570	&	-2.984	&	$^{(a)}$	&	0.828	&	-1.894	&	$^{(b)}$	&	0.726	\\	
$\log(GDP) \times$D(Latin America - Caribbean)	&	-3.522	&	$^{(a)}$	&	0.438	&	-2.804	&	$^{(b)}$	&	1.026	&	-3.028	&	$^{(a)}$	&	0.702	\\	
$\log(GDP) \times$D(Middle East - North Africa)	&	-3.063	&	$^{(a)}$	&	0.563	&	-2.777	&		&	3.143	&	-1.319	&		&	1.117	\\	
$\log(GDP) \times$D(North America)	&	-3.131	&	$^{(a)}$	&	0.645	&	-2.624	&	$^{(a)}$	&	0.598	&	-5.154	&	$^{(a)}$	&	0.328	\\	
$\log(GDP) \times$D(South Asia)	&	-1.446	&	$^{(a)}$	&	0.203	&	-1.672	&	$^{(a)}$	&	0.417	&	-1.402	&	$^{(a)}$	&	0.315	\\	
$\log(GDP) \times$D(Sub-Saharan Africa)	&	-1.251	&	$^{(d)}$	&	0.695	&	-0.669	&		&	2.456	&	-1.411	&		&	1.826	\\[5pt]	
Adj. $R^2$	&	0.144	&		&		&	0.056	&		&		&	0.586	&		&		\\ \midrule	
\textbf{Panel C: Tail index $\xi_t$}	&		&		&		&		&		&		&		&		&		\\[5pt]	
Constant	&	4.108	&	$^{(a)}$	&	0.439	&	4.204	&	$^{(a)}$	&	0.381	&	0.289	&	$^{(a)}$	&	0.051	\\	
$\log(GDP) \times$D(East Asia - Pacific)	&	-0.379	&	$^{(a)}$	&	0.048	&	-0.370	&	$^{(a)}$	&	0.043	&	0.002	&		&	0.006	\\	
$\log(GDP) \times$D(Europe - Central Asia)	&	-0.350	&	$^{(a)}$	&	0.043	&	-0.396	&	$^{(a)}$	&	0.038	&	0.009	&	$^{(d)}$	&	0.005	\\	
$\log(GDP) \times$D(Latin America - Caribbean)	&	-0.366	&	$^{(a)}$	&	0.047	&	-0.325	&	$^{(a)}$	&	0.041	&	0.013	&	$^{(c)}$	&	0.006	\\	
$\log(GDP) \times$D(Middle East - North Africa)	&	-0.384	&	$^{(a)}$	&	0.047	&	-0.387	&	$^{(a)}$	&	0.042	&	-0.096	&	$^{(a)}$	&	0.007	\\	
$\log(GDP) \times$D(North America)	&	-0.375	&	$^{(a)}$	&	0.041	&	-0.355	&	$^{(a)}$	&	0.035	&	-0.114	&	$^{(a)}$	&	0.005	\\	
$\log(GDP) \times$D(South Asia)	&	-0.425	&	$^{(a)}$	&	0.055	&	-0.422	&	$^{(a)}$	&	0.049	&	-0.003	&		&	0.007	\\	
$\log(GDP) \times$D(Sub-Saharan Africa)	&	-0.436	&	$^{(a)}$	&	0.055	&	-0.405	&	$^{(a)}$	&	0.049	&	0.041	&	$^{(a)}$	&	0.006	\\[5pt]	
Adj. $R^2$	&	0.037	&		&		&	0.154	&		&		&	0.995	&		&		\\	[5pt]
Log-likelihood	&	-17825.8	&		&		&	-13833.0	&		&		&	-3513.5	&		&		\\ \bottomrule	
\end{tabular}
}}

\bigskip

\small{
\noindent Note: This table reports the parameter estimates of the Poisson-Generalized Pareto Distribution model for floods, storms, and landslides. For each disaster type, the first column corresponds to the estimated parameter and the second column corresponds to the standard error. $^{(a)}$, $^{(b)}$, $^{(c)}$, and $^{(d)}$ correspond to statistical significance at the $0.1\%$, $1\%$, $5\%$, and $10\%$, respectively. ``nb obs.'' is the number of events with a strictly positive number of deaths.}
\end{table}


\newpage\clearpage

\begin{table}[!ht]
\caption{Parameter estimates (continued)}  \label{tab: Parameter_Estimates_2}

{\scalebox{0.8}[0.8]{   

\begin{tabular}{l r@{\hskip1.5pt}rr |r@{\hskip1.5pt}rr |r@{\hskip1.5pt}rr }
 \toprule
	&	\multicolumn{3}{c|}{\textbf{Wildfires}	}				&	\multicolumn{3}{c|}{\textbf{Heat waves}	}				&	\multicolumn{3}{c}{\textbf{Cold waves}}					\\
	&	\multicolumn{2}{c}{Par. est.}		&	Std. err.	&	\multicolumn{2}{c}{Par. est.}	&	Std. err.	&	\multicolumn{2}{c}{Par. est.}		&	Std. err.	\\ \midrule
\textbf{Panel A: Intensity parameter $\log(\lambda_t)$}	&		&		&		&		&		&		&		&		&		\\[5pt]
Constant	&	-1.394	&	$^{(d)}$	&	0.819	&	-4.146	&	$^{(a)}$	&	1.145	&	-7.780	&	$^{(a)}$	&	0.968	\\
$\log(CO2) \times$D(East Asia - Pacific)	&	1.436	&	$^{(c)}$	&	0.563	&	2.628	&	$^{(a)}$	&	0.781	&	4.693	&	$^{(a)}$	&	0.664	\\
$\log(CO2) \times$D(Europe - Central Asia)	&	1.498	&	$^{(b)}$	&	0.563	&	3.275	&	$^{(a)}$	&	0.773	&	6.247	&	$^{(a)}$	&	0.644	\\
$\log(CO2) \times$D(Latin America - Caribbean)	&	0.973	&	$^{(d)}$	&	0.568	&	1.717	&	$^{(c)}$	&	0.810	&	5.467	&	$^{(a)}$	&	0.650	\\
$\log(CO2) \times$D(Middle East - North Africa)	&	-0.183	&		&	0.603	&	1.654	&	$^{(c)}$	&	0.813	&	4.258	&	$^{(a)}$	&	0.681	\\
$\log(CO2) \times$D(North America)	&	1.417	&	$^{(c)}$	&	0.563	&	2.340	&	$^{(b)}$	&	0.787	&	4.462	&	$^{(a)}$	&	0.672	\\
$\log(CO2) \times$D(South Asia)	&	-0.274	&		&	0.609	&	2.809	&	$^{(a)}$	&	0.778	&	5.545	&	$^{(a)}$	&	0.649	\\
$\log(CO2) \times$D(Sub-Saharan Africa)	&	0.539	&		&	0.576	&	1.026	&		&	0.870	&	3.608	&	$^{(a)}$	&	0.734	\\[5pt]
Adj. $R^2$	&	0.198	&		&		&	0.157	&		&		&	0.268	&		&		\\ \midrule
\textbf{Panel B: Scale parameter $\nu_t$}	&	\multicolumn{3}{c}{(nb obs.: 178)}					&	\multicolumn{3}{c}{(nb obs.: 176)}					&	\multicolumn{3}{c}{(nb obs.: 294)}					\\[5pt]
Constant	&	10.532	&	$^{(c)}$	&	5.253	&	5.365	&		&	10.030	&	18.380	&	$^{(a)}$	&	3.482	\\
$\log(GDP) \times$D(East Asia - Pacific)	&	-0.853	&		&	0.589	&	-0.136	&		&	1.085	&	-1.620	&	$^{(a)}$	&	0.378	\\
$\log(GDP) \times$D(Europe - Central Asia)	&	-0.780	&		&	0.518	&	-0.095	&		&	0.985	&	-1.473	&	$^{(a)}$	&	0.339	\\
$\log(GDP) \times$D(Latin America - Caribbean)	&	-0.824	&		&	0.553	&	-0.029	&		&	1.107	&	-1.520	&	$^{(a)}$	&	0.370	\\
$\log(GDP) \times$D(Middle East - North Africa)	&	-0.845	&		&	0.547	&	-0.222	&		&	1.037	&	-1.811	&	$^{(a)}$	&	0.374	\\
$\log(GDP) \times$D(North America)	&	-0.809	&		&	0.484	&	-0.031	&		&	0.949	&	-1.491	&	$^{(a)}$	&	0.330	\\
$\log(GDP) \times$D(South Asia)	&	-11.452	&	$^{(a)}$	&	0.645	&	0.075	&		&	1.270	&	-1.654	&	$^{(a)}$	&	0.438	\\
$\log(GDP) \times$D(Sub-Saharan Africa)	&	-0.939	&		&	0.656	&	-0.374	&		&	1.266	&	-10.968	&	$^{(a)}$	&	0.449	\\[5pt]
Adj. $R^2$	&	0.988	&		&		&	-0.004	&		&		&	0.967	&		&		\\ \midrule
\textbf{Panel C: Tail index $\xi_t$}	&		&		&		&		&		&		&		&		&		\\[5pt]
Constant	&	1.034	&	$^{(a)}$	&	0.080	&	0.581	&	$^{(a)}$	&	0.044	&	0.566	&	$^{(d)}$	&	0.305	\\
D(Europe - Central Asia)	&	-0.677	&	$^{(a)}$	&	0.094	&	1.964	&	$^{(a)}$	&	0.053	&	0.041	&		&	0.316	\\
D(Latin America - Caribbean)	&	-0.899	&	$^{(a)}$	&	0.141	&	-0.574	&	$^{(a)}$	&	0.122	&	0.066	&		&	0.351	\\
D(Middle East - North Africa)	&	-1.777	&	$^{(a)}$	&	0.084	&	-1.324	&	$^{(a)}$	&	0.053	&	-3.131	&	$^{(a)}$	&	0.308	\\
D(North America)	&	-0.480	&	$^{(a)}$	&	0.100	&	-0.014	&		&	0.064	&	-2.943	&	$^{(a)}$	&	0.306	\\
D(South Asia)	&	-1.978	&	$^{(a)}$	&	0.133	&	-0.162	&	$^{(c)}$	&	0.063	&	-0.360	&		&	0.326	\\
D(Sub-Saharan Africa)	&	-0.790	&	$^{(a)}$	&	0.123	&	-1.455	&	$^{(a)}$	&	0.045	&	-3.367	&	$^{(a)}$	&	0.367	\\	[5pt]
Adj. $R^2$	&	0.809	&		&		&	0.984	&		&	0.163	&	0.824	&		&		\\	[5pt]
Log-likelihood	&	-645.0	&		&		&	-1119.8	&		&		&	-1470.0	&		&		\\ \bottomrule
\end{tabular}
}}

\bigskip

\small{
\noindent Note: This table reports the parameter estimates of the Poisson-Generalized Pareto Distribution model for wildfires, heat waves, and cold waves. For each disaster type, the first column corresponds to the estimated parameter and the second column corresponds to the standard error. $^{(a)}$, $^{(b)}$, $^{(c)}$, and $^{(d)}$ correspond to statistical significance at the $0.1\%$, $1\%$, $5\%$, and $10\%$, respectively. ``nb obs.'' is the number of events with a strictly positive number of deaths.}\end{table}

\newpage\clearpage

\begin{table}[!ht]
\caption{Projections -- Scenario SSP1} \label{tab: Projections_1}
\begin{center}

\vspace{-1cm}
{\scalebox{0.73}[0.73]{   

\begin{tabular}{l rrr| rrr| rrr}
 \toprule
	&	\multicolumn{3}{c|}{\textbf{Floods}}					&	\multicolumn{3}{c|}{\textbf{Storms}}					&	\multicolumn{3}{c}{\textbf{Landslides}}			\\
	&	Nb of	&	Nb of 	&	Ann. nb	&	Nb of	&	Nb of 	&	Ann. nb	&	Nb of	&	Nb of 	&	Ann. nb	\\
	&	disasters	&	deaths	&	of deaths	&	disasters	&	deaths	&	of deaths	&	disasters	&	deaths	&	of deaths	\\
\midrule
\textbf{Panel A: 2015--2019}  &&&&&&&&& 	\\[3pt]																			
East Asia - Pacific	&	30.0	&	30.8	&	925.0	&	24.4	&	19.0	&	464.4	&	5.6	&	31.5	&	176.4	\\
Europe - Central Asia	&	9.8	&	6.9	&	68.0	&	7.6	&	4.0	&	30.6	&	0.8	&	17.8	&	14.2	\\
Latin America - Caribbean	&	18.8	&	12.1	&	228.0	&	10.2	&	28.5	&	290.2	&	2.4	&	77.9	&	187.0	\\
Middle East - North Africa	&	8.2	&	13.9	&	114.0	&	3.6	&	10.4	&	37.4	&	0.4	&	17.5	&	7.0	\\
North America	&	4.0	&	10.1	&	40.4	&	13.0	&	14.3	&	185.8	&	0.2	&	21.0	&	4.2	\\
South Asia	&	20.6	&	63.4	&	1305.4	&	9.8	&	53.8	&	527.4	&	4.4	&	40.1	&	176.4	\\
Sub-Saharan Africa	&	24.6	&	30.7	&	756.4	&	5.8	&	72.0	&	417.6	&	4.6	&	82.4	&	379.0	\\
World	&	116.0	&	29.6	&	3437.2	&	74.4	&	26.3	&	1953.4	&	18.4	&	51.3	&	944.2	\\ \midrule
\textbf{Panel B: 2040}  &&&&&&&&& 	\\[3pt]																			
East Asia - Pacific	&	47.4	&	8.4	&	397.1	&	45.4	&	4.7	&	215.3	&	7.1	&	23.9	&	170.8	\\
Europe - Central Asia	&	26.2	&	1.7	&	43.5	&	14.6	&	1.1	&	16.0	&	2.8	&	10.2	&	29.2	\\
Latin America - Caribbean	&	35.0	&	2.3	&	81.3	&	16.7	&	7.1	&	119.3	&	4.7	&	8.8	&	41.5	\\
Middle East - North Africa	&	10.1	&	4.8	&	48.7	&	1.7	&	4.9	&	8.5	&	0.3	&	13.8	&	3.9	\\
North America	&	7.3	&	2.1	&	15.6	&	17.3	&	4.9	&	85.0	&	0.1	&	0.0	&	0.0	\\
South Asia	&	25.1	&	10.8	&	269.8	&	10.4	&	5.2	&	53.7	&	3.7	&	8.6	&	32.1	\\
Sub-Saharan Africa	&	34.9	&	15.3	&	532.9	&	6.5	&	39.7	&	256.8	&	1.5	&	30.2	&	44.0	\\
World	&	186.0	&	7.5	&	1388.9	&	112.6	&	6.7	&	754.7	&	20.2	&	15.9	&	321.5	\\ \midrule
\textbf{Panel C: 2060}  &&&&&&&&& 	\\[3pt]																			
East Asia - Pacific	&	33.6	&	4.9	&	164.4	&	36.0	&	2.4	&	88.0	&	5.7	&	17.4	&	98.9	\\
Europe - Central Asia	&	19.0	&	0.7	&	12.8	&	12.1	&	0.5	&	6.0	&	2.3	&	6.2	&	14.7	\\
Latin America - Caribbean	&	25.1	&	0.3	&	8.1	&	13.8	&	1.0	&	13.4	&	3.8	&	1.8	&	7.0	\\
Middle East - North Africa	&	7.6	&	1.1	&	8.1	&	1.5	&	1.2	&	1.9	&	0.3	&	0.0	&	0.0	\\
North America	&	5.5	&	1.1	&	6.2	&	14.3	&	2.7	&	38.7	&	0.1	&	0.0	&	0.0	\\
South Asia	&	18.3	&	5.2	&	95.1	&	8.7	&	2.0	&	17.7	&	3.0	&	3.9	&	11.7	\\
Sub-Saharan Africa	&	25.0	&	7.4	&	185.3	&	5.6	&	22.1	&	124.0	&	1.2	&	7.8	&	9.6	\\
World	&	134.2	&	3.6	&	480.0	&	92.0	&	3.1	&	289.8	&	16.4	&	8.7	&	141.9	\\ \midrule
\textbf{Panel D: 2080}  &&&&&&&&& 	\\[3pt]																			
East Asia - Pacific	&	25.2	&	3.3	&	82.3	&	29.6	&	1.5	&	45.0	&	4.7	&	12.8	&	60.0	\\
Europe - Central Asia	&	14.5	&	0.3	&	4.1	&	10.4	&	0.2	&	2.5	&	2.0	&	3.8	&	7.5	\\
Latin America - Caribbean	&	19.0	&	0.1	&	1.4	&	11.7	&	0.2	&	2.8	&	3.2	&	0.5	&	1.7	\\
Middle East - North Africa	&	6.0	&	0.3	&	2.0	&	1.4	&	0.4	&	0.6	&	0.2	&	0.0	&	0.0	\\
North America	&	4.4	&	0.6	&	2.6	&	12.1	&	1.5	&	18.5	&	0.1	&	0.0	&	0.0	\\
South Asia	&	13.9	&	3.2	&	44.8	&	7.5	&	1.1	&	8.1	&	2.5	&	2.0	&	5.1	\\
Sub-Saharan Africa	&	18.9	&	4.7	&	89.3	&	4.9	&	15.4	&	76.0	&	1.1	&	1.6	&	1.7	\\
World	&	101.9	&	2.2	&	226.5	&	77.6	&	2.0	&	153.4	&	13.8	&	5.5	&	76.1	\\ \midrule
\textbf{Panel E: 2100}  &&&&&&&&& 	\\[3pt]																			
East Asia - Pacific	&	7.5	&	2.2	&	16.4	&	13.2	&	1.0	&	12.6	&	2.1	&	9.0	&	18.9	\\
Europe - Central Asia	&	4.7	&	0.1	&	0.6	&	5.3	&	0.1	&	0.6	&	1.0	&	2.2	&	2.2	\\
Latin America - Caribbean	&	5.9	&	0.0	&	0.1	&	5.9	&	0.1	&	0.4	&	1.5	&	0.2	&	0.3	\\
Middle East - North Africa	&	2.2	&	0.2	&	0.3	&	1.0	&	0.2	&	0.2	&	0.2	&	0.0	&	0.0	\\
North America	&	1.7	&	0.3	&	0.5	&	6.1	&	0.8	&	5.2	&	0.1	&	0.0	&	0.0	\\
South Asia	&	4.5	&	2.0	&	9.0	&	4.1	&	0.6	&	2.4	&	1.2	&	1.0	&	1.3	\\
Sub-Saharan Africa	&	5.9	&	3.5	&	20.5	&	2.8	&	12.4	&	34.9	&	0.6	&	0.3	&	0.2	\\
World	&	32.4	&	1.5	&	47.4	&	38.4	&	1.5	&	56.3	&	6.7	&	3.4	&	22.9	\\
 \bottomrule
\end{tabular}
}}

\end{center}
\vspace{-0.25cm}

\footnotesize{
\noindent Note: This table reports projections of the number of disasters per year, the number of deaths per disaster, and the annual number of deaths for floods, storms, and landslides in Scenario SSP1. The number of deaths corresponds to the population projected for the given year.}

\end{table}


\newpage\clearpage

\begin{table}[!ht]
\caption{Projections -- Scenario SSP1 (continued)} \label{tab: Projections_2}
\begin{center}

\vspace{-1cm}
{\scalebox{0.73}[0.73]{   

\begin{tabular}{l rrr| rrr |rrr}
 \toprule
	&\multicolumn{3}{c|}{\textbf{Wildfires}}&	\multicolumn{3}{c|}{\textbf{Heat waves}}					&	\multicolumn{3}{c}{\textbf{Cold waves}}					\\	
	&	Nb of	&	Nb of 	&	Ann. nb	&	Nb of	&	Nb of 	&	Ann. nb	&	Nb of	&	Nb of 	&	Ann. nb	\\
	&	disasters	&	deaths	&	of deaths	&	disasters	&	deaths	&	of deaths	&	disasters	&	deaths	&	of deaths	\\
\midrule
\textbf{Panel A: 2015--2019}  &&&&&&&&& 	\\[3pt]																		
East Asia - Pacific	&	1.2	&	14.3	&	17.2	&	1.4	&	47.7	&	66.8	&	1.0	&	23.6	&	23.6	\\
Europe - Central Asia	&	1.2	&	41.0	&	49.2	&	2.2	&	505.7	&	1112.6	&	3.0	&	17.3	&	52.0	\\
Latin America - Caribbean	&	0.4	&	6.0	&	2.4	&	0.0	&	0.0	&	0.0	&	0.2	&	21.0	&	4.2	\\
Middle East - North Africa	&	0.0	&	0.0	&	0.0	&	0.2	&	110.0	&	22.0	&	0.2	&	8.0	&	1.6	\\
North America	&	2.4	&	13.9	&	33.4	&	0.2	&	70.0	&	14.0	&	0.0	&	0.0	&	0.0	\\
South Asia	&	0.6	&	11.7	&	7.0	&	1.4	&	619.0	&	866.6	&	0.8	&	44.5	&	35.6	\\
Sub-Saharan Africa	&	0.6	&	20.7	&	12.4	&	0.4	&	13.5	&	5.4	&	0.0	&	0.0	&	0.0	\\
World	&	6.4	&	19.0	&	121.6	&	5.8	&	359.9	&	2087.4	&	5.2	&	22.5	&	117.0	\\ \midrule
\textbf{Panel B: 2040}  &&&&&&&&& 	\\[3pt]																			
East Asia - Pacific	&	2.9	&	3.5	&	10.3	&	1.1	&	76.0 &	83.2	&	0.8	&	3.3	&	2.6	\\
Europe - Central Asia	&	3.2	&	9.6	&	30.8	&	3.0	&	270.7$^*$	&	823.5	&	9.9	&	35.8	&	359.4	\\
Latin America - Caribbean	&	1.4	&	7.9	&	11.0	&	0.3	&	145.5	&	39.2	&	2.8	&	32.9	&	91.9	\\
Middle East - North Africa	&	0.2	&	9.1	&	1.6	&	0.2	&	34.3	&	6.7	&	0.4	&	3.5	&	1.3	\\
North America	&	2.8	&	6.2	&	17.7	&	0.7	&	198.3	&	140.2	&	0.6	&	14.6	&	7.7	\\
South Asia	&	0.2	&	0.0	&	0.0	&	1.5	&	589.5	&	866.8	&	3.2	&	8.5	&	27.6	\\
Sub-Saharan Africa	&	0.7	&	9.5	&	6.7	&	0.1	&	8.2	&	0.8	&	0.2	&	0.0	&	0.0	\\
World	&	11.5	&	6.8	&	78.1	&	6.9	&	284.3	&	1960.3	&	17.9	&	27.3	&	490.6	\\ \midrule
\textbf{Panel C: 2060}  &&&&&&&&& 	\\[3pt]																			
East Asia - Pacific	&	2.4	&	2.1	&	5.2	&	0.9	&	68.3	&	58.1	&	0.6	&	1.4	&	0.8	\\
Europe - Central Asia	&	2.7	&	7.9	&	21.1	&	2.3	&	48.9$^*$	&	420.6	&	6.2	&	25.7	&	159.0	\\
Latin America - Caribbean	&	1.2	&	4.8	&	5.7	&	0.2	&	137.0	&	29.3	&	1.9	&	14.5	&	26.5	\\
Middle East - North Africa	&	0.2	&	4.9	&	0.9	&	0.2	&	28.2	&	4.8	&	0.3	&	1.4	&	0.4	\\
North America	&	2.4	&	5.5	&	13.1	&	0.5	&	207.0	&	116.1	&	0.4	&	10.5	&	3.9	\\
South Asia	&	0.2	&	0.0	&	0.0	&	1.1	&	619.4	&	701.9	&	2.1	&	3.0	&	6.3	\\
Sub-Saharan Africa	&	0.6	&	4.9	&	3.0	&	0.1	&	6.3	&	0.5	&	0.1	&	0.0	&	0.0	\\
World	&	9.6	&	5.1	&	48.8	&	5.3	&	250.9	&	1331.3	&	11.4	&	17.2	&	196.9	\\ \midrule
\textbf{Panel D: 2080}  &&&&&&&&& 	\\[3pt]																			
East Asia - Pacific	&	2.1	&	1.4	&	2.9	&	0.7	&	56.5	&	38.8	&	0.4	&	0.8	&	0.3	\\
Europe - Central Asia	&	2.3	&	6.3	&	14.4	&	1.8	&	125.7$^*$	&	223.6	&	4.1	&	18.0	&	74.5	\\
Latin America - Caribbean	&	1.0	&	3.0	&	3.1	&	0.2	&	113.5	&	21.2	&	1.3	&	7.4	&	9.3	\\
Middle East - North Africa	&	0.2	&	2.8	&	0.5	&	0.2	&	21.9	&	3.2	&	0.2	&	0.7	&	0.1	\\
North America	&	2.0	&	4.7	&	9.6	&	0.4	&	207.0	&	94.6	&	0.3	&	7.5	&	2.0	\\
South Asia	&	0.1	&	0.0	&	0.0	&	0.9	&	591.5	&	536.2	&	1.4	&	1.3	&	1.9	\\
Sub-Saharan Africa	&	0.5	&	2.5	&	1.4	&	0.1	&	4.4	&	0.3	&	0.1	&	0.0	&	0.0	\\
World	&	8.3	&	3.8	&	31.9	&	4.2	&	216.3	&	918.0	&	7.8	&	11.3	&	88.2	\\ \midrule
\textbf{Panel E: 2100}  &&&&&&&&& 	\\[3pt]																			
East Asia - Pacific	&	1.1	&	0.7	&	0.7	&	0.3	&	43.3	&	12.2	&	0.1	&	0.4	&	0.0	\\
Europe - Central Asia	&	1.2	&	4.8	&	5.7	&	0.6	&	12.7$^*$	&	8.1	&	0.8	&	12.2	&	9.4	\\
Latin America - Caribbean	&	0.6	&	1.9	&	1.1	&	0.1	&	90.8	&	8.5	&	0.3	&	3.8	&	1.1	\\
Middle East - North Africa	&	0.1	&	1.8	&	0.2	&	0.1	&	16.6	&	1.3	&	0.1	&	0.4	&	0.0	\\
North America	&	1.1	&	3.9	&	4.2	&	0.2	&	198.9	&	39.6	&	0.1	&	5.1	&	0.4	\\
South Asia	&	0.1	&	0.0	&	0.0	&	0.4	&	519.7	&	187.1	&	0.3	&	0.6	&	0.2	\\
Sub-Saharan Africa	&	0.3	&	1.4	&	0.5	&	0.0	&	3.1	&	0.1	&	0.0	&	0.0	&	0.0	\\
World	&	4.5	&	2.7	&	12.5	&	1.7	&	151.9	&	256.9	&	1.6	&	6.9	&	11.1	\\
\bottomrule
\end{tabular}
}}

\end{center}
\vspace{-0.25cm}

\footnotesize{
\noindent Note: This table reports projections of the number of disasters per year, the number of deaths per disaster, and the annual number of deaths for wildfires, heat waves, and cold waves in Scenario SSP1. The number of deaths corresponds to the population projected for the given year. For the projection of the number of deaths due to heat waves in Europe and Central Asia, we report the median instead of the expected value because the tail index is greater than 1.}
\end{table}


\newpage\clearpage

\begin{table}[!ht]
\caption{Projections -- Scenario SSP3} \label{tab: Projections_3}
\begin{center}

\vspace{-1cm}
{\scalebox{0.73}[0.73]{   

\begin{tabular}{l rrr| rrr |rrr}
 \toprule
	&	\multicolumn{3}{c|}{\textbf{Floods}}					&	\multicolumn{3}{c|}{\textbf{Storms}}					&	\multicolumn{3}{c}{\textbf{Landslides}}					\\
	&	Nb of	&	Nb of 	&	Ann. nb	&	Nb of	&	Nb of 	&	Ann. nb	&	Nb of	&	Nb of 	&	Ann. nb	\\
	&	disasters	&	deaths	&	of deaths	&	disasters	&	deaths	&	of deaths	&	disasters	&	deaths	&	of deaths	\\
 \midrule
\textbf{Panel A: 2015--2019}  &&&&&&&&& 	\\[3pt]																			
East Asia - Pacific	&	30.0	&	30.8	&	925.0	&	24.4	&	19.0	&	464.4	&	5.6	&	31.5	&	176.4	\\
Europe - Central Asia	&	9.8	&	6.9	&	68.0	&	7.6	&	4.0	&	30.6	&	0.8	&	17.8	&	14.2	\\
Latin America - Caribbean	&	18.8	&	12.1	&	228.0	&	10.2	&	28.5	&	290.2	&	2.4	&	77.9	&	187.0	\\
Middle East - North Africa	&	8.2	&	13.9	&	114.0	&	3.6	&	10.4	&	37.4	&	0.4	&	17.5	&	7.0	\\
North America	&	4.0	&	10.1	&	40.4	&	13.0	&	14.3	&	185.8	&	0.2	&	21.0	&	4.2	\\
South Asia	&	20.6	&	63.4	&	1305.4	&	9.8	&	53.8	&	527.4	&	4.4	&	40.1	&	176.4	\\
Sub-Saharan Africa	&	24.6	&	30.7	&	756.4	&	5.8	&	72.0	&	417.6	&	4.6	&	82.4	&	379.0	\\
World	&	116.0	&	29.6	&	3437.2	&	74.4	&	26.3	&	1953.4	&	18.4	&	51.3	&	944.2	\\ \midrule
\textbf{Panel B: 2040}  &&&&&&&&& 	\\[3pt]																			
East Asia - Pacific	&	76.3	&	16.7	&	1276.6	&	62.5	&	11.1	&	692.9	&	9.8	&	34.3	&	337.7	\\
Europe - Central Asia	&	40.9	&	2.2	&	91.2	&	18.9	&	1.4	&	26.5	&	3.7	&	11.5	&	42.9	\\
Latin America - Caribbean	&	55.5	&	11.3	&	624.8	&	21.8	&	24.0	&	519.6	&	6.3	&	30.1	&	189.9	\\
Middle East - North Africa	&	14.9	&	15.0	&	224.7	&	2.0	&	14.1	&	28.2	&	0.3	&	31.0	&	10.5	\\
North America	&	10.6	&	2.9	&	30.4	&	22.7	&	6.3	&	143.6	&	0.2	&	0.2	&	0.0	\\
South Asia	&	39.1	&	24.6	&	960.0	&	13.2	&	13.9	&	184.2	&	4.9	&	18.7	&	91.5	\\
Sub-Saharan Africa	&	55.2	&	31.8	&	1757.8	&	8.1	&	81.1	&	656.6	&	1.8	&	63.7	&	118.8	\\
World	&	292.6	&	17.0	&	4965.6	&	149.2	&	15.1	&	2251.6	&	27.0	&	29.4	&	791.4	\\ \midrule
\textbf{Panel C: 2060}  &&&&&&&&& 	\\[3pt]																			
East Asia - Pacific	&	74.2	&	16.8	&	1250.9	&	61.3	&	11.1	&	678.7	&	9.6	&	35.2	&	341.0	\\
Europe - Central Asia	&	39.9	&	1.1	&	44.6	&	18.7	&	0.8	&	14.2	&	3.6	&	7.5	&	27.7	\\
Latin America - Caribbean	&	54.0	&	7.1	&	386.0	&	21.5	&	20.0	&	430.5	&	6.2	&	22.1	&	137.4	\\
Middle East - North Africa	&	14.6	&	14.0	&	203.3	&	2.0	&	13.5	&	26.5	&	0.3	&	33.4	&	11.3	\\
North America	&	10.4	&	1.8	&	19.1	&	22.4	&	4.2	&	94.2	&	0.2	&	0.0	&	0.0	\\
South Asia	&	38.1	&	21.0	&	798.0	&	13.0	&	11.0	&	144.3	&	4.8	&	16.5	&	79.8	\\
Sub-Saharan Africa	&	53.8	&	31.4	&	1689.4	&	8.0	&	79.9	&	638.6	&	1.8	&	63.1	&	116.7	\\
World	&	284.9	&	15.4	&	4391.3	&	146.8	&	13.8	&	2026.9	&	26.5	&	26.9	&	713.9	\\ \midrule
\textbf{Panel D: 2080}  &&&&&&&&& 	\\[3pt]																			
East Asia - Pacific	&	76.2	&	16.9	&	1289.6	&	62.4	&	11.1	&	694.7	&	9.8	&	35.3	&	347.5	\\
Europe - Central Asia	&	40.9	&	0.6	&	24.4	&	18.9	&	0.4	&	8.2	&	3.7	&	5.0	&	18.8	\\
Latin America - Caribbean	&	55.4	&	4.5	&	247.1	&	21.8	&	13.5	&	294.9	&	6.3	&	15.8	&	99.9	\\
Middle East - North Africa	&	14.9	&	9.6	&	143.8	&	2.0	&	9.6	&	19.2	&	0.3	&	27.1	&	8.9	\\
North America	&	10.6	&	1.2	&	13.2	&	22.7	&	2.9	&	66.5	&	0.2	&	0.0	&	0.0	\\
South Asia	&	39.0	&	18.2	&	710.5	&	13.2	&	9.0	&	119.6	&	4.9	&	14.6	&	71.8	\\
Sub-Saharan Africa	&	55.1	&	26.9	&	1484.1	&	8.1	&	71.8	&	575.8	&	1.8	&	49.3	&	90.9	\\
World	&	292.1	&	13.4	&	3912.8	&	149.1	&	11.9	&	1778.9	&	26.9	&	23.7	&	637.9	\\ \midrule
\textbf{Panel E: 2100}  &&&&&&&&& 	\\[3pt]																			
East Asia - Pacific	&	86.4	&	17.1	&	1480.3	&	67.9	&	11.3	&	770.1	&	10.6	&	35.3	&	377.3	\\
Europe - Central Asia	&	46.0	&	0.4	&	16.2	&	20.3	&	0.3	&	5.5	&	4.0	&	3.6	&	14.2	\\
Latin America - Caribbean	&	62.5	&	3.0	&	189.4	&	23.4	&	9.3	&	217.7	&	6.8	&	11.9	&	81.4	\\
Middle East - North Africa	&	16.6	&	5.8	&	96.2	&	2.0	&	6.0	&	12.5	&	0.3	&	15.6	&	4.9	\\
North America	&	11.7	&	0.9	&	10.8	&	24.4	&	2.2	&	54.0	&	0.2	&	0.0	&	0.0	\\
South Asia	&	43.9	&	16.2	&	708.5	&	14.1	&	7.6	&	107.3	&	5.3	&	13.0	&	69.1	\\
Sub-Saharan Africa	&	62.3	&	21.8	&	1360.1	&	8.5	&	62.0	&	526.7	&	1.9	&	30.1	&	60.6	\\
World	&	329.3	&	11.7	&	3861.4	&	160.6	&	10.5	&	1693.8	&	29.1	&	20.9	&	607.5	\\
 \bottomrule
\end{tabular}
}}

\end{center}
\vspace{-0.25cm}

\footnotesize{
\noindent Note: This table reports projections of the number of disasters per year, the number of deaths per disaster, and the annual number of deaths for floods, storms, and landslides in Scenario SSP3. The number of deaths corresponds to the population projected for the given year.}
\end{table}


\newpage\clearpage

\begin{table}[!ht]
\caption{Projections -- Scenario SSP3 (continued)} \label{tab: Projections_4}
\begin{center}

\vspace{-1cm}
{\scalebox{0.73}[0.73]{   

\begin{tabular}{l rrr| rrr| rrr}
 \toprule
	&\multicolumn{3}{c|}{\textbf{Wildfires}}&	\multicolumn{3}{c|}{\textbf{Heat waves}}					&	\multicolumn{3}{c}{\textbf{Cold waves}}					\\	
	&	Nb of	&	Nb of 	&	Ann. nb	&	Nb of	&	Nb of 	&	Ann. nb	&	Nb of	&	Nb of 	&	Ann. nb	\\
	&	disasters	&	deaths	&	of deaths	&	disasters	&	deaths	&	of deaths	&	disasters	&	deaths	&	of deaths	\\
\midrule
\textbf{Panel A: 2015--2019}  &&&&&&&&& 	\\[3pt]		
East Asia - Pacific	&	1.2	&	14.3	&	17.2	&	1.4	&	47.7	&	66.8	&	1.0	&	23.6	&	23.6	\\
Europe - Central Asia	&	1.2	&	41.0	&	49.2	&	2.2	&	505.7	&	1112.6	&	3.0	&	17.3	&	52.0	\\
Latin America - Caribbean	&	0.4	&	6.0	&	2.4	&	0.0	&	0.0	&	0.0	&	0.2	&	21.0	&	4.2	\\
Middle East - North Africa	&	0.0	&	0.0	&	0.0	&	0.2	&	110.0	&	22.0	&	0.2	&	8.0	&	1.6	\\
North America	&	2.4	&	13.9	&	33.4	&	0.2	&	70.0	&	14.0	&	0.0	&	0.0	&	0.0	\\
South Asia	&	0.6	&	11.7	&	7.0	&	1.4	&	619.0	&	866.6	&	0.8	&	44.5	&	35.6	\\
Sub-Saharan Africa	&	0.6	&	20.7	&	12.4	&	0.4	&	13.5	&	5.4	&	0.0	&	0.0	&	0.0	\\
World	&	6.4	&	19.0	&	121.6	&	5.8	&	359.9	&	2087.4	&	5.2	&	22.5	&	117.0	\\ \midrule
\textbf{Panel B: 2040}  &&&&&&&&& 	\\[3pt]																			
East Asia - Pacific	&	3.8	&	6.2	&	23.2	&	1.5	&	85.1	&	132.7	&	1.4	&	8.3	&	11.6	\\
Europe - Central Asia	&	4.2	&	9.4	&	39.6	&	4.5	&	426.3$^*$	&	1939.3	&	19.3	&	37.8	&	739.7	\\
Latin America - Caribbean	&	1.7	&	12.6	&	21.9	&	0.3	&	173.1	&	60.2	&	5.2	&	64.9	&	330.2	\\
Middle East - North Africa	&	0.2	&	12.1	&	3.0	&	0.3	&	34.5	&	10.8	&	0.7	&	7.2	&	4.4	\\
North America	&	3.6	&	6.2	&	23.1	&	1.0	&	184.3	&	179.7	&	0.9	&	16.4	&	14.4	\\
South Asia	&	0.2	&	0.0	&	0.0	&	2.1	&	632.1	&	1348.3	&	5.9	&	22.2	&	132.0	\\
Sub-Saharan Africa	&	0.8	&	17.3	&	14.6	&	0.1	&	12.1	&	1.5	&	0.3	&	0.0	&	0.0	\\
World	&	14.6	&	8.6	&	125.4	&	9.9	&	369.4	&	3672.6	&	33.6	&	36.7	&	1232.3	\\ \midrule
\textbf{Panel C: 2060}  &&&&&&&&& 	\\[3pt]																			
East Asia - Pacific	&	3.7	&	6.3	&	23.3	&	1.5	&	88.7	&	135.3	&	1.4	&	8.3	&	11.2	\\
Europe - Central Asia	&	4.1	&	7.5	&	31.0	&	4.4	&	380.5$^*$	&	1690.4	&	18.6	&	27.4	&	517.2	\\
Latin America - Caribbean	&	1.7	&	12.6	&	21.5	&	0.3	&	194.4	&	66.8	&	5.0	&	59.5	&	292.1	\\
Middle East - North Africa	&	0.2	&	13.5	&	3.3	&	0.3	&	41.5	&	12.8	&	0.6	&	7.5	&	4.5	\\
North America	&	3.6	&	5.5	&	19.9	&	0.9	&	176.5	&	168.7	&	0.9	&	12.8	&	10.9	\\
South Asia	&	0.2	&	0.0	&	0.0	&	2.1	&	754.9	&	1564.4	&	5.7	&	18.1	&	104.1	\\
Sub-Saharan Africa	&	0.8	&	18.8	&	15.6	&	0.1	&	15.1	&	1.8	&	0.2	&	0.0	&	0.0	\\
World	&	14.4	&	8.0	&	114.6	&	9.7	&	374.0	&	3640.2	&	32.4	&	29.0	&	940.0	\\ \midrule
\textbf{Panel D: 2080}  &&&&&&&&& 	\\[3pt]																			
East Asia - Pacific	&	3.8	&	6.3	&	23.7	&	1.5	&	88.6	&	138.1	&	1.4	&	8.4	&	11.6	\\
Europe - Central Asia	&	4.2	&	5.9	&	24.6	&	4.5	&	349.1$^*$	&	1585.6	&	19.3	&	19.9	&	388.9	\\
Latin America - Caribbean	&	1.7	&	12.1	&	20.9	&	0.3	&	209.8	&	73.4	&	5.2	&	52.6	&	265.6	\\
Middle East - North Africa	&	0.2	&	12.5	&	3.1	&	0.3	&	43.5	&	13.6	&	0.7	&	6.3	&	3.9	\\
North America	&	3.6	&	4.7	&	17.4	&	1.0	&	163.8	&	159.3	&	0.9	&	10.0	&	8.8	\\
South Asia	&	0.2	&	0.0	&	0.0	&	2.1	&	860.3	&	1817.5	&	5.9	&	15.0	&	88.4	\\
Sub-Saharan Africa	&	0.8	&	17.4	&	14.5	&	0.1	&	16.5	&	2.0	&	0.3	&	0.0	&	0.0	\\
World	&	14.6	&	7.2	&	104.3	&	9.9	&	381.7	&	3789.5	&	33.5	&	22.9	&	767.1	\\ \midrule
\textbf{Panel E: 2100}  &&&&&&&&& 	\\[3pt]																		
East Asia - Pacific	&	4.0	&	6.4	&	25.8	&	1.7	&	87.9	&	149.8	&	1.6	&	8.5	&	13.6	\\
Europe - Central Asia	&	4.5	&	4.8	&	21.4	&	5.0	&	363.4$^*$	&	1834.8	&	22.9	&	15.2	&	352.9	\\
Latin America - Caribbean	&	1.8	&	11.7	&	21.4	&	0.4	&	224.0	&	84.3	&	6.0	&	47.5	&	280.7	\\
Middle East - North Africa	&	0.3	&	10.3	&	2.6	&	0.3	&	41.3	&	13.9	&	0.8	&	4.7	&	3.4	\\
North America	&	3.9	&	4.2	&	16.6	&	1.0	&	154.6	&	163.6	&	1.1	&	8.3	&	8.4	\\
South Asia	&	0.2	&	0.0	&	0.0	&	2.3	&	969.7	&	2257.6	&	6.9	&	12.5	&	86.4	\\
Sub-Saharan Africa	&	0.9	&	14.5	&	12.6	&	0.2	&	16.5	&	2.1	&	0.3	&	0.0	&	0.0	\\
World	&	15.5	&	6.5	&	100.4	&	10.9	&	412.1	&	4506.1	&	39.6	&	18.8	&	745.3	\\
\bottomrule
\end{tabular}
}}

\end{center}
\vspace{-0.25cm}

\footnotesize{
\noindent Note: This table reports projections of the number of disasters per year, the number of deaths per disaster, and the annual number of deaths for wildfires, heat waves, and cold waves in Scenario SSP3. The number of deaths corresponds to the population projected for the given year. For the projection of the number of deaths due to heat waves in Europe and Central Asia, we report the median instead of the expected value because the tail index is greater than 1.}
\end{table}

\clearpage\newpage

\vspace{-.5cm}
\footnotesize{
\noindent Note: This figure reports the world population, the aggregate PPP GDP per capita, the aggregate CO$_2$ emission, and the aggregate CO$_2$ emission per capita. The vertical line corresponds to 2019, i.e., the last year of observable variables. Projections are those from IPCC, adapted to updated data until 2019.}

\clearpage\newpage

%
%
%
%
%
%
%
%
%
%
%
%
%
%
%
%
%
%
%
%

\clearpage\newpage

\vspace{-.5cm}
\footnotesize{
\noindent Note: This figure reports the average number of disasters per year and the $95\%$ uncertainty intervals for the six disaster types. }

\clearpage\newpage

\vspace{-.5cm}
\noindent Note: This figure reports the average number of deaths per disaster and the $95\%$ uncertainty intervals for the six disaster types. 

\clearpage\newpage

\vspace{-.5cm}
\footnotesize{

\noindent Note: This figure reports the average annual number of deaths and the $95\%$ uncertainty intervals for the six disaster types. }

\clearpage\newpage

\vspace{-.5cm}
\noindent Note: This figure reports the average number of disasters per year and the $95\%$ uncertainty intervals for the six disaster types.

\clearpage\newpage

\vspace{-.5cm}
\footnotesize{

\noindent Note: This figure reports the average number of deaths per disaster and the $95\%$ uncertainty intervals for the six disaster types. }

\clearpage\newpage

\vspace{-.5cm}
\footnotesize{

\noindent Note: This figure reports the average annual number of deaths and the $95\%$ uncertainty intervals for the six disaster types. }


\newpage\clearpage
\appendix
\appendixpage
\setcounter{table}{0}
\setcounter{equation}{0}
\setcounter{figure}{0}
\renewcommand{\theequation}{A.\arabic{equation}}
\renewcommand{\thetable}{A.\arabic{table}}
\renewcommand{\thefigure}{A.\arabic{figure}}

\renewcommand{\baselinestretch}{1.1}
\selectfont














\normalsize

\section{Likelihood Ratio Tests for Model Selection}\label{app: LRTests}

The specification of GPD parameters has been chosen as follows. 

\medskip

\noindent \textbf{Specification for $\nu_t^{(r)}$ (Table \ref{tab: LRT1}):}

\begin{itemize}

\item For floods, storms, and landslides, the specification for $\nu_t^{(r)}$ is unambiguous: the combination of regional dummies and log(GDP)$\times$regional dummies provides the highest likelihood. Parameters associated with log(GDP) have the expected negative sign.

\item For wildfires and heat waves, the specification for $\nu_t^{(r)}$ is again unambiguous: the model with log(GDP)$\times$regional dummies provides the highest likelihood. For cold waves, the model with regional dummies and log(GDP)$\times$regional dummies would provide the highest likelihood. However, parameters associated with log(GDP) do not have the expected (negative) sign. For consistency, we kept the same model for all three disaster types with log(GDP)$\times$regional dummies.

\end{itemize}

\medskip

\noindent \textbf{Specification for $\xi_t^{(r)}$ (Table \ref{tab: LRT2}):}

\begin{itemize}

\item For floods and storms, the LR statistics point in favor of the model with log(GDP)$\times$ regional dummies. For landslides, the models with regional dummies only and with log(GDP)$\times$ regional dummies have similar likelihoods. We keep the same model as for floods and storms. For storms, the model with regional dummies and log(GDP)$\times$ regional dummies does not converge. 

\item For wildfires, heat waves, and cold waves, all LR statistics point in favor of the model with regional dummies only. In all cases, in the model with log(GDP)$\times$regional dummies, parameters associated with log(GDP) do not have the expected (negative) sign. 

\end{itemize}

\medskip

The final specifications for $\nu_t^{(r)}$ and $\xi_t^{(r)}$ are for floods, storms, and landslides:
\begin{eqnarray*}
\nu_t &=& \nu_0 +  \sum_{r=1}^7 \nu_{1,r} \log(GDP_{t}^{(r)}) \times R_{t}^{(r)} +  \sum_{r=1}^6 \nu_{2,r} R_{t}^{(r)} + \epsilon_{\nu,t} ,\\
\xi_t &=& \xi_0 +  \sum_{r=1}^7 \xi_{1,r} \log(GDP_{t}^{(r)}) \times R_{t}^{(r)} + \epsilon_{\xi,t}, 
\end{eqnarray*}

and for wildfires, heat waves, and cold waves:
\begin{eqnarray*}
\nu_t &=& \nu_0 +   \sum_{r=1}^7 \nu_{1,r} \log(GDP_{t}^{(r)}) \times R_{t}^{(r)} + \epsilon_{\nu,t} ,\\
\xi_t &=& \xi_0 +  \sum_{r=1}^6 \xi_{1,r} R_{t}^{(r)}+ \epsilon_{\xi,t},
\end{eqnarray*}
where $R_{t}^{(r)}= 1$ if the disaster that occurs on day $t$ is located in region $r$, and 0 otherwise.


\bigskip


\begin{table}[h!]
\caption{Likelihood Ratio Tests for $\nu_t^{(r)}$}  \label{tab: LRT1}

\vspace{-.5cm}
{\scalebox{0.85}[0.85]{   

\begin{tabular}{l cccccc}
 \toprule
Specification	&	Flood	&	Storm	&	Landslide	&	Wildfire	&	Heat wave	&	Cold wave	\\ \midrule
(1) Constant	&	-18468.0	&	-14363.1	&	-3586.8	&	-666.8	&	-1153.6	&	-1539.7	\\[5pt]
(2) GDP	&	-18127.5	&	-14108.9	&	-3567.7	&	-656.3	&	-1148.2	&	-1489.0	\\
wrt Constant	&	(0.000)	&	(0.000)	&	(0.000)	&	(0.000)	&	(0.001)	&	(0.000)	\\[5pt]
(3) Regions	&	-18139.5	&	-14155.8	&	-3568.7	&	-655.9	&	-1139.8	&	-1492.1	\\
wrt Constant	&	(0.001)	&	(0.000)	&	(0.000)	&	(0.000)	&	(0.000)	&	(0.000)	\\[5pt]
(4) GDP by Regions	&	-17912.3	&	-13932.3	&	-3535.0	&	\textbf{-652.3}	&	\textbf{-1139.8}&	\textbf{-1479.4}	\\
wrt GDP	&	(0.000)	&	(0.000)	&	(0.000)	&	(0.233)	&	(0.010)	&	(0.004)	\\[5pt]
(5) GDP by Regions+Regions	&	\textbf{-17849.1}	&	\textbf{-13897.7}	&	\textbf{-3519.7}	&	-651.5	&	-1138.1	&	-1471.3	\\
wrt Regions	&	(0.000)	&	(0.000)	&	(0.000)	&	(0.258)	&	(0.832)	&	(0.000)	\\
wrt GDP by Regions	&	(0.000)	&	(0.000)	&	(0.000)	&	(0.949)	&	(0.743)	&	(0.012)	\\ \bottomrule
\end{tabular}
}}
\end{table}
%
%

\begin{table}[h!]
\caption{Likelihood Ratio Tests for $\xi_t^{(r)}$}  \label{tab: LRT2}

\vspace{-.5cm}
{\scalebox{0.85}[0.85]{   

\begin{tabular}{l cccccc}
 \toprule
Specification	&	Flood	&	Storm	&	Landslide	&	Wildfire	&	Heat wave	&	Cold wave	\\ \midrule
(1) Constant	&	-17849.1	&	-13897.7	&	-3519.7	&	-652.3	&	-1139.8	&	-1479.4	\\[5pt]
(2) GDP	&	-17839.6	&	-13862.2	&	-3519.5	&	-652.3	&	-1130.5	&	-1474.6	\\
wrt Constant	&	(0.000)	&	(0.000)	&	(0.567)	&	(0.805)	&	(0.000)	&	(0.000)	\\[5pt]
(3) Regions	&	-17839.1	&	-13842.0	&	-3513.8	&	\textbf{-645.0}	&	\textbf{-1119.6}	&	\textbf{-1470.0}	\\
wrt Constant	&	(0.003)	&	(0.000)	&	(0.069)	&	(0.023)	&	(0.000)	&	(0.100)	\\[5pt]
(4) GDP by Regions	&	\textbf{-17825.8}	&	\textbf{-13833.0}	&	\textbf{-3513.5}	&	-644.5	&	-1119.5	&	-1470.0	\\
wrt GDP	&	(0.000)	&	(0.000)	&	(0.062)	&	(0.017)	&	(0.001)	&	(0.009)	\\[5pt]
(5) GDP by Regions+Regions	&	-17820.5	&	no cvg	&	-3510.7	&	-640.4	&	-1117.8	&	-1464.5	\\
wrt Regions	&	(0.000)	&	(--)	&	(0.508)	&	(0.235)	&	(0.838)	&	(0.136)	\\
wrt GDP by Regions	&	(0.104)	&	(--)	&	(0.464)	&	(0.217)	&	(0.869)	&	(0.086)	\\ \bottomrule
\end{tabular}
}}

\bigskip
\small{
\noindent Note: These tables report results of the likelihood ratio tests. For each specification, the first row corresponds to the log likelihood of the model, the subsequent rows correspond to the $p$-value of the likelihood ratio test. 
}
\end{table}


\section{Additional Regional Results}\label{app: Regional Results}

In this section, we represent the historical and projected number of disasters and annual number of deaths for each region and each disaster type. We indicate the world number at the top of the bars. We consider the data for Scenarios SSP1 and SSP3.

\clearpage\newpage

\begin{figure}[h!]
\caption{Number of disasters per year -- Scenario SSP1} \label{fig: Number of disasters}

\vspace{-1.25cm}

\begin{center}
\bigskip
\includegraphics[scale=.95,trim= 2.25cm 8.5cm 2.25cm 8.75cm,clip]{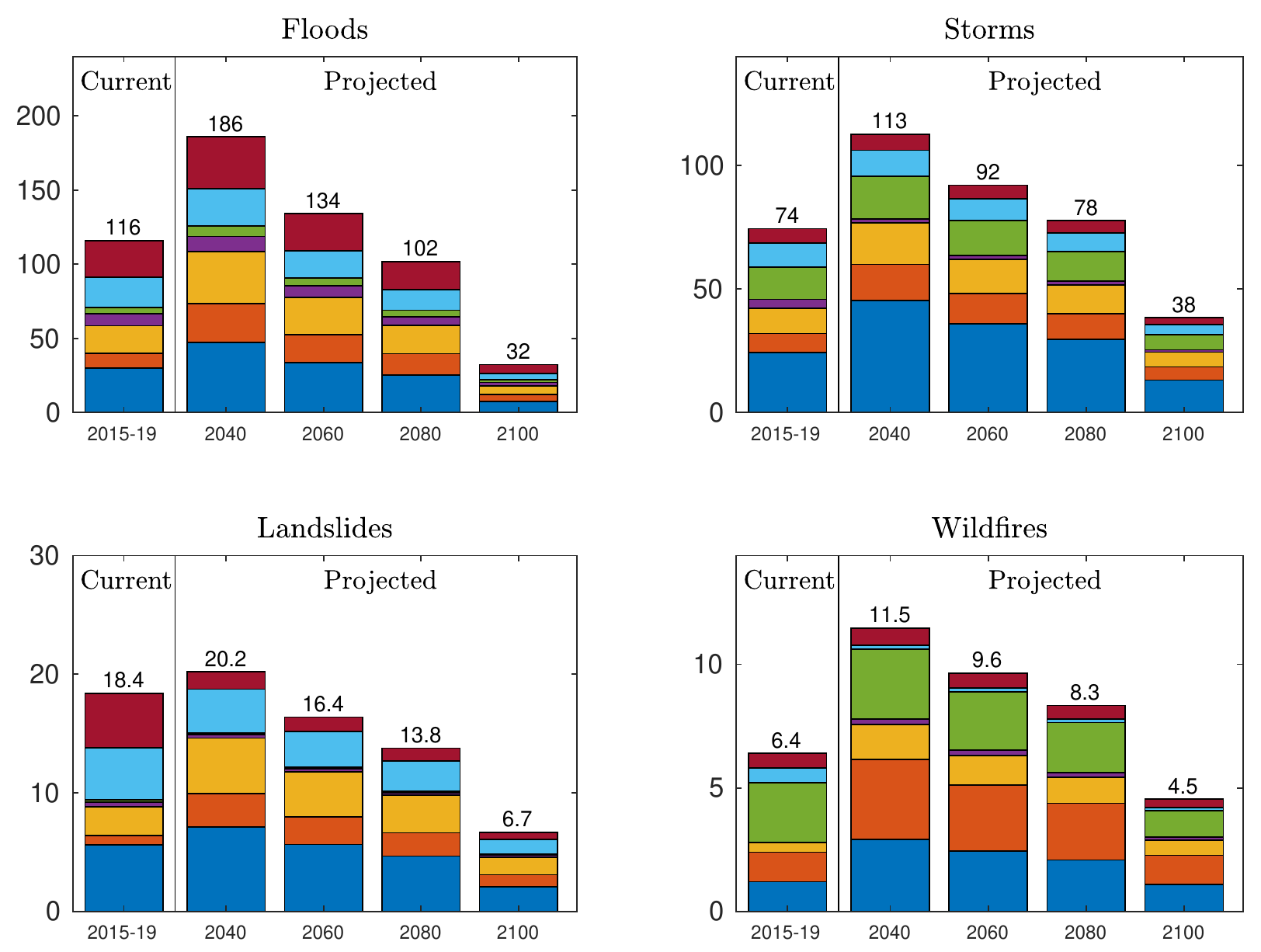} 
\includegraphics[scale=.95,trim= 2.25cm 8.5cm 2.25cm 7.25cm,clip]{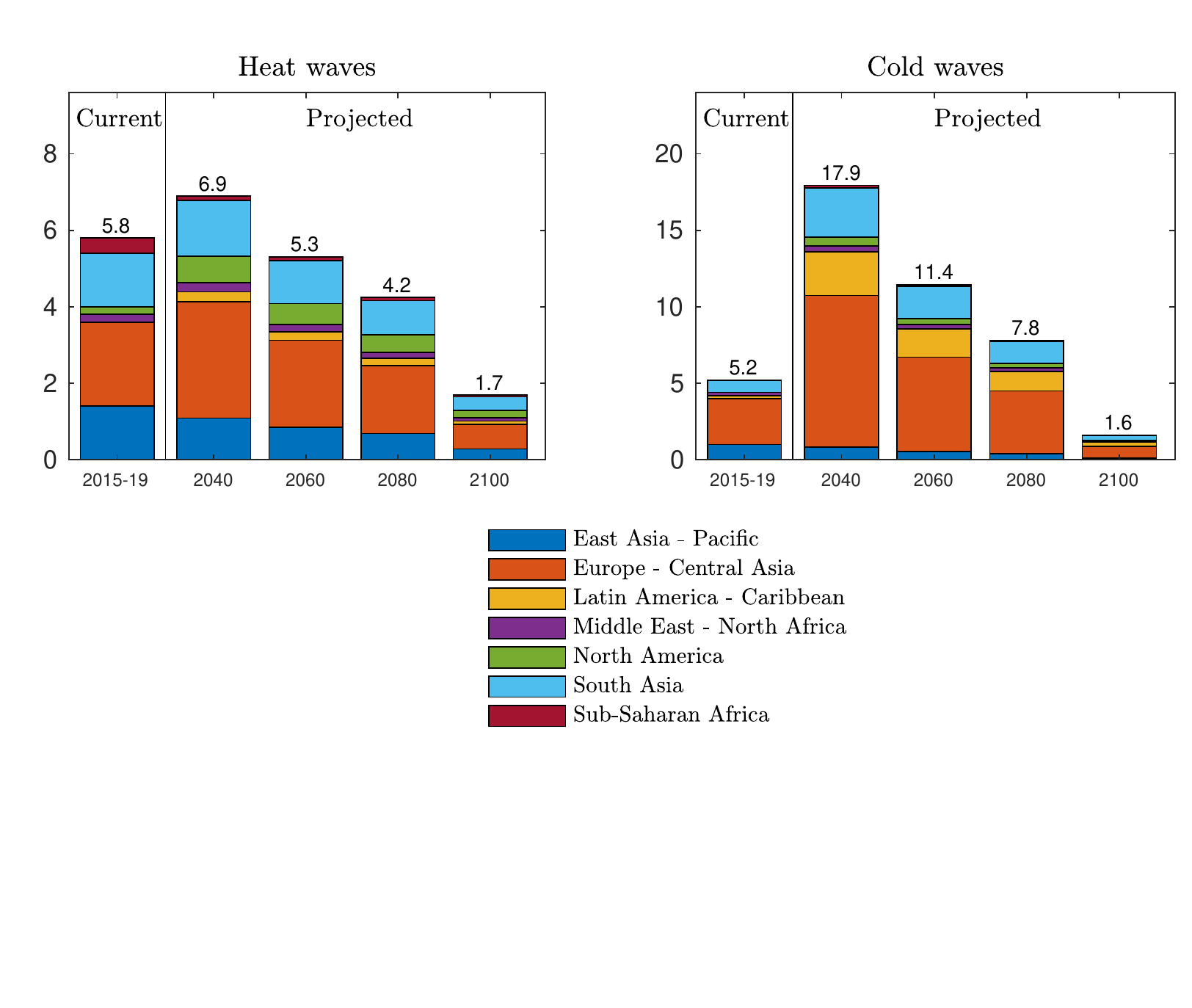} \end{center}

\end{figure}






\clearpage\newpage

\begin{figure}[h!]
\caption{Annual number of deaths -- Scenario SSP1} \label{fig: Number of disasters}

\vspace{-1.25cm}

\begin{center}
\bigskip
\includegraphics[scale=.95,trim= 2.25cm 8.5cm 2.25cm 9cm,clip]{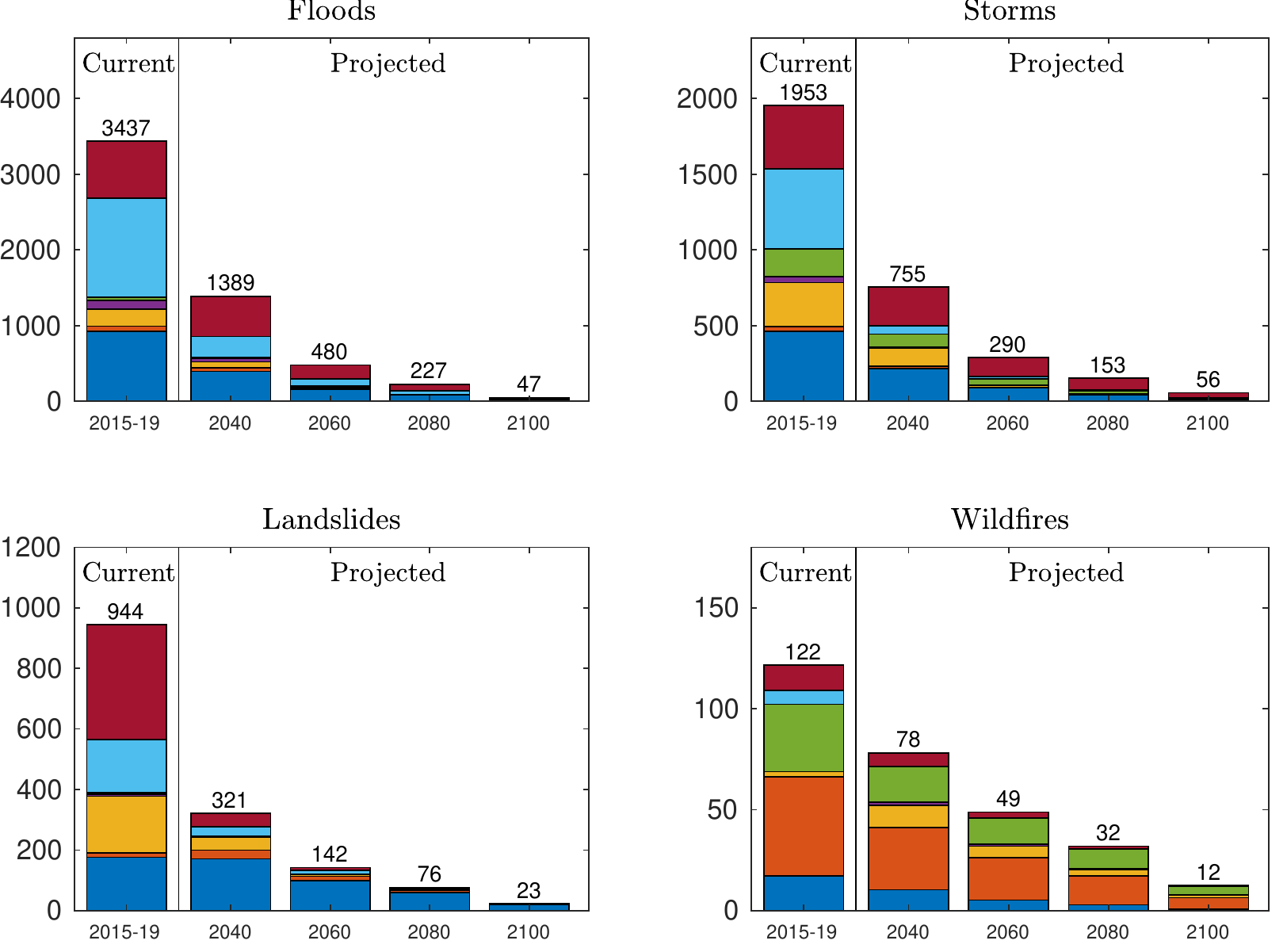} 
\includegraphics[scale=.95,trim= 2.25cm 8.5cm 2.25cm 7.25cm,clip]{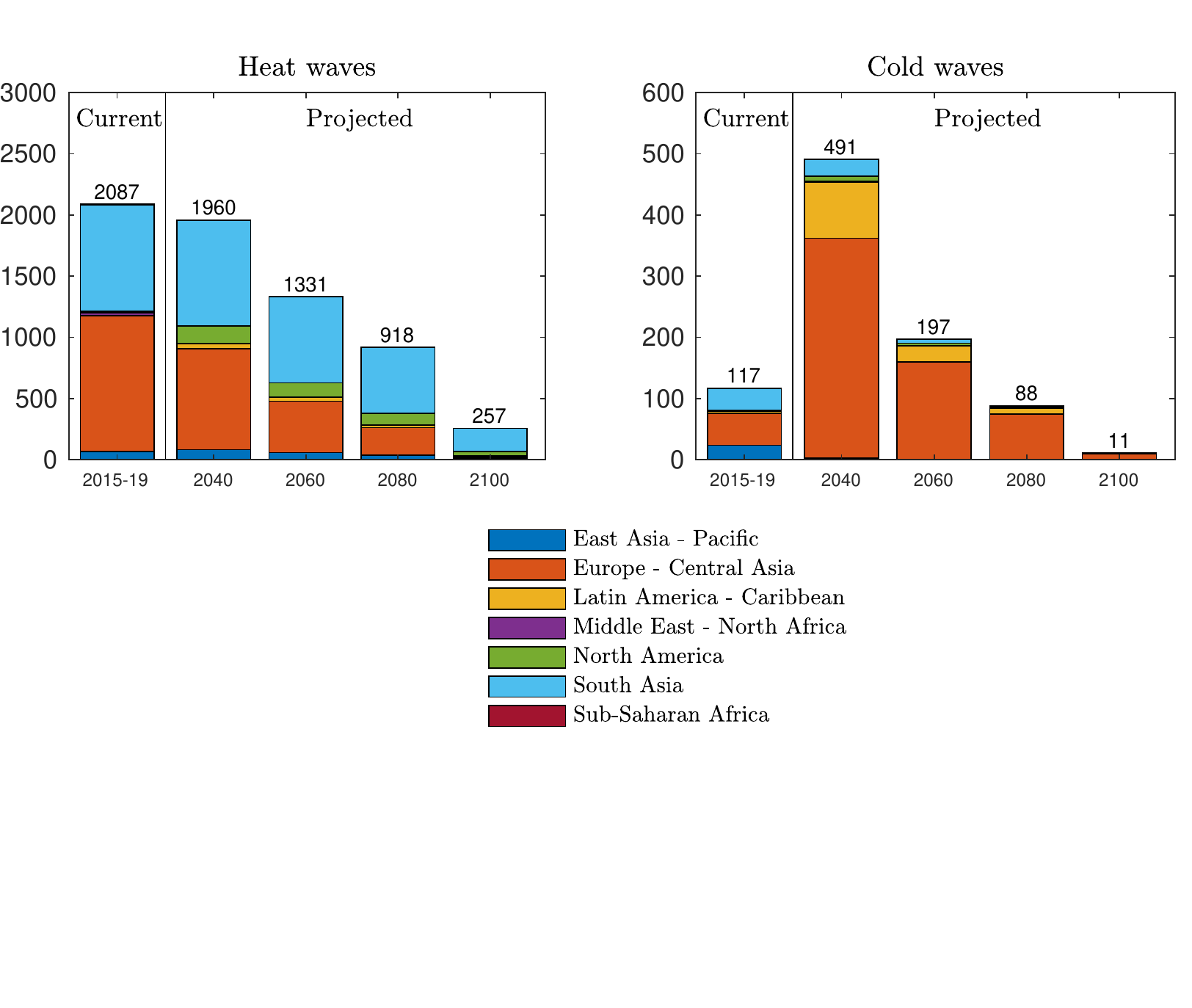} \end{center}

\end{figure}

\clearpage\newpage

\begin{figure}[h!]
\caption{Number of disasters per year -- Scenario SSP3} \label{fig: Number of disasters}

\vspace{-1.25cm}

\begin{center}
\bigskip
\includegraphics[scale=.95,trim= 2.25cm 8.5cm 2.25cm 9cm,clip]{NbDis_Sce1_Model_0_1.pdf} 
\includegraphics[scale=.95,trim= 2.25cm 8.5cm 2.25cm 7.25cm,clip]{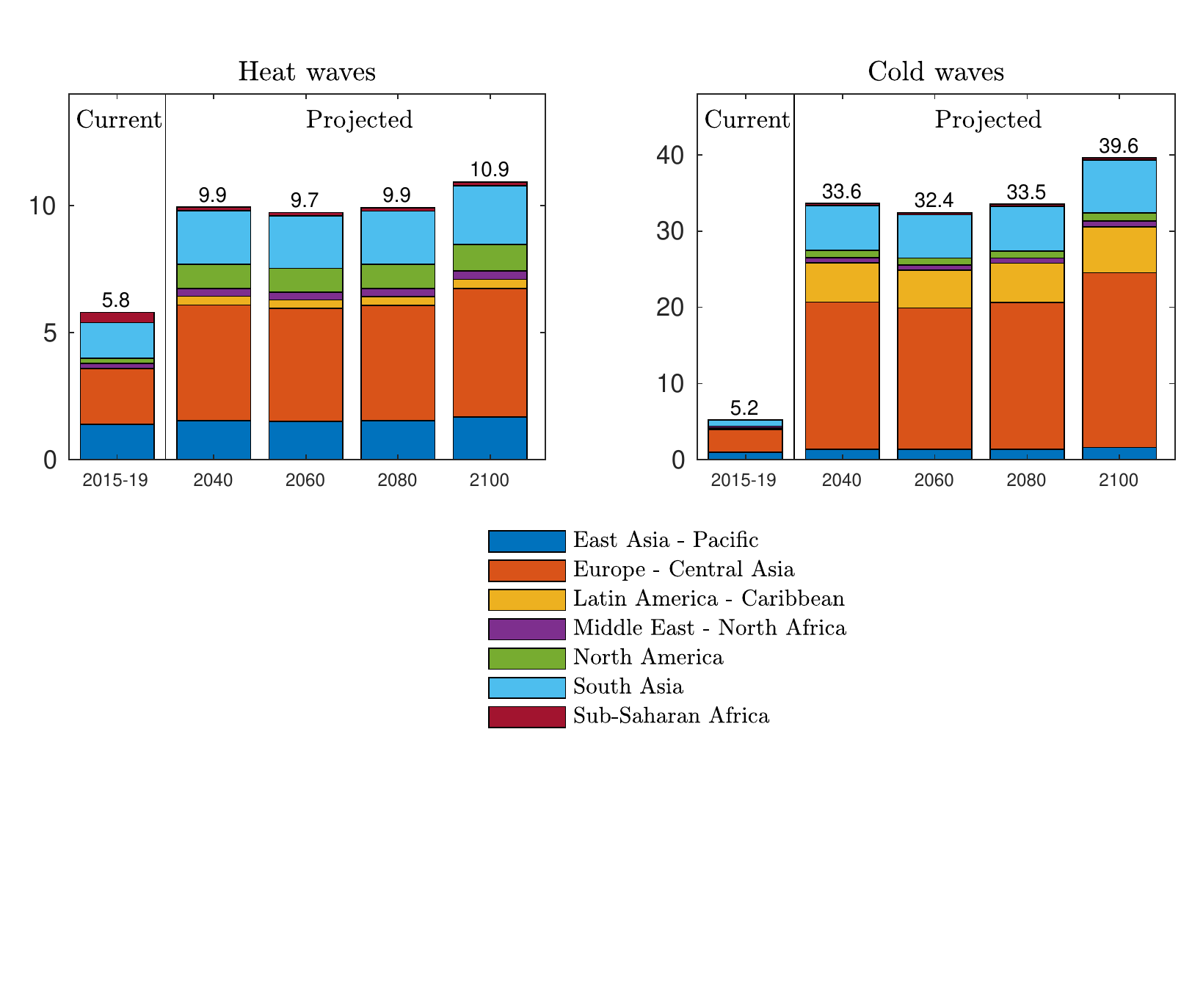} \end{center}

\end{figure}






\clearpage\newpage

\begin{figure}[h!]
\caption{Annual number of deaths -- Scenario SSP3} \label{fig: Number of disasters}

\vspace{-1.25cm}

\begin{center}
\bigskip
\includegraphics[scale=.95,trim= 2.25cm 8.5cm 2.25cm 9cm,clip]{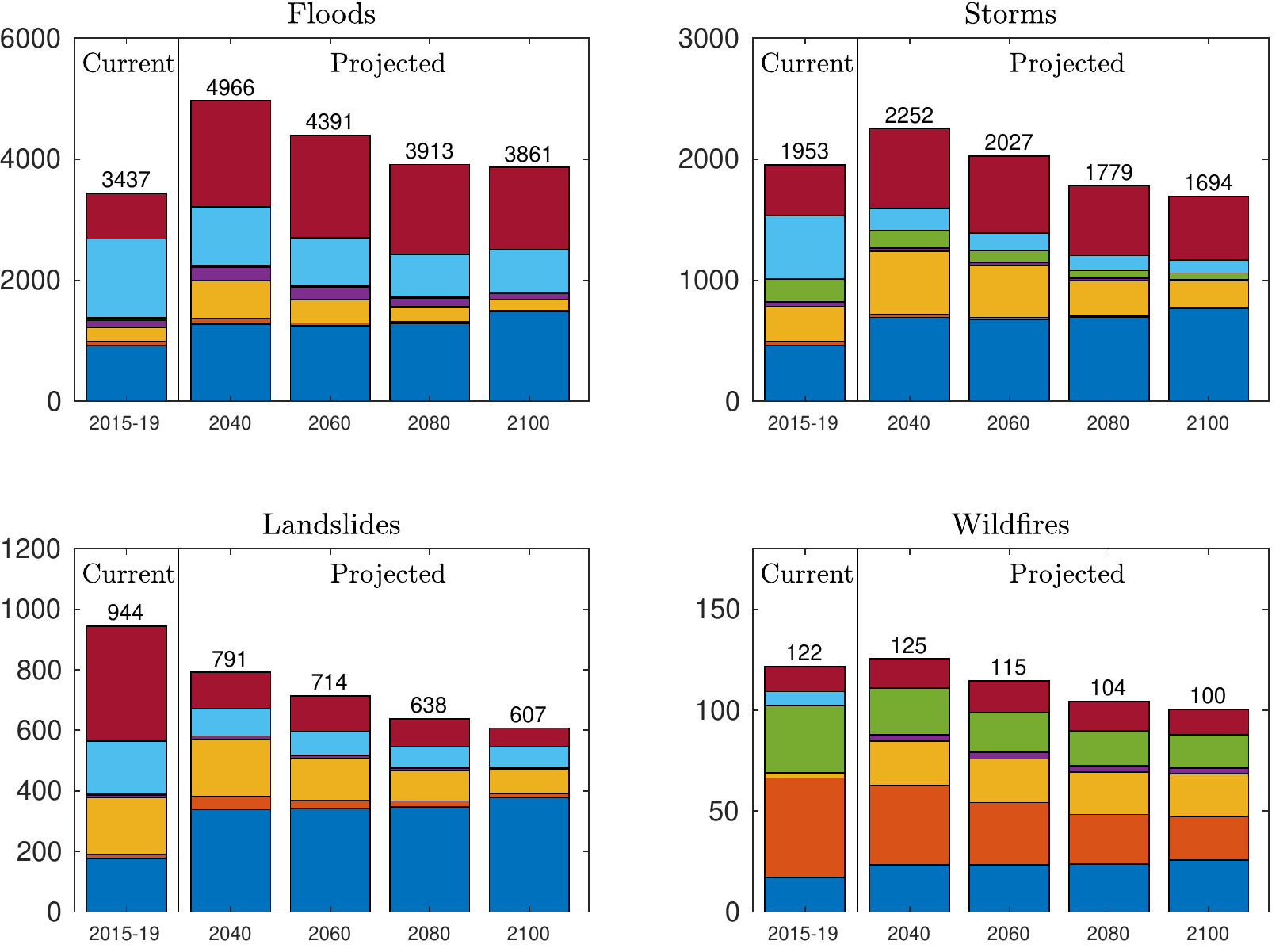} 
\includegraphics[scale=.95,trim= 2.25cm 8.5cm 2.25cm 7.25cm,clip]{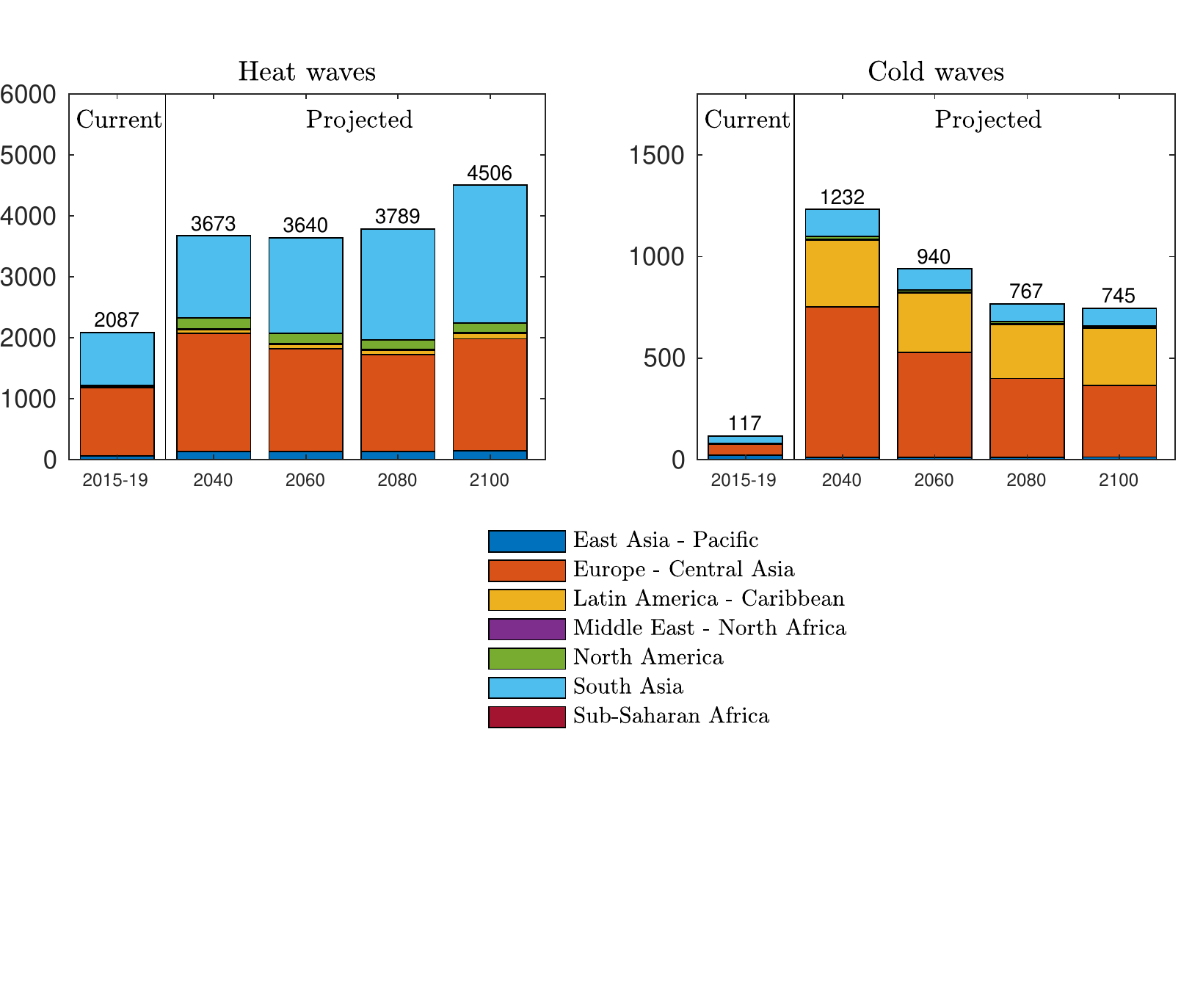} \end{center}

\end{figure}


\clearpage\newpage

\section{Bivariate Analysis}\label{app: Bivariate Analysis}

\begin{figure}[h!]
\caption{Correlation Analysis of the Number of Disasters } \label{fig: Correlation Number of disasters1}


\begin{center}
\bigskip

 \quad \quad Flood and Storm  \ \  \quad\quad\quad\quad\quad\quad\quad\quad\quad\quad\quad Flood and Landslide
\medskip

\includegraphics[scale=.45,trim= 0cm 0cm 0cm 0cm,clip]{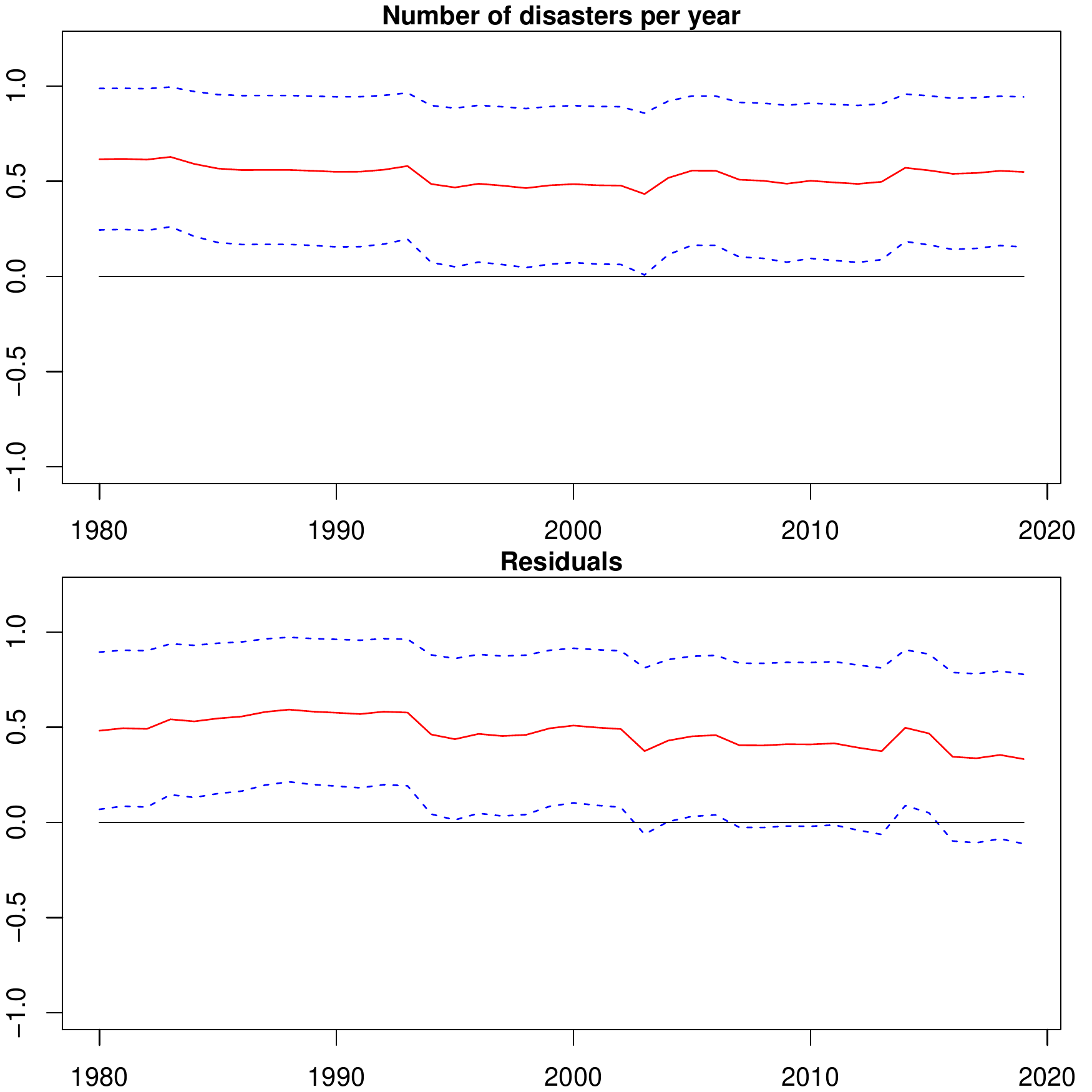} 
\includegraphics[scale=.45,trim= 0cm 0cm 0cm 0cm,clip]{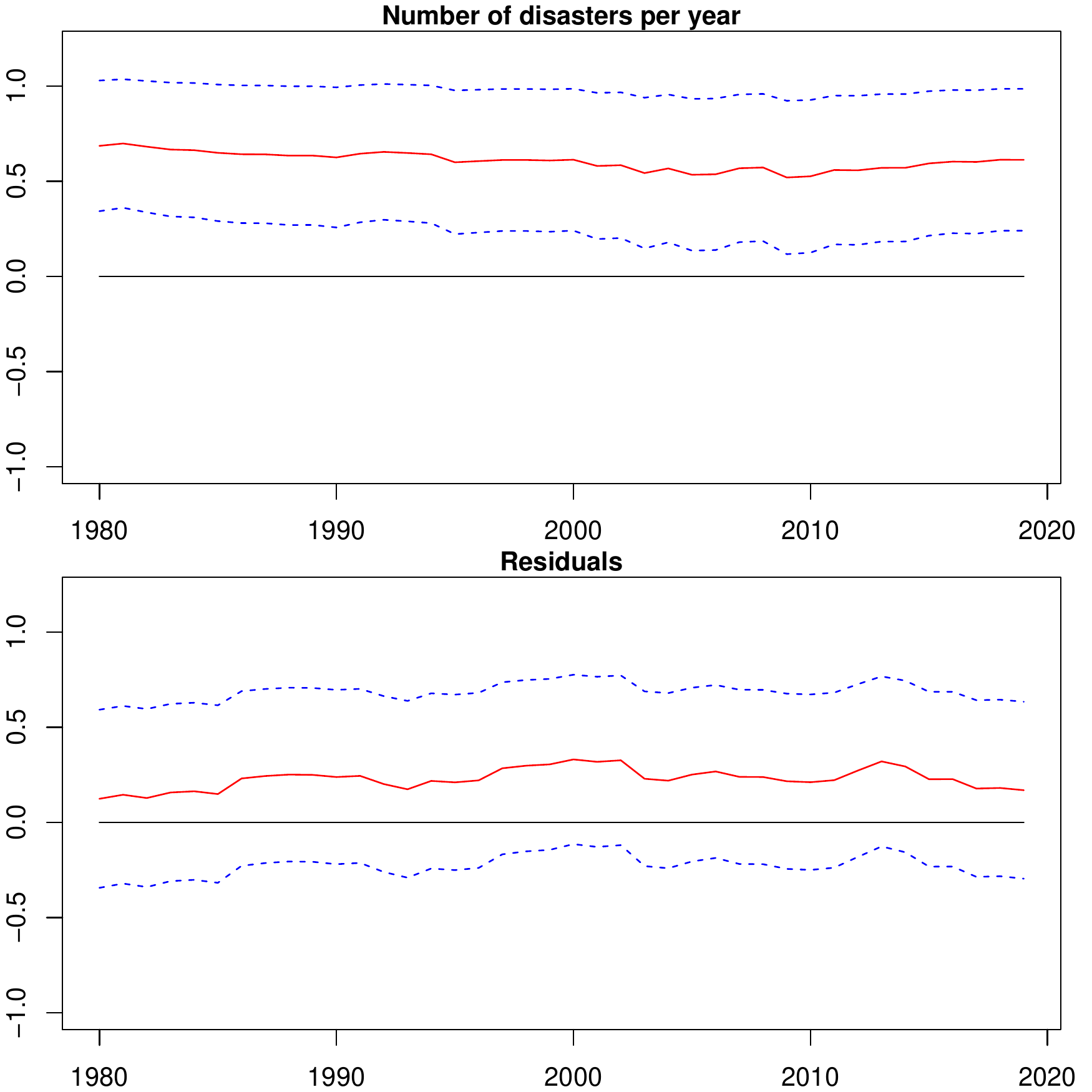} 

\bigskip
\quad Storm and Landslide
\medskip

\includegraphics[scale=.45,trim= 0cm 0cm 0cm 0cm,clip]{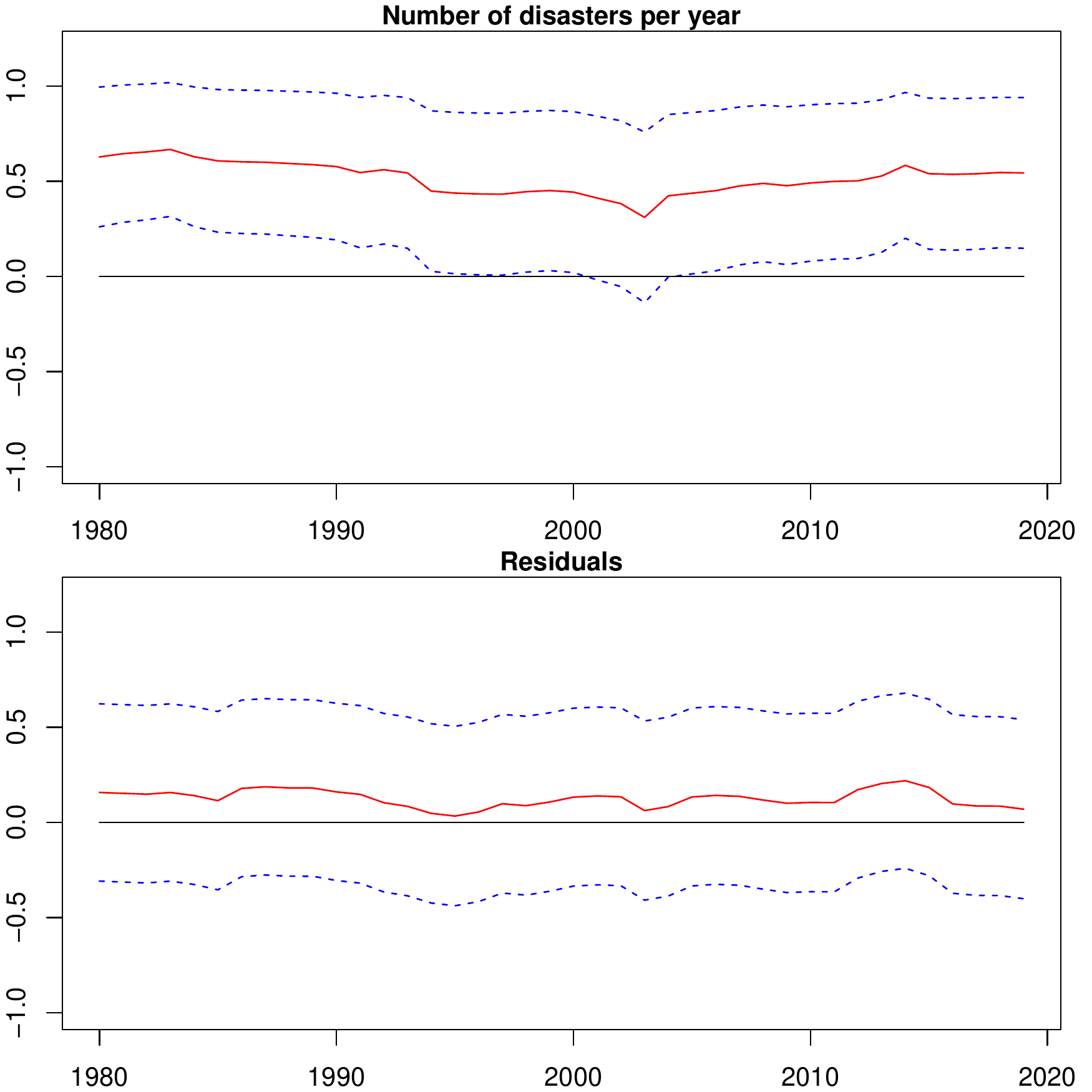} 
\end{center}

\end{figure}

\begin{figure}[h!]
\caption{Correlation Analysis of the Number of Disasters (Continued) } \label{fig: Correlation Number of disasters2}


\begin{center}
\bigskip

 Wildfire and Heat wave \ \quad\quad\quad\quad\quad\quad\quad\quad Wildfire and Cold wave
\medskip

\includegraphics[scale=.45,trim= 0cm 0cm 0cm 0cm,clip]{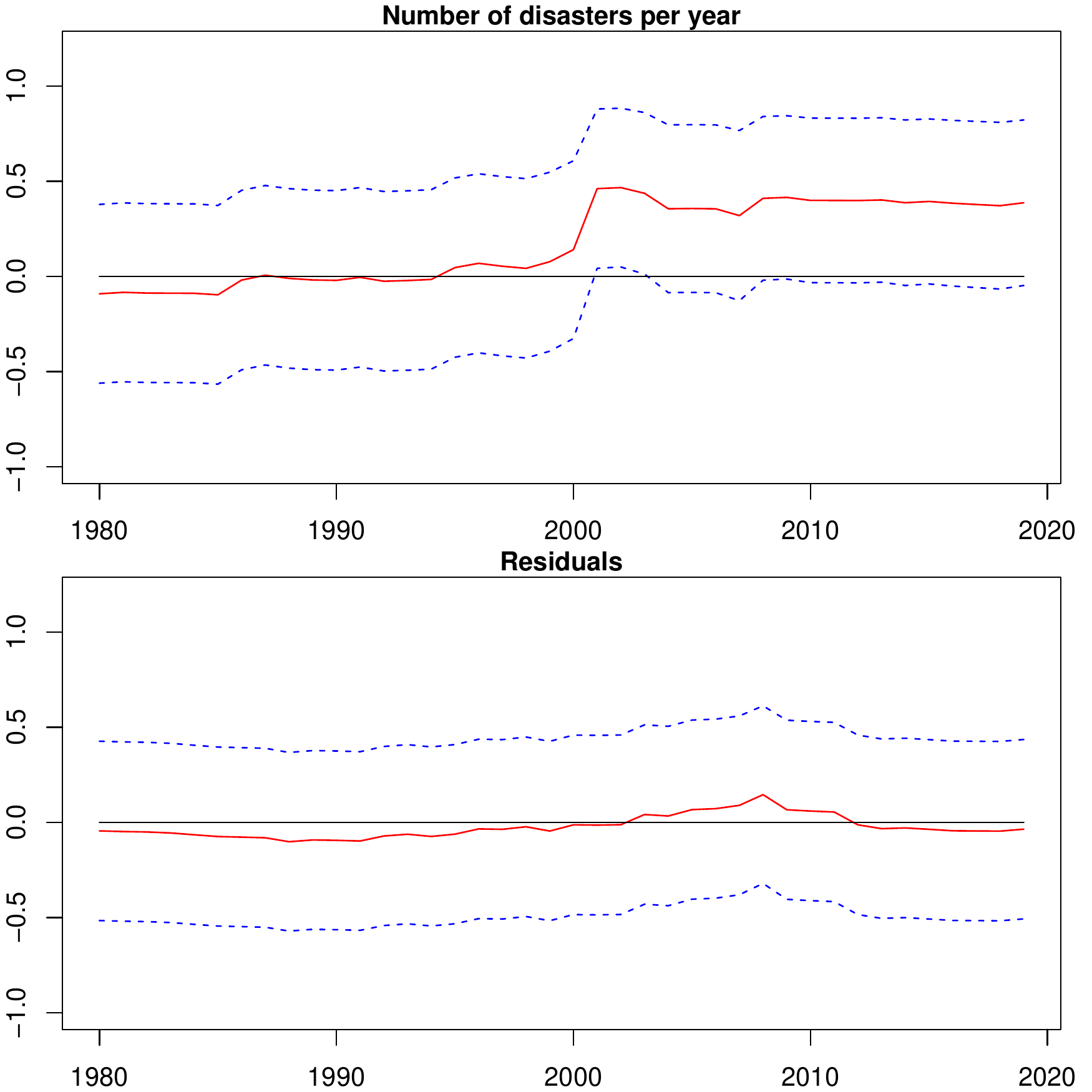} 
\includegraphics[scale=.45,trim= 0cm 0cm 0cm 0cm,clip]{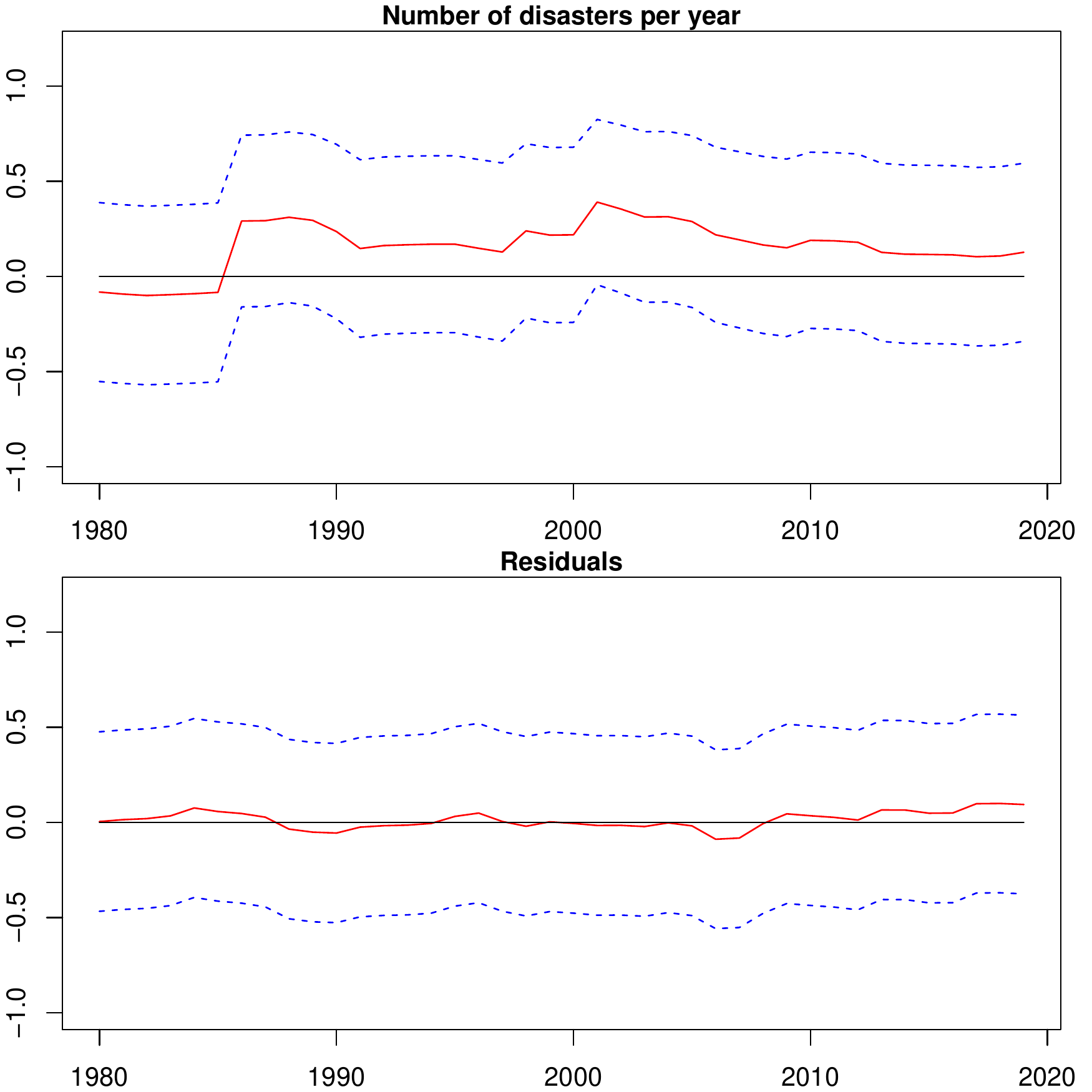} 

\bigskip
\quad Heat wave and Cold wave
\medskip

\includegraphics[scale=.45,trim= 0cm 0cm 0cm 0cm,clip]{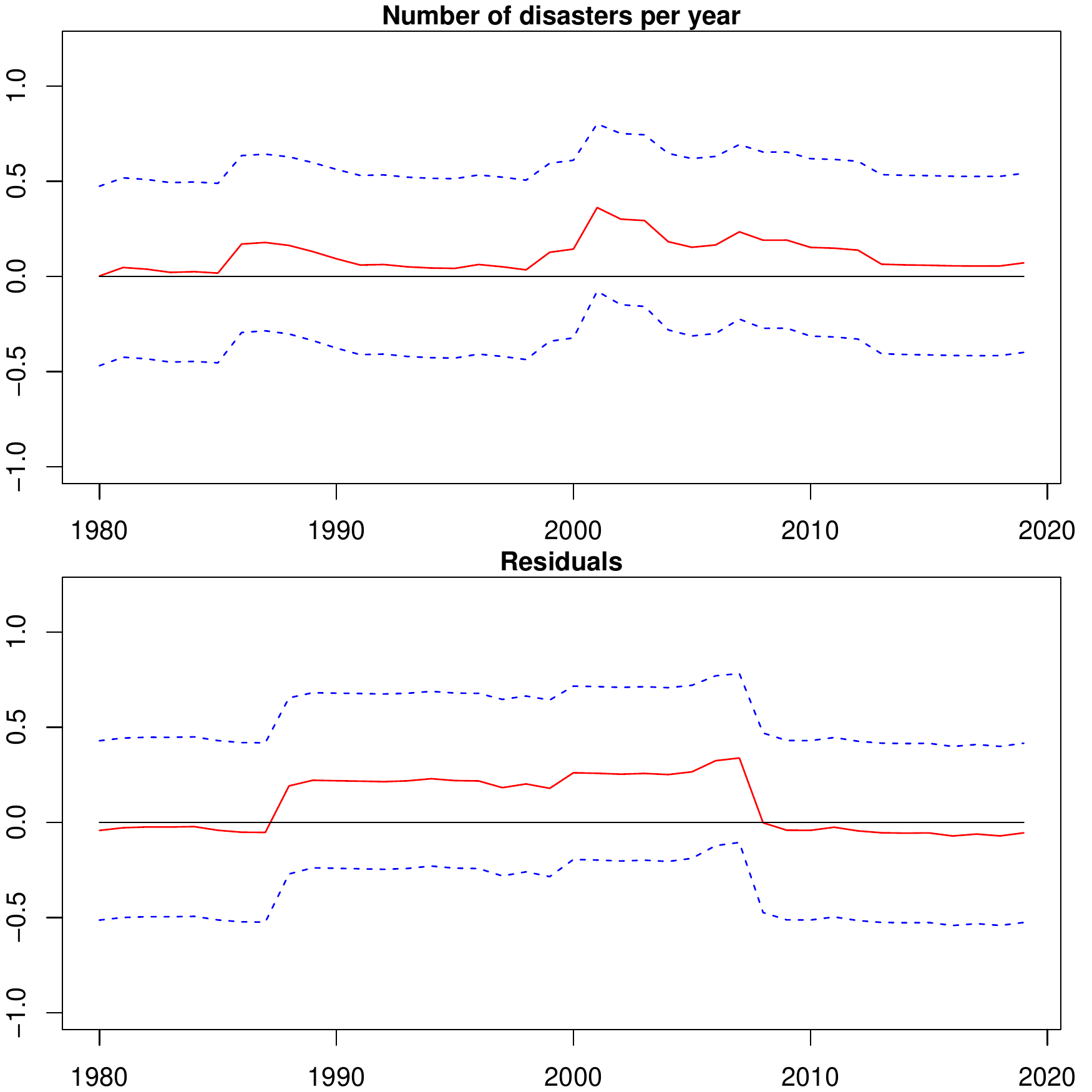} 
\end{center}

\end{figure}

In this Appendix, we report results on the bivariate analysis based on residuals of the univariate models. In order to obtain a time-varying dependence measure, the bivariate analysis is conducted over moving windows of $20$ years. For the frequency (Figures \ref{fig: Correlation Number of disasters1} and \ref{fig: Correlation Number of disasters2}), we compute the correlation between the annual number of disasters for pairs of disaster's types filtered by the univariate Poisson model. If the dependence between the two types of disasters is driven by the region or the aggregate CO$_2$ emission, the dependence should be filtered out by the model as these variables are used as covariates of the model. If the dependence is driven by other covariates, there should remain some residual dependence. The figures reveal that the only case with some residual correlation is the pair flood-storm, with a correlation of approximately $0.4$ between residuals of the number of floods and the number of storms. For other pairs, the residual correlation ranges between $0$ and $0.2$.

\begin{figure}[h!]
\caption{Dependence Analysis of the Number of Deaths } \label{fig: Correlation Number of deaths1}


\begin{center}
\includegraphics[scale=.45]{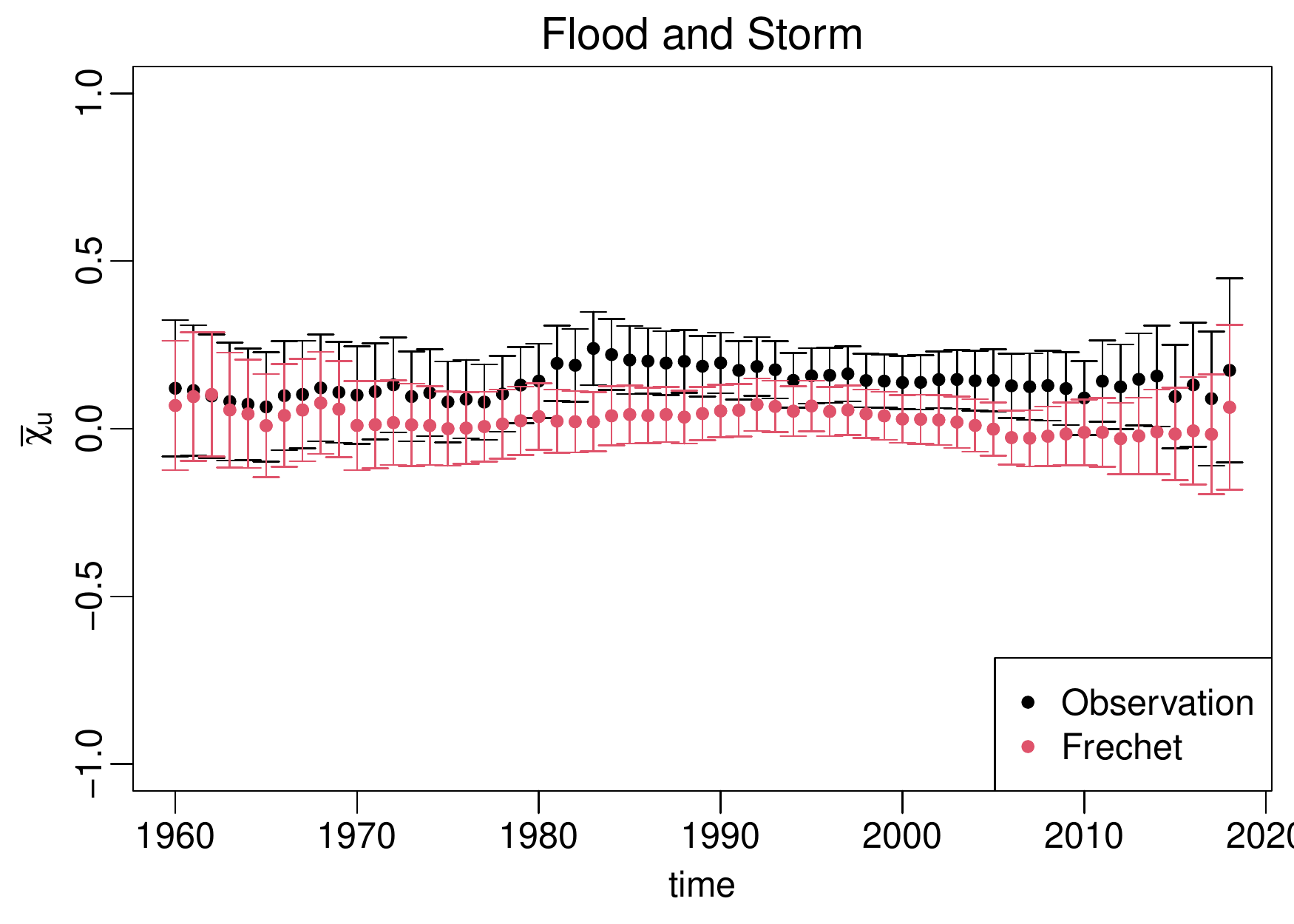} 
\includegraphics[scale=.45]{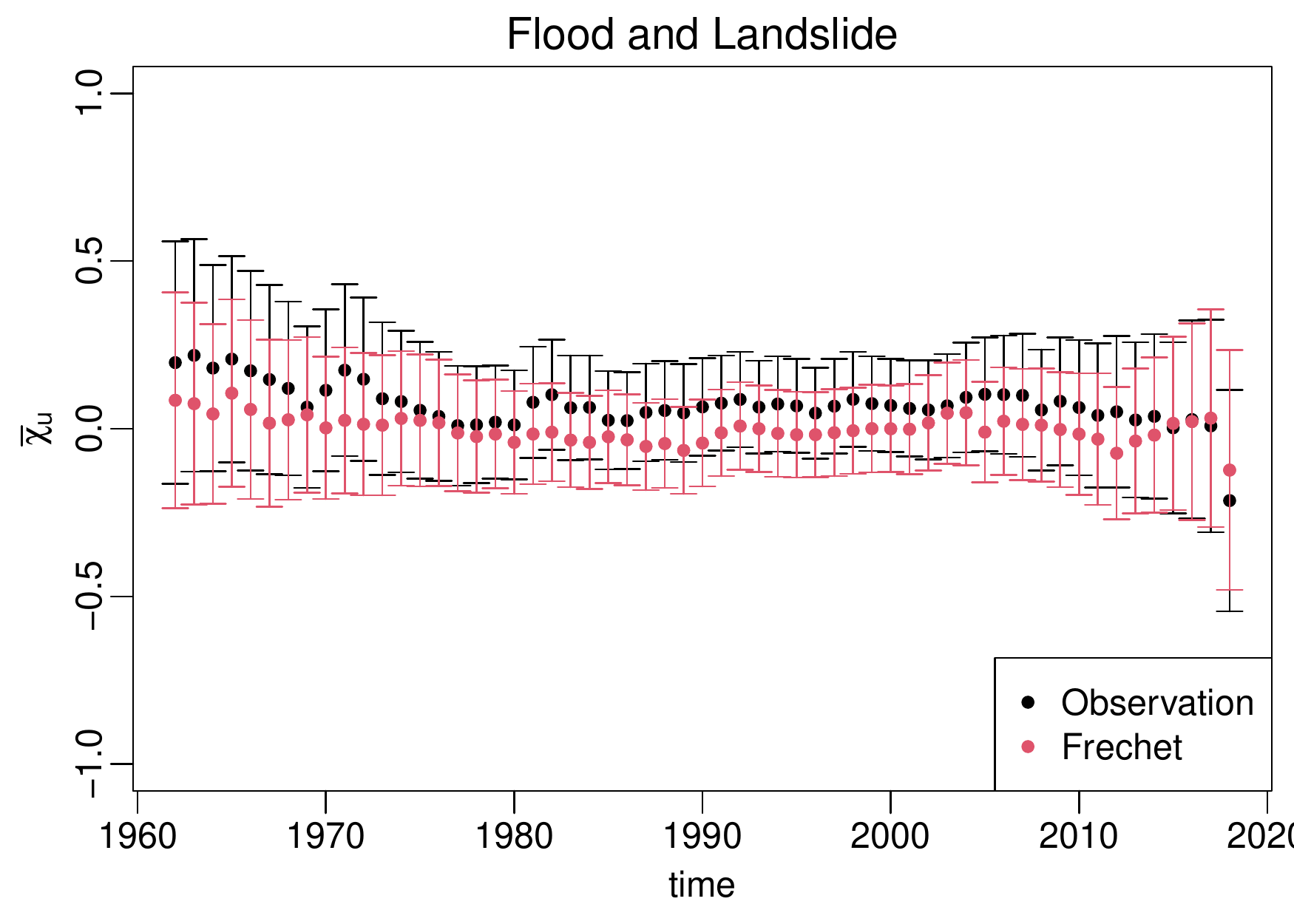} \\
\includegraphics[scale=.45]{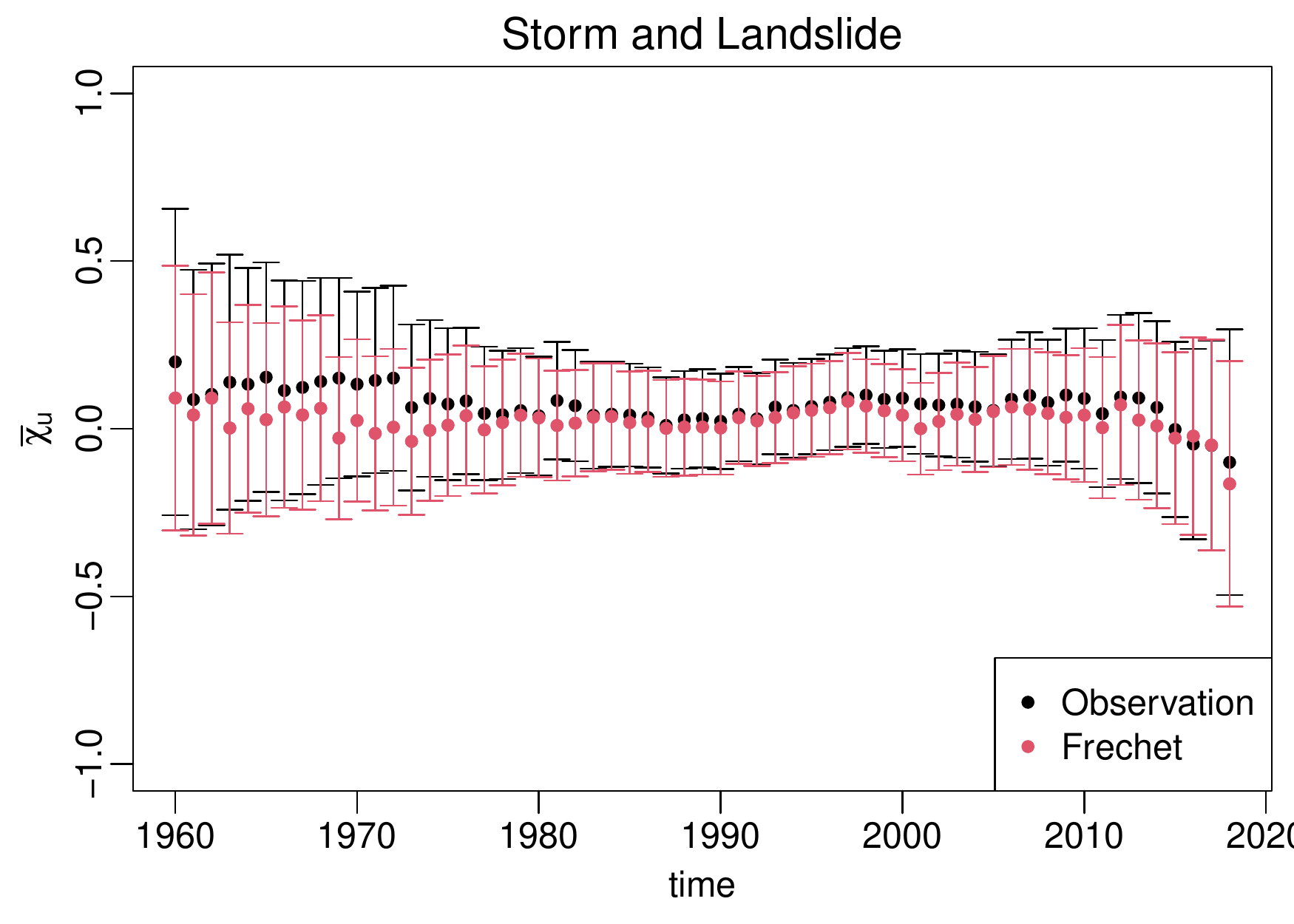} 

\end{center}

\end{figure}

\begin{figure}[h!]
\caption{Dependence Analysis of the Number of Deaths (Continued) } \label{fig: Correlation Number of deaths2}


\begin{center}
\includegraphics[scale=.45]{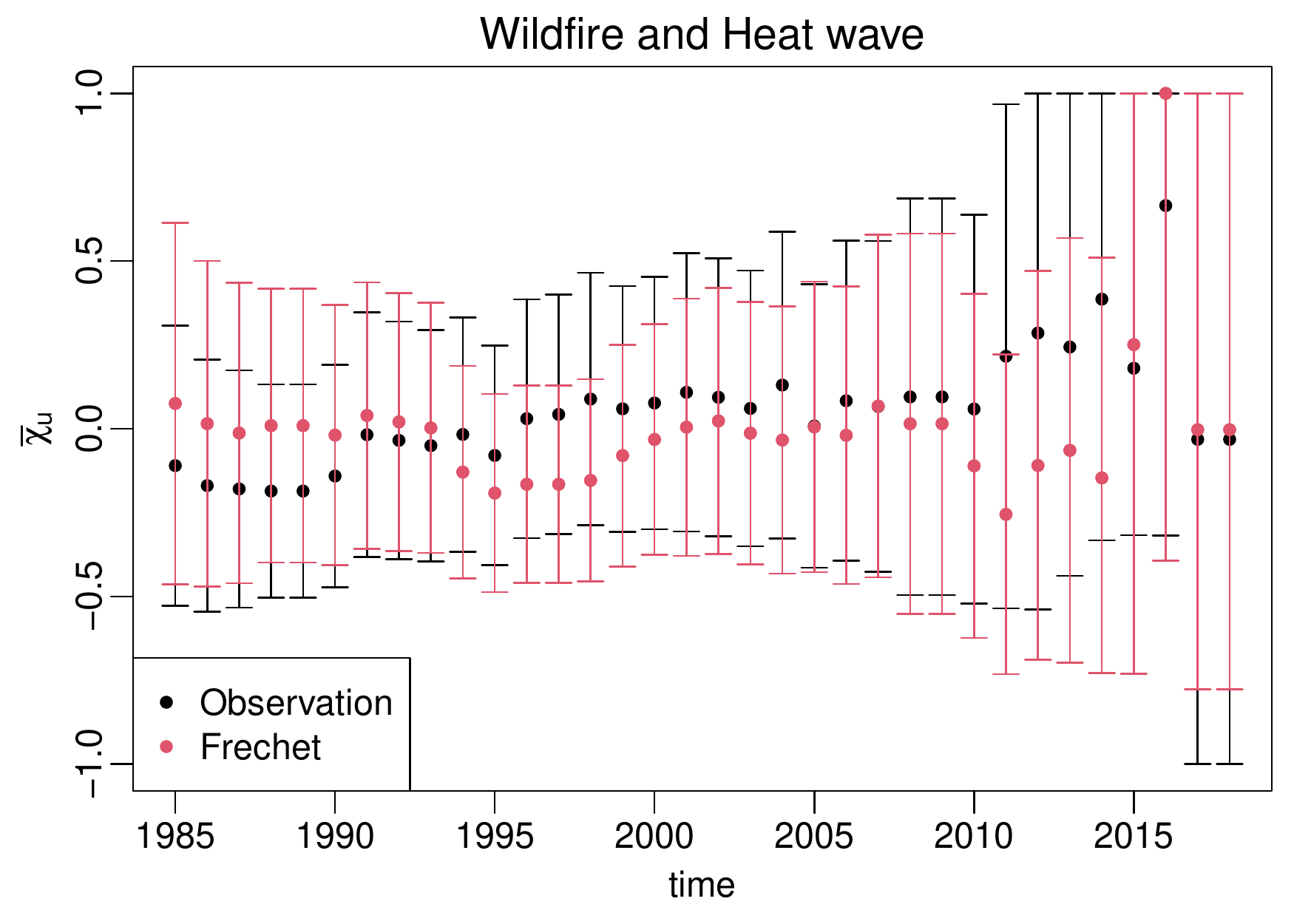} 
\includegraphics[scale=.45]{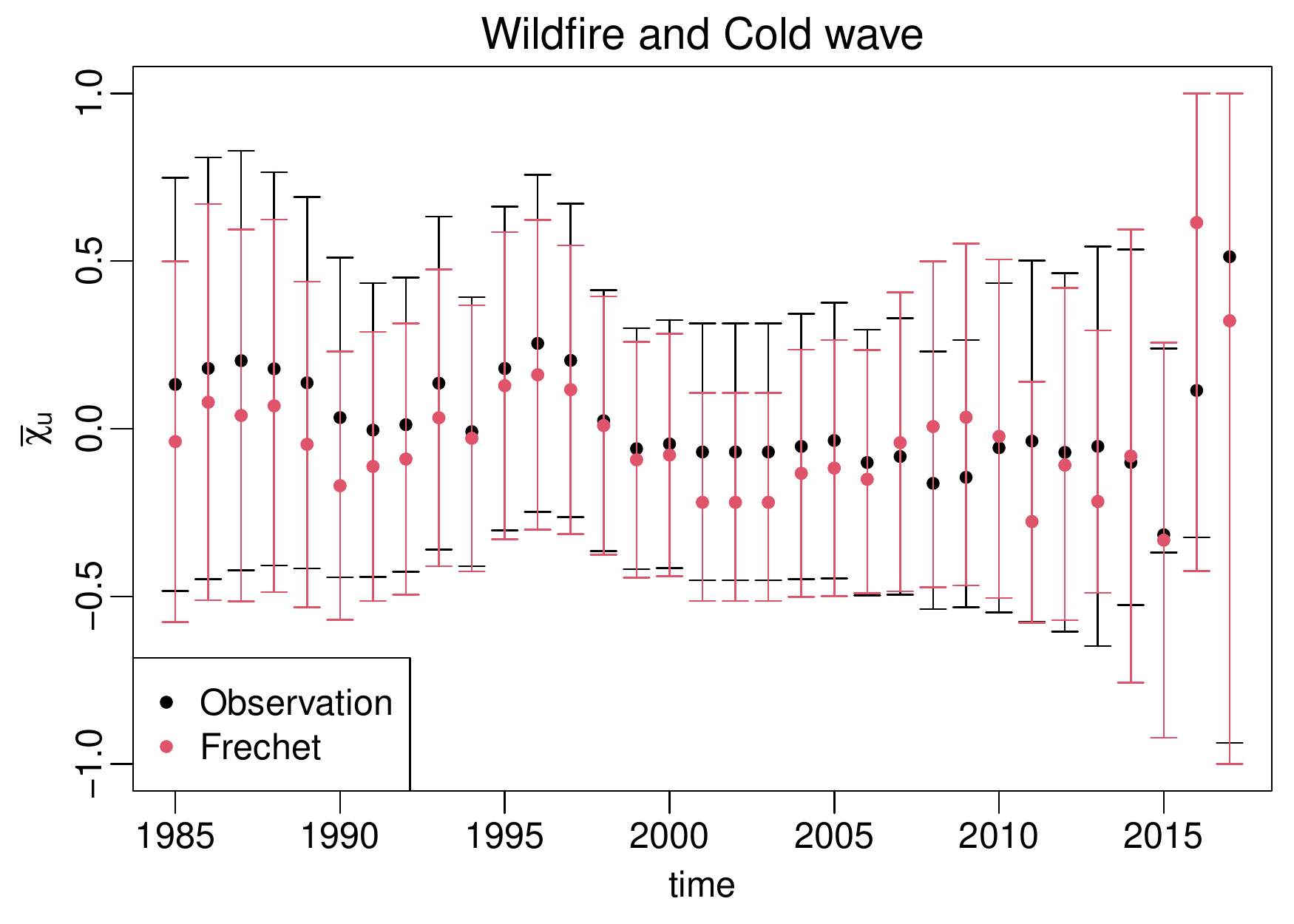} \\
\includegraphics[scale=.45]{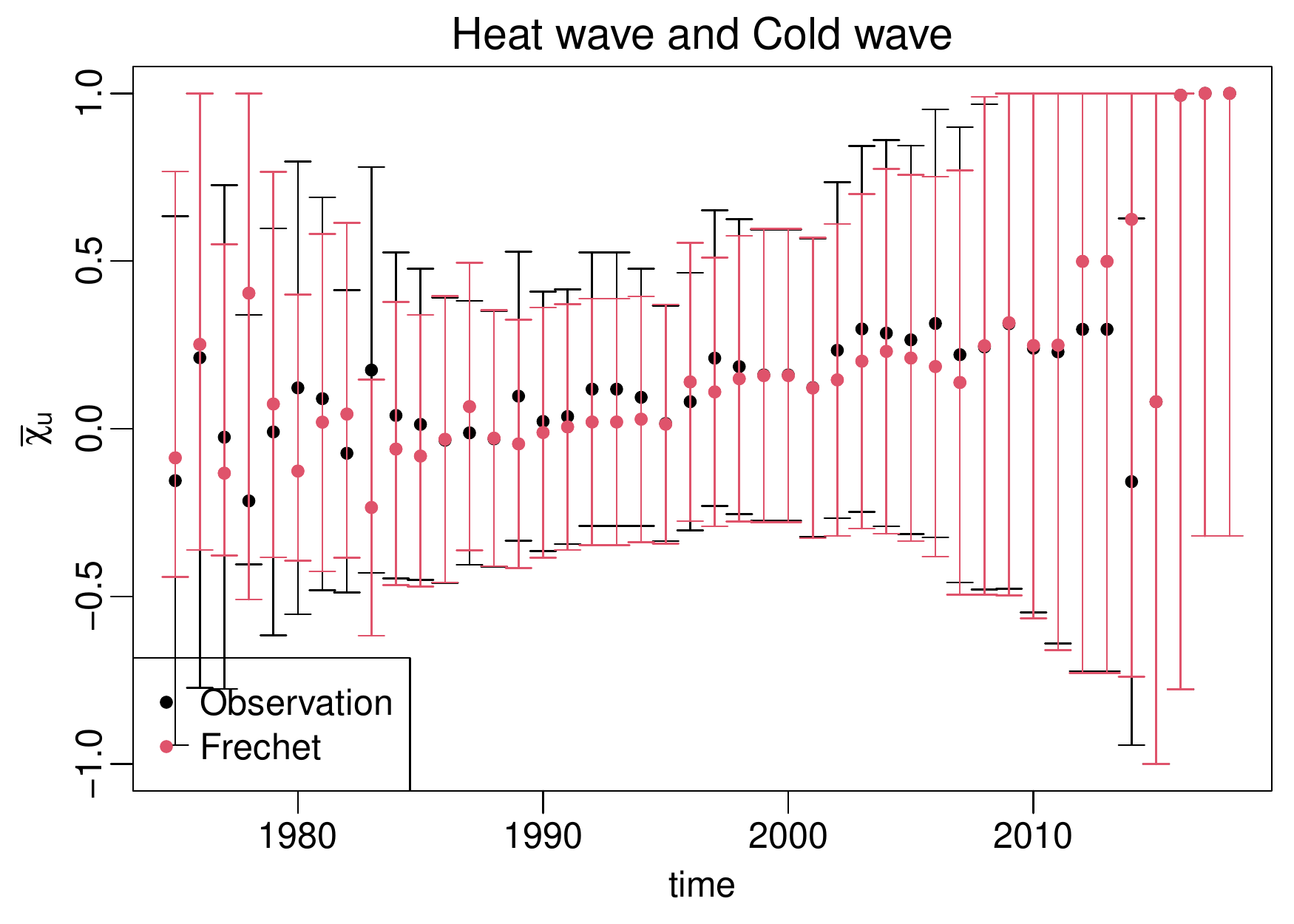} 

\end{center}

\end{figure}

We proceed in the same way to measure the dependence between the number of deaths per disaster for pairs of disasters' types filtered by the univariate GPD (Figures \ref{fig: Correlation Number of deaths1} and \ref{fig: Correlation Number of deaths2}). Again, if the dependence between the two types of disasters is driven by the region or the regional GDP, the dependence should be filtered out by the model. Here, dependence in a random vector $(X,Y)$ is measured through the tail quantity
\begin{equation*}
    \bar{\chi}(u) = \dfrac{2 \log \Pr \{ F_X(X)>u\}}{\log \Pr \{ F_X(X)>u, F_Y(Y)>u \}} \quad \text{for } 0 \leq u \leq 1,
\end{equation*}
where $F_X$ and $F_Y$ are the marginal distributions of $X$ and $Y$, respectively. Then, we compute empirical versions of $\bar{\chi}(u)$ for both the filtered observations, i.e., observations transformed (marginally) to the unit Fr\'echet scale using the fitted dynamic GPD, as well as the raw observations. Discrepancies between both estimates reflect a proportion of the dependence that is attributable to the exposure of the margins to the same covariates, included in the marginal models. The figures reveal that for both raw and filtered observations, dependence between the number of deaths is negligible (taking uncertainties into account). Note that a significant discrepancy is observed for the pair flood-storm between 1980 and 1990 where a significant positive dependence is observed at the raw scale but vanishes when treating for the marginal effects in the univariate GPD.


\clearpage\newpage

\clearpage\newpage

\clearpage\newpage

\clearpage\newpage

\end{document}